\documentclass[aps,prd,twocolumn,english,superscriptaddress]{revtex4-2}

\usepackage[T1]{fontenc}
\usepackage[utf8]{inputenc}
\usepackage{babel}

\usepackage{amsmath,amssymb,amsfonts,bm}
\usepackage{mathrsfs}
\usepackage{braket}
\usepackage{graphicx}
\usepackage{tikz} 
\usepackage{xcolor}
\usetikzlibrary{arrows.meta,calc,positioning,decorations.pathreplacing}

\begin{document}

\title{Modular Self-Duality, Symmetrized Relative Entropy, and Bogoliubov--Kubo--Mori Susceptibility in Quantum Field Theory}

\author{Rupak Chatterjee}
\affiliation{Department of Applied Physics, New York University, 2 MetroTech Center, Brooklyn, New York 11201, USA}
\email{Rupak.Chatterjee@nyu.edu}

\date{\today}

\begin{abstract}
We develop an operator-algebraic framework for modular self-duality,
symmetrized relative entropy, and Bogoliubov--Kubo--Mori (BKM)
susceptibility of local states in quantum field theory. In finite
dimensions, modular self-duality singles out fixed points at which a
state coincides with its modularly reflected partner. At such points,
the natural comparison functional is the symmetrized Umegaki relative
entropy. It vanishes at coincidence, and its Hessian is governed by the
BKM quantum Fisher information along the reflected tangent direction.
We then extend this fixed-point construction to the local type~III von
Neumann algebras that arise in quantum field theory. Here, a local state
is compared with the modular pullback of its commutant restriction, and
the intrinsic comparison functional is the symmetrized Araki relative
entropy. For sufficiently regular state deformations, the
fixed-localization Hessian at the self-dual point defines a type~III
BKM susceptibility. This coefficient is obtained by evaluating the BKM
bilinear form on the tangent selected by the modular pairing. Exact
coherent-state realizations are obtained for the free scalar field on
wedge algebras and for the chiral $U(1)$ current on half-line
algebras. In both examples, the comparison functional is exactly
quadratic in the deformation parameter, and the susceptibility
coefficients admit explicit boost-energy, stress-tensor, or half-line
integral representations. We also apply the construction to a ball-shaped boundary region in the
vacuum state of a conformal field theory with a semiclassical
anti-de Sitter (AdS) gravitational dual. For linear bulk coherent
excitations in an exact complementary-recovery model of this sector,
bulk--boundary modular intertwining identifies the modular partner with
the AdS--Rindler-reflected coherent state, and the symmetrized relative
entropy is exactly quadratic. Its Hessian is twice the bulk canonical
energy of the reflected-difference perturbation.
\end{abstract}

\maketitle

\section{INTRODUCTION}

A basic question in quantum theory is how to quantify state
distinguishability and second-order response under smooth deformations
of a state. In finite-dimensional quantum systems, this question is
usually formulated in terms of density matrices, relative entropy, and
quantum Fisher metrics. In local quantum field theory, however, the
analogous problem is subtler because observables localized in a
spacetime region are typically described by type~III von Neumann
algebras, for which intrinsic density matrices do not exist
\cite{Haag1,Witt2018}. One therefore needs a formulation of comparison
and response that does not rely on finite-dimensional matrix language.
The aim of this paper is to develop such a formulation using a common
fixed-point structure that treats type~I and type~III von Neumann
algebraic systems within a single framework. This operator-algebraic
perspective builds on earlier work by the author
\cite{ChatterjeePRD2024Concurrence,ChatterjeePRD2025HarvestCurv},
where modular self-duality and related entanglement questions were
studied in finite-dimensional and quantum-field-theoretic settings
using Tomita--Takesaki modular theory.

In the finite-dimensional part of the paper, we consider smooth
one-parameter families of states together with their modularly
reflected partners. Modular self-duality singles out special parameter
values at which a state coincides with its reflected partner. The
natural comparison functional near this locus is the symmetrized
Umegaki relative entropy. It vanishes at coincidence, differentiability
forces its first variation to vanish, and its Hessian is governed by
the Bogoliubov--Kubo--Mori (BKM) quantum Fisher information along the
reflected tangent direction
\cite{Petz1985Quasi,LesniewskiRuskai1999,Jencova2005}. In this sense,
the finite-dimensional type~I theory identifies a canonical BKM
susceptibility selected by modular self-duality and symmetrized
relative entropy. A related but distinct finite-dimensional connection
between quantum Fisher information and entanglement curvature has
recently been studied in two-qubit systems
\cite{SaleemEtAl2025}, although without the fixed-point comparison
principle used here.

The local quantum-field-theoretic problem starts from a family of
ambient states on the full theory and, for each localization region
$O_t$, compares its restriction to the local algebra
$\mathcal{A}(O_t)$ with the modular pullback of its commutant
restriction. The resulting intrinsic comparison functional is the
symmetrized Araki relative entropy
\cite{Araki1975,Araki1977}. At the self-dual locus, the paired local
states coincide, so the first variation again vanishes. At fixed
localization, the corresponding Hessian is twice the BKM
Fisher-information scalar obtained by evaluating the type~III BKM
bilinear form on the tangent direction selected by the modularly
paired family. This scalar is the local operator-algebraic BKM
susceptibility. The identification is made under the standard
second-order regularity assumption that Araki relative entropy admits
a diagonal BKM expansion for the local state curves under
consideration.

A central part of the paper is the construction of exact
quantum-field-theoretic examples. For coherent states of the free
scalar field on wedge algebras, known relative-entropy formulas
\cite{CasiniGrilloPontello2019,LongoEntropyDistribution2020,
LongoEntropyCoherent2019,BostelmannCadamuroDelVecchio2022,
HollandsChiralCFT2020,GarbarzPalau2023}, together with the
Bisognano--Wichmann identification of wedge modular flow
\cite{BisognanoWichmann1975,BisognanoWichmann1976}, yield an exact
two-parameter comparison surface and explicit local susceptibility
coefficients. These coefficients are determined by the
reflected-difference profile selected by the modular pairing and admit
boost-energy and stress-tensor representations. For the chiral
$U(1)$ current, the same construction leads to explicit half-line
formulas in the vacuum representation and to thermal half-line
formulas in which the entropy kernel is explicit while the reflected
profile is determined by the Borchers--Yngvason standard-subspace
modular conjugation.

Beyond these exact quantum-field-theoretic realizations, we examine
the physical interpretation of the construction in the
anti-de Sitter/conformal field theory (AdS/CFT) correspondence. The
AdS/CFT correspondence relates a conformal field theory (CFT) to a
quantum-gravitational theory in an asymptotically anti-de Sitter (AdS)
spacetime
\cite{Maldacena1998,GubserKlebanovPolyakov1998,Witten1998}. In this setting, a boundary region and a semiclassical reference state
determine a corresponding entanglement wedge in the bulk.  This wedge
is the domain of dependence of a bulk hypersurface bounded by the
boundary region and its associated extremal surface
\cite{RyuTakayanagi2006,DongHarlowWall}. For a vacuum boundary ball,
subregion relative entropy and quantum Fisher information are related
to canonical energy in the corresponding AdS--Rindler wedge
\cite{CasiniHuertaMyers,FaulknerEtAl,
LashkariVanRaamsdonk,JLMS,HollandsWald2013}. The purpose of the holographic part of the paper is to determine how,
for a chosen ambient perturbation, modular conjugation fixes its
reflected comparison and how the resulting reflected-difference
tangent is represented in the bulk.

This question is close to, but distinct from, holographic entanglement
asymmetry. In that construction, a reduced state is compared with its
average under an internal symmetry, and the leading relative-entropy
coefficient is represented by Fisher information and bulk canonical
energy \cite{BeniniGodetSingh2025}. The comparison considered here is
instead fixed by the vacuum modular conjugation. It exchanges the ball
algebra with its commutant, and the commutant restriction is pulled
back before the two local states are compared. Although both
quadratic coefficients admit a bulk canonical-energy representation
in their respective regimes, they resolve different notions of
asymmetry. The former probes internal-symmetry breaking, whereas the
latter probes modular-reflection asymmetry.

There are independent physical reasons for regarding the modular
pairing associated with the vacuum standard pair as distinguished. For
a Minkowski wedge, the Bisognano--Wichmann theorem identifies modular
flow with Lorentz boosts and modular conjugation with the corresponding
charge conjugation, parity, and time reversal (CPT)-type wedge
reflection
\cite{BisognanoWichmann1975,BisognanoWichmann1976}. For a ball in the
vacuum of a conformal field theory, the Casini--Huerta--Myers conformal
map sends the restricted vacuum to a thermal
Kubo--Martin--Schwinger (KMS) state on the hyperbolic cylinder, whose
holographic description is the AdS--Rindler wedge, equivalently the
exterior of the massless hyperbolic black hole
\cite{CasiniHuertaMyers}. In this thermal frame, for regular
exponential source deformations that admit a Kubo--Mori score
representation, the BKM quadratic form is the Kubo--Mori equilibrium
susceptibility
\cite{PetzToth1993,Bratteli1997}. The reflected-difference direction
therefore has the physical meaning of the response direction that
breaks the reflection symmetry of a geometrically selected
equilibrium reference state.

The operator-algebraic input needed to identify the reflected tangent
is supplied by the exact complementary-recovery setting of
~\cite{FaulknerConditional2020}. Standard complementary reconstructability
(\(c\)-reconstructability) preserves relative entropy and intertwines
the boundary and bulk modular conjugations. For the linear bulk
coherent family considered here, the modular partner is therefore the
AdS--Rindler-reflected coherent state, and the comparison depends on
the difference between the original bulk perturbation and its modular
reflection. The symmetrized relative entropy is exactly quadratic,
and its Hessian equals twice the canonical energy of this
reflected-difference perturbation.

The significance of these examples is therefore not that the
underlying relative-entropy formulas are new in isolation. Rather, the
point is that these known formulas become exact realizations of a
single modular comparison principle. For each localization region, the
vacuum modular conjugation selects a canonical comparison partner by
pulling back the commutant restriction to the original local algebra.
The self-dual point is then a coincidence point in the local state
space, and the first nonvanishing departure from this coincidence is
measured by a Bogoliubov--Kubo--Mori susceptibility. This
susceptibility is not associated with an arbitrary perturbation of the
state, but with the reflected-difference tangent selected by the
modularly paired family. It therefore depends simultaneously on the
physical state deformation and on the localization geometry through
the modular data of the local algebra. In this way, the wedge, chiral, and encoded AdS--Rindler coherent
families realize the type~III local version of the same fixed-point
mechanism that appears in the finite-dimensional type~I prototype.

\section{Modular Self-Duality and Symmetrized Relative Entropy in Finite-Dimensional (Type I) Systems}
\label{sec:FiniteTypeI}

\subsection{The need for operator algebras}

Consider a simple spin Hamiltonian on \(N\) sites,
\begin{equation}
H_N
=
-\sum_{i=1}^{N}\sigma_i^z ,
\label{eq:finite_spin_hamiltonian_intro}
\end{equation}
with noninteracting sites and the standard Pauli matrix algebra at each
site. If \(N\) is finite, the algebra of observables generated by the
Pauli spin operators is a finite-dimensional matrix algebra $\mathfrak A_N$. More
explicitly,
\begin{equation}
\mathfrak A_N
=
\bigotimes_{i=1}^{N}M_2(\mathbb C)
\cong
M_{2^N}(\mathbb C).
\label{eq:finite_spin_matrix_algebra_intro}
\end{equation}
Up to unitary equivalence, this algebra has a unique irreducible
representation. Every normal state is represented by a density matrix on
a fixed finite-dimensional Hilbert space, and state comparison can be
formulated directly in terms of density matrices and Umegaki relative
entropy.

This uniqueness no longer persists when the number of degrees of freedom
is infinite. For an infinite spin chain, a formal Hilbert space
containing all possible infinite spin configurations as mutually
orthogonal basis vectors would be nonseparable. Instead, one usually
works in a separable Hilbert space generated from a chosen reference
state. This choice of reference state selects a physical sector, and the
weak operator closure of the represented local spin algebra depends on
that sector.

Let \(\mathfrak A_0\) denote the unital \(^*\)-algebra generated by
finite linear combinations of finite products of local Pauli operators.
Let \(\pi_\uparrow\) be the representation of \(\mathfrak A_0\) generated
from the all-spin-up reference vector
\begin{equation}
\Omega_\uparrow
=
\ket{\uparrow}\otimes\ket{\uparrow}\otimes\ket{\uparrow}\otimes\cdots .
\label{eq:all_up_reference_intro}
\end{equation}
The finite-excitation domain is
\begin{equation}
\mathcal D_\uparrow
:=
\pi_\uparrow(\mathfrak A_0)\Omega_\uparrow .
\label{eq:up_sector_domain_intro}
\end{equation}
The corresponding Hilbert space is the completion
\begin{equation}
\mathcal H_\uparrow
:=
\overline{\mathcal D_\uparrow}.
\label{eq:up_sector_Hilbert_space_intro}
\end{equation}
Equivalently, \(\mathcal H_\uparrow\) has an orthonormal basis labelled
by finite subsets of spin sites that have been flipped relative to
\(\Omega_\uparrow\), while \(\mathcal D_\uparrow\) is the finite linear
span of these basis vectors. Thus \(\mathcal D_\uparrow\), not the full
completion \(\mathcal H_\uparrow\), consists of vectors differing from
\(\Omega_\uparrow\) at only finitely many sites.

The von Neumann algebra associated with the all-spin-up sector is the
weak operator topology (WOT) closure
\begin{equation}
\mathfrak M_\uparrow
:=
\pi_\uparrow(\mathfrak A_0)''
=
\overline{\pi_\uparrow(\mathfrak A_0)}^{\,{\rm WOT}}
\subset
\mathcal B(\mathcal H_\uparrow).
\label{eq:up_sector_von_neumann_intro}
\end{equation}
For the sequence used below, weak operator convergence means convergence
of all matrix elements: a sequence \(A_i\) converges weakly to \(A\) if
\begin{equation}
\langle \psi,A_i\phi\rangle_{\mathcal H_\uparrow}
\longrightarrow
\langle \psi,A\phi\rangle_{\mathcal H_\uparrow},
\qquad
\psi,\phi\in\mathcal H_\uparrow .
\label{eq:weak_operator_convergence_intro}
\end{equation}

The sector dependence can be seen explicitly. Define the spin raising
and lowering operators by
\begin{equation}
\sigma_i^+
=
\frac{1}{2}\left(\sigma_i^x+i\sigma_i^y\right),
\qquad
\sigma_i^-
=
\frac{1}{2}\left(\sigma_i^x-i\sigma_i^y\right),
\label{eq:spin_raising_lowering_intro}
\end{equation}
so that
\begin{equation}
\begin{array}{c}
\displaystyle
\sigma_i^+\ket{\downarrow}_i=\ket{\uparrow}_i,
\qquad
\sigma_i^+\ket{\uparrow}_i=0,
\\[1ex]
\displaystyle
\sigma_i^-\ket{\uparrow}_i=\ket{\downarrow}_i,
\qquad
\sigma_i^-\ket{\downarrow}_i=0.
\end{array}
\label{eq:spin_raising_lowering_action_intro}
\end{equation}
The projection onto a down spin at site \(i\) is therefore
\begin{equation}
P_i^\downarrow
:=
\sigma_i^-\sigma_i^+
=
\ket{\downarrow}_i\bra{\downarrow}_i .
\label{eq:down_spin_projection_intro}
\end{equation}
In the all-spin-up sector, every vector in the dense domain
\(\mathcal D_\uparrow\) has \(\ket{\uparrow}\) at all sufficiently far
sites. Hence, for \(\psi,\phi\in\mathcal D_\uparrow\),
\begin{equation}
\langle\psi,P_i^\downarrow\phi\rangle_{\mathcal H_\uparrow}
=
0
\label{eq:up_sector_matrix_element_zero_intro}
\end{equation}
for all sufficiently large \(i\). Since \(\|P_i^\downarrow\|\leq 1\),
density extends this weak convergence to all
\(\psi,\phi\in\mathcal H_\uparrow\). Moreover, because
\(P_i^\downarrow\) is a projection, the same density argument applied to
\(\|P_i^\downarrow\psi\|\) gives
\begin{equation}
P_i^\downarrow
\longrightarrow
0
\qquad
\text{strongly, and hence weakly, on } \mathcal H_\uparrow .
\label{eq:up_sector_weak_limit_intro}
\end{equation}

If instead one starts from the all-spin-down reference vector
\begin{equation}
\Omega_\downarrow
=
\ket{\downarrow}\otimes\ket{\downarrow}\otimes\ket{\downarrow}\otimes\cdots ,
\label{eq:all_down_reference_intro}
\end{equation}
and lets \(\pi_\downarrow\) denote the corresponding representation of
\(\mathfrak A_0\), then the finite-excitation domain and Hilbert space
are
\begin{equation}
\mathcal D_\downarrow
:=
\pi_\downarrow(\mathfrak A_0)\Omega_\downarrow,
\qquad
\mathcal H_\downarrow
:=
\overline{\mathcal D_\downarrow}.
\label{eq:down_sector_Hilbert_space_intro}
\end{equation}
The associated von Neumann algebra is
\begin{equation}
\mathfrak M_\downarrow
:=
\pi_\downarrow(\mathfrak A_0)''
=
\overline{\pi_\downarrow(\mathfrak A_0)}^{\,{\rm WOT}}
\subset
\mathcal B(\mathcal H_\downarrow).
\label{eq:down_sector_von_neumann_intro}
\end{equation}
In this sector, every vector in \(\mathcal D_\downarrow\) has
\(\ket{\downarrow}\) at all sufficiently far sites. Therefore, for
\(\psi,\phi\in\mathcal D_\downarrow\),
\begin{equation}
\langle\psi,P_i^\downarrow\phi\rangle_{\mathcal H_\downarrow}
=
\langle\psi,\phi\rangle_{\mathcal H_\downarrow}
\label{eq:down_sector_matrix_element_intro}
\end{equation}
for all sufficiently large \(i\). Again by density and uniform
boundedness,
\begin{equation}
P_i^\downarrow
\longrightarrow
\mathbf 1
\qquad
\text{weakly on } \mathcal H_\downarrow .
\label{eq:down_sector_weak_limit_intro}
\end{equation}
Since \(\mathbf 1-P_i^\downarrow\) is the projection onto an up spin at
site \(i\), the same density argument gives strong convergence,
\begin{equation}
P_i^\downarrow
\longrightarrow
\mathbf 1
\qquad
\text{strongly on } \mathcal H_\downarrow .
\label{eq:down_sector_strong_limit_intro}
\end{equation}

Thus the same sequence of finite-site observables has different weak
limits in the two representations:
\begin{equation}
\begin{array}{c}
\displaystyle
P_i^\downarrow\longrightarrow 0
\quad
\text{in the }\Omega_\uparrow\text{ sector},
\\[1ex]
\displaystyle
P_i^\downarrow\longrightarrow \mathbf 1
\quad
\text{in the }\Omega_\downarrow\text{ sector}.
\end{array}
\label{eq:different_weak_limits_intro}
\end{equation}
This illustrates the basic need for the operator-algebraic point of view. In an infinite
system, the represented algebra and its weak closure retain information
about the physical sector or state used to realize the observables. The
example is not meant to assert that these spin-chain sector closures are
type~III; it is meant to show why representations and weak closures
become essential once infinitely many degrees of freedom are present.

This operator-algebraic viewpoint treats observables as elements of a
\(C^*\)-algebra and states as positive normalized linear functionals on
that algebra. If \(\mathfrak A\) is the algebra of observables, then a
state \(\omega\) assigns expectation values by the pairing
\begin{equation}
A
\longmapsto
\omega(A),
\qquad
A\in\mathfrak A .
\label{eq:state_functional_pairing_intro}
\end{equation}
Dynamics is described algebraically by a one-parameter group of
\(C^*\)-automorphisms
\begin{equation}
\begin{array}{c}
\alpha_t:\mathfrak A\to\mathfrak A,
\\\\
\alpha_t(AB)=\alpha_t(A)\alpha_t(B),
\\\\
\alpha_t(A^*)=\alpha_t(A)^* .
\end{array}
\label{eq:Cstar_dynamics_intro}
\end{equation}
This point of view separates the abstract observables from the particular
Hilbert-space representation selected by a state.

The same structural issue appears in quantum field theory. One first
specifies an abstract algebra of local observables before choosing a
particular Hilbert-space representation. Let \(\mathfrak A\) denote the
abstract \(C^*\)-algebra of observables, and let
\(\mathfrak A(\mathcal O)\subset\mathfrak A\) denote the abstract local
algebra associated with a spacetime region \(\mathcal O\). A state on
\(\mathfrak A\) is again a positive normalized linear functional.
Choosing a state determines a Hilbert-space representation through the
Gelfand--Naimark--Segal (GNS) construction. For the vacuum state \(\omega_0\),
this gives a triple
\begin{equation}
(\pi_0,\mathcal H,\Omega).
\label{eq:GNS_vacuum_triple_intro}
\end{equation}
Here \(\mathcal H\) is the vacuum Hilbert space,
\(\mathcal B(\mathcal H)\) is the algebra of bounded linear operators on
\(\mathcal H\), and
\begin{equation}
\pi_0:\mathfrak A\longrightarrow \mathcal B(\mathcal H)
\label{eq:GNS_vacuum_representation_intro}
\end{equation}
is the representation that turns each abstract observable into a bounded
operator on \(\mathcal H\). The vector \(\Omega\in\mathcal H\) is the
vacuum vector. It represents the vacuum state by
\begin{equation}
\omega_0(A)
=
\langle\Omega,\pi_0(A)\Omega\rangle,
\qquad
A\in\mathfrak A,
\label{eq:GNS_vacuum_vector_state_intro}
\end{equation}
and is cyclic for the represented algebra,
\begin{equation}
\overline{\pi_0(\mathfrak A)\Omega}
=
\mathcal H .
\label{eq:GNS_vacuum_cyclicity_intro}
\end{equation}

For each region \(\mathcal O\), the represented local von Neumann
algebra is
\begin{equation}
\mathcal A(\mathcal O)
:=
\pi_0\bigl(\mathfrak A(\mathcal O)\bigr)''
\subset
\mathcal B(\mathcal H).
\label{eq:represented_local_algebra_intro}
\end{equation}
Equivalently, \(\mathcal A(\mathcal O)\) is the weak operator closure of
\(\pi_0(\mathfrak A(\mathcal O))\). The double-prime notation denotes
the bicommutant, and the equivalence with weak closure is the content of
von Neumann's bicommutant theorem (see Sec.~\ref{sec:vN_TT} below). The result is the concrete von
Neumann algebra generated by the bounded local observables associated
with \(\mathcal O\) in the vacuum sector.

The local algebras obtained in relativistic quantum field theory are
typically type~III von Neumann algebras. They do not possess an
intrinsic trace, do not admit canonical density matrices for local state
restrictions, and do not come with a canonical tensor-product
factorization into a region and its complement. Consequently,
finite-dimensional notions such as reflected density matrices and
Umegaki relative entropy must be replaced by modular-theoretic notions:
standard pairs, modular conjugations, relative modular operators, and
Araki relative entropy.

In the present paper, the finite-dimensional type~I construction serves
as a prototype. It explains the fixed-point mechanism in the familiar
language of density matrices. The local type~III construction then
replaces density matrices by states on local von Neumann algebras and
replaces ordinary relative entropy by Araki relative entropy.
\subsection{Modular reflection and self-dual points}
\label{subsec:modular_self_duality}

We start with a modular-theoretic idea that will be used in the
finite-dimensional type I construction. The Tomita--Takesaki construction
needed below is briefly reviewed in Sec.~\ref{sec:vN_TT}. Let
\(\mathcal H\) be a complex Hilbert space, and let
\(\mathcal B(\mathcal H)\) denote the algebra of bounded linear
operators on \(\mathcal H\). A von Neumann algebra
\(\mathfrak M\subset\mathcal B(\mathcal H)\) is a unital
\(^*\)-subalgebra closed in the weak operator topology. Its commutant is
\begin{equation}
\mathfrak M'
:=
\{B\in\mathcal B(\mathcal H):BA=AB,\ \forall\,A\in\mathfrak M\}.
\label{eq:commutant_intro}
\end{equation}
A vector \(\Omega\in\mathcal H\) is cyclic for \(\mathfrak M\) if
\(\mathfrak M\Omega\) is dense in \(\mathcal H\), and separating for
\(\mathfrak M\) if
$A\Omega=0,\; A\in\mathfrak M
\rightarrow
A=0$.
A pair \((\mathfrak M,\Omega)\) with \(\Omega\) cyclic and separating is
called a standard pair. Tomita--Takesaki theory assigns to such a pair an
antiunitary modular conjugation \(J\). One of its basic structural
consequences is
\begin{equation}
J\mathfrak M J=\mathfrak M'.
\label{eq:modular_conjugation_commutant_intro}
\end{equation}
Thus \(J\) implements an algebraic reflection in the sense that it
exchanges the algebra with its commutant. For the finite-dimensional
discussion below, only the idea of an antiunitary reflection is needed;
the antiunitary \(J\) is fixed as part of the comparison data.

 Let \(J\) be a fixed antiunitary involution on the finite-dimensional
Hilbert space \(\mathcal H\). Given a density matrix \(\rho\), we
define its reflected partner by
\begin{equation}
\rho_J:=J\rho J .
\label{eq:finite_reflected_density}
\end{equation}
A state is called modularly self-dual, relative to this reflection, if
\begin{equation}
\rho=\rho_J .
\label{eq:finite_self_dual_density}
\end{equation}
Equivalently, \(\rho\) is a fixed point of the reflected state map
\(\rho\mapsto J\rho J\).

For a smooth one-parameter family of density matrices \(\rho(g)\), we
assume that the reflection induces \textit{an involution \(r\) of parameter
space},
\begin{equation}
\rho_J(g):=J\rho(g)J=\rho(r(g)),
\qquad
r(r(g))=g .
\label{eq:parameter_reflection_intro}
\end{equation}
A parameter value \(g_\star\) is then a modular self-dual point if
\begin{equation}
r(g_\star)=g_\star,
\qquad
\rho(g_\star)=\rho_J(g_\star).
\label{eq:parameter_self_dual_intro}
\end{equation}
The next subsection studies the symmetrized Umegaki relative entropy
between \(\rho(g)\) and its reflected partner \(\rho_J(g)\) near such a
fixed point. The finite-dimensional construction below should be read as the
type~I prototype of the local modular-pullback construction developed
later in the paper.

\subsection{Symmetrized Umegaki relative entropy and the BKM metric at modular fixed points}

The natural comparison functional for a state and its reflected pair is the symmetrized Umegaki relative entropy \cite{Umegaki1962}
\begin{equation}
\mathfrak S_J(g)
=
S\!\left(\rho(g)\middle\|\rho_J(g)\right)
+
S\!\left(\rho_J(g)\middle\|\rho(g)\right),
\label{eq:typeI_symrel}
\end{equation}
with
\begin{equation}
S(\rho\|\sigma)=\operatorname{Tr}\,\rho(\log\rho-\log\sigma).
\label{eq:typeI_umegaki}
\end{equation}
At the modular self-dual point,
\begin{equation}
\mathfrak S_J(g_\star)=0,
\label{eq:typeI_symrel_zero}
\end{equation}
because the two states coincide. Since \(\mathfrak S_J(g)\ge 0\), differentiability gives
\begin{equation}
\partial_g \mathfrak S_J(g)\Big|_{g=g_\star}=0.
\label{eq:typeI_first_variation_zero}
\end{equation}
The first nontrivial local datum is therefore the Hessian
\begin{equation}
I_J(g_\star)
=
\partial_g^2 \mathfrak S_J(g)\Big|_{g=g_\star},
\label{eq:typeI_modular_hessian}
\end{equation}
so that
\begin{equation}
\mathfrak S_J(g)
=
\frac12\,I_J(g_\star)\,(g-g_\star)^2
+
o\!\left((g-g_\star)^2\right).
\label{eq:typeI_quadratic_expansion}
\end{equation}

To further identify the quadratic coefficient, write
\begin{equation}
\begin{array}{c}
\delta=g-g_\star,
\qquad
\rho_\star=\rho(g_\star),
\\\\
X_\star=\partial_g\rho(g)\Big|_{g=g_\star},
\qquad
Y_\star=\partial_g^2\rho(g)\Big|_{g=g_\star}.
\end{array}
\label{eq:typeI_tangent_data}
\end{equation}
Near a fixed point \(g_\star\), we choose the local coordinate
\(\delta:=g-g_\star\) so that this involution acts as reflection about
\(g_\star\):
\begin{equation}
r(g_\star+\delta)=g_\star-\delta .
\end{equation}
Then a \(C^2\)-differentiability assumption gives
\begin{equation}
\begin{aligned}
\rho(g_\star+\delta)
&=
\rho_\star+\delta X_\star+\frac12\,\delta^2 Y_\star+o(\delta^2),\\
\rho_J(g_\star+\delta)
&=
\rho(r(g_\star+\delta))
=
\rho(g_\star-\delta)
\\&=
\rho_\star-\delta X_\star+\frac12\,\delta^2 Y_\star+o(\delta^2).
\end{aligned}
\label{eq:typeI_reflected_tangent_expansion}
\end{equation}

Before applying the logarithm, it is useful to spell out the
support convention used in this subsection. The operator
logarithm and the BKM kernel are singular on zero eigenvalues.
Thus the formulas below are meant either for strictly positive
density matrices on the full Hilbert space, or more generally on
a fixed common support. Concretely, let
\begin{equation}
P := \operatorname{supp}\rho_\star .
\end{equation}
We assume that, for \(|\delta|\) sufficiently small,
\begin{equation}
\operatorname{supp}\rho(g_\star+\delta)
=
\operatorname{supp}\rho_J(g_\star+\delta)
=
P .
\end{equation}
All logarithms, Kubo--Mori maps, traces, and spectral sums are
then understood after restricting the density matrices and
tangent operators to the smaller Hilbert space \(P\mathcal H\).
On this reduced Hilbert space the reference state \(\rho_\star\)
is faithful, so the ordinary finite-dimensional formulas apply.
The discarded sector is a parameter-independent null sector:
no probability weight moves into or out of it along the family.
If the support were allowed to change with \(g\), additional
singular or non-smooth contributions to the relative entropy
could appear, and the Hessian formula would require a separate
analysis.

Let \(D(\log)_{\rho_\star}[X]\) denote the Fr\'echet derivative of the
operator logarithm map \(\rho\mapsto \log\rho\), evaluated at the
reference state \(\rho_\star\) and applied to the perturbation direction
\(X\). After restricting to the fixed support \(P\mathcal H\),
the reference state \(\rho_\star\) has no zero eigenvalues.
Therefore the following Fr\'echet derivative of the logarithm is
well defined on \(P\mathcal H\),
\begin{equation}
D(\log)_{\rho_\star}[X]
=
\int_0^\infty du\,
(\rho_\star+uI)^{-1}X(\rho_\star+uI)^{-1}.
\label{eq:typeI_log_frechet}
\end{equation}
Equivalently, for small \(\varepsilon\),
\begin{equation}
\log(\rho_\star+\varepsilon X)
=
\log\rho_\star
+
\varepsilon\,D(\log)_{\rho_\star}[X]
+
O(\varepsilon^2).
\label{eq:typeI_log_frechet_linearization}
\end{equation}
It is convenient to package this in the Kubo--Mori map
\begin{equation}
\mathcal K_{\rho_\star}(X)
=
\int_0^1 ds\,
\rho_\star^{\,s}X\rho_\star^{\,1-s},
\label{eq:typeI_kubo_map}
\end{equation}
whose inverse is exactly the Fr\'echet derivative of the logarithm $\mathcal K_{\rho_\star}^{-1}(X)
=
D(\log)_{\rho_\star}[X]$.

The BKM bilinear form at \(\rho_\star\) is then defined by
\begin{equation}
\gamma_{\rho_\star}^{\mathrm{BKM}}(X,Y)
=
\operatorname{Tr}\!\left[
X\,\mathcal K_{\rho_\star}^{-1}(Y)
\right].
\label{eq:typeI_BKM_bilinear}
\end{equation}
The associated BKM quantum Fisher information along the family \(g\mapsto \rho(g)\) is \cite{PetzToth1993,LesniewskiRuskai1999}
\begin{equation}
\begin{aligned}
F_{\mathrm{BKM}}(g_\star)
&=
\gamma_{\rho_\star}^{\mathrm{BKM}}(X_\star,X_\star)=\operatorname{Tr}\!\left[
X_\star\,\mathcal K_{\rho_\star}^{-1}(X_\star)
\right]
\\
&=
\int_0^\infty du\,
\operatorname{Tr}\!\left[
X_\star(\rho_\star+uI)^{-1}
X_\star(\rho_\star+uI)^{-1}
\right].
\end{aligned}
\label{eq:typeI_BKM_QFI}
\end{equation}

If \(\rho_\star=\sum_i p_i \ket{i}\!\bra{i}\) is the spectral decomposition, then the BKM form is given by \cite{Petz1996,MichorPetzAndai2000}
\begin{equation}
\gamma_{\rho_\star}^{\mathrm{BKM}}(X,Y)
=
\sum_{i,j}
c_{\mathrm{BKM}}(p_i,p_j)\,
X_{ij}Y_{ji},
\label{eq:typeI_BKM_spectral}
\end{equation}
with kernel
\begin{equation}
c_{\mathrm{BKM}}(x,y)
=
\frac{\log x-\log y}{x-y}
\quad
(x\neq y),
\;
c_{\mathrm{BKM}}(x,x)=\frac1x.
\label{eq:typeI_BKM_kernel}
\end{equation}
This makes positivity manifest, since \(\gamma_{\rho_\star}^{\mathrm{BKM}}(X,X)\ge 0\) for Hermitian \(X\).

To extract the quadratic term, we expand the difference of the two logarithms directly. Because \(\log\) is analytic on positive definite operators, and because the quadratic terms in \eqref{eq:typeI_reflected_tangent_expansion} coincide, one obtains
\begin{equation}
\log\rho(g_\star+\delta)-\log\rho_J(g_\star+\delta)
=
2\delta\,\mathcal K_{\rho_\star}^{-1}(X_\star)
+
o(\delta^2).
\label{eq:typeI_log_expansion_pm}
\end{equation}
Substituting \eqref{eq:typeI_reflected_tangent_expansion} and \eqref{eq:typeI_log_expansion_pm} into the first relative entropy yields
\begin{equation}
\begin{array}{c}
S\!\left(\rho(g_\star+\delta)\middle\|\rho_J(g_\star+\delta)\right)
\\\\=
\operatorname{Tr}\!\left[
\left(\rho_\star+\delta X_\star+\frac12\,\delta^2Y_\star+o(\delta^2)\right)
\left(
2\delta\,\mathcal K_{\rho_\star}^{-1}(X_\star)+o(\delta^2)
\right)
\right]
\\\\
=
2\delta\,
\operatorname{Tr}\!\left[
\rho_\star \mathcal K_{\rho_\star}^{-1}(X_\star)
\right]
+
2\delta^2\,
\operatorname{Tr}\!\left[
X_\star \mathcal K_{\rho_\star}^{-1}(X_\star)
\right]
+
o(\delta^2).
\end{array}
\label{eq:typeI_first_relent_expansion}
\end{equation}
Since \(\operatorname{Tr}X_\star=0\) and, in the spectral basis of \(\rho_\star\),
\begin{equation}
\begin{array}{c}
\operatorname{Tr}\!\left[
\rho_\star \mathcal K_{\rho_\star}^{-1}(X_\star)
\right]
=
\sum_i p_i\,c_{\mathrm{BKM}}(p_i,p_i)\,(X_\star)_{ii}
\\\\=
\sum_i (X_\star)_{ii}
=
\operatorname{Tr}X_\star
=
0,
\end{array}
\label{eq:typeI_trace_identity}
\end{equation}
the linear term vanishes and we obtain
\begin{equation}
S\!\left(\rho(g_\star+\delta)\middle\|\rho_J(g_\star+\delta)\right)
=
2\,F_{\mathrm{BKM}}(g_\star)\,\delta^2
+
o(\delta^2).
\label{eq:typeI_one_sided_relent_BKM}
\end{equation}
The same expansion holds after exchanging the two arguments, so
\begin{equation}
\mathfrak S_J(g_\star+\delta)
=
4\,F_{\mathrm{BKM}}(g_\star)\,\delta^2
+
o(\delta^2).
\label{eq:typeI_symrel_BKM}
\end{equation}
Comparing with \eqref{eq:typeI_quadratic_expansion} gives
\begin{equation}
I_J(g_\star)=8\,F_{\mathrm{BKM}}(g_\star).
\label{eq:typeI_hessian_BKM}
\end{equation}
Thus, with the normalization used in \eqref{eq:typeI_quadratic_expansion}, the Hessian of the
symmetrized Umegaki relative entropy is the BKM quantum Fisher
information of the single tangent \(X_\star\), multiplied by the
universal factor \(8\).
The factor \(8\) comes from two sources. First, the reflected comparison
uses the tangent difference \(X_\star-(-X_\star)=2X_\star\). Second, the
functional is symmetrized, so the two relative-entropy terms contribute
equally. This is the direct type-I precursor of the local type-III
Hessian built later from symmetrized Araki relative entropy.

\subsection{Phase-damped single-excitation \(X\)-state family}

As a concrete mixed-state illustration, consider the phase-damped single-excitation family
\begin{equation}
\ket{\psi(g)}
=
\cos\!\left(\frac{\pi}{4}+g\right)\ket{01}
+
\sin\!\left(\frac{\pi}{4}+g\right)\ket{10},
\;
|g|<\frac{\pi}{4},
\label{eq:pd_pure_family_rewritten}
\end{equation}
with even dephasing profile
\begin{equation}
\Lambda(g)=e^{-(\gamma_0+\gamma_2 g^2)\tau},
\;
\gamma_0>0,
\;
\gamma_2\ge 0,
\;
\tau> 0.
\label{eq:pd_lambda_rewritten}
\end{equation}
The resulting density matrix in the computational basis
\(\{\ket{00},\ket{01},\ket{10},\ket{11}\}\) is
\begin{equation}
\begin{array}{c}
\rho(g)=
\begin{pmatrix}
0 & 0 & 0 & 0\\
0 & \cos^2\theta & \Lambda(g)\sin\theta\cos\theta & 0\\
0 & \Lambda(g)\sin\theta\cos\theta & \sin^2\theta & 0\\
0 & 0 & 0 & 0
\end{pmatrix},
\\\\
\theta=\frac{\pi}{4}+g.
\end{array}
\label{eq:pd_rho_rewritten}
\end{equation}
Its support is the fixed two-dimensional subspace
\begin{equation}
P\mathcal H=\operatorname{span}\{\ket{01},\ket{10}\}.
\label{eq:pd_support}
\end{equation}
For \(\tau>0\), one has \(0<\Lambda(g)<1\), and since
\(|g|<\pi/4\) also \(\sin\theta\cos\theta>0\). Hence the
restricted state \(P\rho(g)P\), viewed as a \(2\times 2\)
density matrix on \(P\mathcal H\), is faithful for all such
\(g\). The two zero rows and columns in \eqref{eq:pd_rho_rewritten} therefore do
not enter the logarithms. They represent a fixed null sector
outside \(P\mathcal H\), not eigenvalues that are varied through
zero. All Umegaki entropies and BKM quantities in this example
are computed on \(P\mathcal H\).

On \(P\mathcal H\), let \(K_P\) denote complex conjugation in the ordered basis \(\{|01\rangle,|10\rangle\}\), and let \(S_P\) be the unitary swap satisfying \(S_P|01\rangle=|10\rangle\) and \(S_P|10\rangle=|01\rangle\). Define the antiunitary involution \(J:=S_PK_P\). Since \(\Lambda(g)\) is even, this reflection satisfies
\begin{equation}
\rho_J(g):=J\rho(g)J=\rho(-g),\qquad g_\star=0 .
\label{eq:pd_reflection_rewritten}
\end{equation}
Thus \(g=0\) is the modular self-dual point.
The family of states takes the effective qubit form
\begin{equation}
\rho(g)
=
\frac12
\left(
\mathbf 1
+
\Lambda(g)\cos(2g)\,\sigma_x
-
\sin(2g)\,\sigma_z
\right),
\label{eq:pd_bloch_form}
\end{equation}
where \(\sigma_x\) and \(\sigma_z\) are the Pauli matrices. At the fixed point,
\begin{equation}
\begin{array}{c}
\lambda_0=e^{-\gamma_0 \tau}\in(0,1),
\\\\
\rho_0=\rho(0)=\frac12\left(\mathbf 1+\lambda_0\sigma_x\right),
\\\\
X_0=\partial_g\rho(g)\Big|_{g=0}=-\sigma_z,
\end{array}
\label{eq:pd_fixed_tangent}
\end{equation}
where the last identity uses \(\Lambda'(0)=0\) by evenness.

In the eigenbasis of \(\sigma_x\), the fixed-point state is diagonal,
\begin{equation}
\rho_0=
\begin{pmatrix}
p_+ & 0\\
0 & p_-
\end{pmatrix},
\qquad
p_\pm=\frac{1\pm\lambda_0}{2},
\label{eq:pd_fixed_eigs}
\end{equation}
while the tangent takes the form
\begin{equation}
X_0=
\begin{pmatrix}
0 & -1\\
-1 & 0
\end{pmatrix}.
\label{eq:pd_fixed_tangent_matrix}
\end{equation}
The BKM spectral formula \eqref{eq:typeI_BKM_spectral}--\eqref{eq:typeI_BKM_kernel} then gives \cite{Petz1996,MichorPetzAndai2000}
\begin{equation}
\begin{aligned}
F_{\mathrm{BKM}}(0)
&=
2\,c_{\mathrm{BKM}}(p_+,p_-)
\\\\
&=
2\,\frac{\log p_+-\log p_-}{p_+-p_-}
=
\frac{2}{\lambda_0}\,
\log\!\frac{1+\lambda_0}{1-\lambda_0}.
\end{aligned}
\label{eq:pd_BKM_value}
\end{equation}
Hence the reflected Hessian is
\begin{equation}
I_J(0)
=
8\,F_{\mathrm{BKM}}(0)
=
\frac{16}{\lambda_0}\,
\log\!\frac{1+\lambda_0}{1-\lambda_0}.
\label{eq:pd_IJ_value}
\end{equation}

The same coefficient can also be extracted directly from the symmetrized
Umegaki functional \eqref{eq:typeI_symrel} \cite{Umegaki1962}. Writing
\begin{equation}
\rho(g)=\frac12\left(\mathbf 1+\vec R(g)\cdot\vec\sigma\right),
\;
\vec R(g)=\bigl(\Lambda(g)\cos 2g,\ 0,\ -\sin 2g\bigr),
\label{eq:pd_bloch_vector_rewritten}
\end{equation}
one has \(|\vec R(g)|=|\vec R(-g)|\). Let
\begin{equation}
\mathcal R(g)=|\vec R(g)|
=
\sqrt{\Lambda(g)^2\cos^2(2g)+\sin^2(2g)}.
\label{eq:pd_bloch_radius}
\end{equation}
For a faithful qubit state
\(\rho=\frac12(\mathbf 1+\vec R\cdot\vec\sigma)\) with
\(0<\mathcal R=|\vec R|<1\), one has
\begin{equation}
\log\rho
=
a(\mathcal R)\,\mathbf 1
+
b(\mathcal R)\,\hat R\cdot\vec\sigma,
\label{eq:pd_qubit_log}
\end{equation}
with
\begin{equation}
a(\mathcal R)=\frac12\log\!\frac{1-\mathcal R^2}{4},
\;
b(\mathcal R)=\operatorname{arctanh}(\mathcal R),
\;
\hat R=\frac{\vec R}{\mathcal R}.
\label{eq:pd_qubit_log_coeffs}
\end{equation}
Since \(\rho(g)\) and \(\rho(-g)\) have the same Bloch radius, the scalar
part cancels in the relative entropy. Using
\(\operatorname{Tr}\!\left[(\vec a\cdot\vec\sigma)(\vec b\cdot\vec\sigma)\right]
=2\,\vec a\cdot\vec b\), a direct calculation gives
\begin{equation}
S\!\left(\rho(g)\middle\|\rho(-g)\right)
=
\frac{\operatorname{arctanh}\mathcal R(g)}{\mathcal R(g)}
\left(
\mathcal R(g)^2-\vec R(g)\cdot\vec R(-g)
\right).
\label{eq:pd_one_sided_exact}
\end{equation}
Now
\begin{equation}
\mathcal R(g)^2-\vec R(g)\cdot\vec R(-g)
=
2\sin^2(2g),
\label{eq:pd_dot_difference}
\end{equation}
so the symmetrized Umegaki entropy is exactly
\begin{equation}
\mathfrak S_J(g)
=
\frac{4\,\operatorname{arctanh}\mathcal R(g)}{\mathcal R(g)}\,
\sin^2(2g).
\label{eq:pd_symrel_exact}
\end{equation}
Since \(\mathcal R(g)\) is even and \(\mathcal R(0)=\lambda_0\), one has
\begin{equation}
\begin{array}{c}
\dfrac{\operatorname{arctanh}\mathcal R(g)}{\mathcal R(g)}
=
\dfrac{\operatorname{arctanh}\lambda_0}{\lambda_0}
+
O(g^2),
\\\\
\sin^2(2g)=4g^2+O(g^4).
\end{array}
\label{eq:pd_small_g_expansions}
\end{equation}
Therefore
\begin{equation}
\begin{aligned}
\mathfrak S_J(g)
&=
\frac{16\,\operatorname{arctanh}\lambda_0}{\lambda_0}\,g^2
+
O(g^4)
\\
&=
\frac{8}{\lambda_0}\,
\log\!\frac{1+\lambda_0}{1-\lambda_0}\,g^2
+
O(g^4)
\\
&=
\frac12\,I_J(0)\,g^2
+
O(g^4).
\end{aligned}
\label{eq:pd_symrel_expansion}
\end{equation}
This example therefore shows explicitly that the coefficient of \(g^2\) in
the reflected Umegaki functional is \(4\,F_{\mathrm{BKM}}(0)\) \cite{PetzToth1993,LesniewskiRuskai1999}, or equivalently
that the Hessian is \(I_J(0)=8\,F_{\mathrm{BKM}}(0)\). It also shows that the
parameter \(\gamma_2\) in the even dephasing profile affects only quartic and
higher even orders. The Hessian itself is fixed entirely by the self-dual
state \(\rho_0\) and the tangent \(X_0\).

\section{Von Neumann Algebras and Modular Theory}
\label{sec:vN_TT}

We now recast the finite-dimensional discussion of Sec.~\ref{sec:FiniteTypeI}
in the operator-algebraic language appropriate to local quantum field theory.
The essential point is that local observables in relativistic quantum field
theory are not described by finite-dimensional matrix algebras, or even by
type~I factors with intrinsic density matrices. They are described by local
von Neumann algebras, which are generically type~III. Consequently, the
finite-dimensional notions of density matrices, reflected density matrices,
and Umegaki relative entropy must be replaced by modular-theoretic objects:
standard pairs, modular conjugations, relative modular operators, and Araki
relative entropy. The relative modular and Araki-entropy conventions used below are fixed in
Appendix~\ref{sec:RelativeModularAraki}.

\subsection{Smeared Fields, Weyl Operators, and Local Nets}
\label{smearedfields}

For a free real scalar field on Minkowski spacetime $\mathcal M$, it is convenient to work with smeared fields $\Phi(f)$, where $f\in C_0^\infty(\mathcal M)$ is a smooth compactly supported test function. Formally,
\begin{equation}
\Phi(f)=\int_{\mathcal M}\phi(x)f(x)\,d\mu(x),
\end{equation}
where $\phi$ is the operator-valued distribution associated with the scalar field. The smeared fields satisfy linearity,
\begin{equation}
\Phi(\alpha f_1+\beta f_2)=\alpha \Phi(f_1)+\beta \Phi(f_2),
\qquad
\alpha,\beta\in\mathbb C,
\end{equation}
and the adjoint relation
\begin{equation}
\Phi(f)^*=\Phi(\bar f).
\end{equation}
In particular, if $f$ is real-valued, then $\Phi(f)$ is symmetric; in the vacuum representation it is essentially self-adjoint on the standard finite-particle domain.

Let
\begin{equation}
P:=\Box+m^2
\end{equation}
denote the Klein--Gordon operator, and let
\begin{equation}
E:=E_{\mathrm{ret}}-E_{\mathrm{adv}}
\end{equation}
be the causal propagator. The field equation and the covariant commutation relations are
\begin{equation}
\Phi(Pf)=0,
\end{equation}
\begin{equation}
[\Phi(f),\Phi(g)]=iE(f,g)\mathbf 1,
\end{equation}
where
\begin{equation}
E(f,g):=\int_{\mathcal M\times\mathcal M} f(x)\,E(x,y)\,g(y)\,d\mu(x)\,d\mu(y)
\end{equation}
is the induced antisymmetric bilinear form. The relation $\Phi(Pf)=0$ means that the field depends only on the equivalence class of $f$ modulo $PC_0^\infty(\mathcal M)$. Equivalently, one may regard the basic real test-function space as the quotient
\begin{equation}
\mathcal K_{\mathbb R}:=C_0^\infty(\mathcal M,\mathbb R)\big/PC_0^\infty(\mathcal M,\mathbb R),
\end{equation}
equipped with the symplectic form induced by $E$.

Because the smeared fields are generally unbounded, it is standard to pass to the bounded Weyl operators. For real-valued $f$, we define
\begin{equation}
W(f):=e^{i\Phi(f)}.
\end{equation}
These unitaries satisfy the Weyl relations
\begin{equation}
W(f)W(g)=e^{-\,\frac{i}{2}E(f,g)}\,W(f+g),
\end{equation}
or equivalently
\begin{equation}
W(f)W(g)=e^{-iE(f,g)}\,W(g)W(f),
\label{WeylCCR}
\end{equation}
together with
\begin{equation}
W(f)^*=W(-f).
\end{equation}
Equation~\eqref{WeylCCR} is the Weyl form of the canonical commutation relations (CCR).
Since $\Phi(Pf)=0$, the Weyl operator $W(f)$ also depends only on the equivalence class of $f$ in $\mathcal K_{\mathbb R}$, although we continue to write $W(f)$ for simplicity.

Let $\mathfrak A$ denote the global Weyl $C^*$-algebra, that is, the universal $C^*$-algebra generated by the symbols $W(f)$ subject to the Weyl relations above. For an open region $\mathcal O\subset\mathcal M$, we define the local $C^*$-subalgebra
\begin{equation}
\mathfrak A(\mathcal O)\subset\mathfrak A
\end{equation}
to be the $C^*$-algebra generated by those Weyl operators $W(f)$ with $\operatorname{supp}f\subset\mathcal O$. The assignment
\begin{equation}
\mathcal O\longmapsto \mathfrak A(\mathcal O)
\end{equation}
is the abstract local Weyl net. Here ``net'' simply means a region-by-region assignment of algebras, ordered by inclusion of regions. In particular, if
\begin{equation}
\mathcal O_1\subset\mathcal O_2,
\end{equation}
then isotony gives
\begin{equation}
\mathfrak A(\mathcal O_1)\subset\mathfrak A(\mathcal O_2).
\end{equation}
Moreover, if $\mathcal O'$ denotes the causal complement of $\mathcal O$, then locality means
\begin{equation}
\mathcal O_1\subset\mathcal O_2'
\quad\Longrightarrow\quad
[A_1,A_2]=0,
\qquad
A_i\in\mathfrak A(\mathcal O_i).
\end{equation}
This is the algebraic expression of Einstein causality: observables localized in spacelike separated regions commute. The regions relevant later in the paper include both bounded regions and unbounded ones such as wedges.

To obtain concrete operators on a Hilbert space, one must choose a state and pass to its GNS representation. Recall that a state on $\mathfrak A$ is a positive normalized linear functional. For the free scalar field there is a distinguished vacuum state, denoted $\omega_0$, and its GNS construction yields a triple
\begin{equation}
(\pi_0,\mathcal H,\Omega),
\end{equation}
called the vacuum representation. Here $\pi_0:\mathfrak A\to\mathcal B(\mathcal H)$ is a $*$-representation of the abstract algebra by bounded operators on a Hilbert space $\mathcal H$, and $\Omega\in\mathcal H$ is a cyclic vector satisfying
\begin{equation}
\omega_0(A)=\langle\Omega,\pi_0(A)\Omega\rangle,
\qquad
A\in\mathfrak A.
\end{equation}
Thus the vacuum state is realized as a vector state in the Hilbert-space representation, and $\Omega$ is the vacuum vector.

For each region $\mathcal O$, we then define the represented local von Neumann algebra by
$\mathcal A(\mathcal O):=\pi_0\bigl(\mathfrak A(\mathcal O)\bigr)''\subset\mathcal B(\mathcal H)$.
The double commutant here is equivalent, by von Neumann's double commutant theorem, to the weak operator closure of $\pi_0(\mathfrak A(\mathcal O))$. In other words, $\mathcal A(\mathcal O)$ is the concrete algebra of bounded local observables associated with $\mathcal O$ in the vacuum sector.

This distinction between the abstract net and its vacuum realization is important throughout the paper. The notation
$\mathfrak A(\mathcal O)$
refers to the abstract local Weyl $C^*$-algebra, while
$\mathcal A(\mathcal O)$ refers to the corresponding represented local von Neumann algebra acting on the vacuum Hilbert space. The former is part of the model-independent $C^*$-algebraic description, whereas the latter is the concrete operator algebra used in the vacuum representation. Throughout the main text, local algebras are understood in this represented sense. In the free scalar theory this includes, in particular, the wedge algebras $\mathcal A(W_t)$ used later, and in the chiral setting the analogous notation is used for local algebras associated with intervals or half-lines.

\subsection{Tomita--Takesaki Theory}
\label{sec:TomitaTakesakiTheory}
For extensive reviews of the topics summarized below, see
\cite{Haag1,Guido2011,Witt2018,Bratteli1997}.
As above, we work in the vacuum representation \((\pi_0,\mathcal H,\Omega)\) of the
abstract local net and write
\begin{equation}
\mathcal A(\mathcal O)
:=
\pi_0\bigl(\mathfrak A(\mathcal O)\bigr)''
\subset \mathcal B(\mathcal H)
\label{eq:main_represented_local_algebra}
\end{equation}
for the represented local von Neumann algebra associated with an open
spacetime region \(\mathcal O\subset\mathcal M\). Here \(\mathcal M\)
denotes Minkowski spacetime, while
\(\mathfrak M\subset\mathcal B(\mathcal H)\) denotes a generic von Neumann
algebra. All modular and relative modular objects used below are attached to
these represented local algebras.

A von Neumann algebra \(\mathfrak M\) is a unital \(^{*}\)-subalgebra of
\(\mathcal B(\mathcal H)\) that is closed in the weak operator topology, or
equivalently satisfies the bicommutant relation
\begin{equation}
\mathfrak M=\mathfrak M''.
\label{eq:main_bicommutant}
\end{equation}
In finite-dimensional quantum mechanics, the basic example is the type~I
factor
\begin{equation}
\mathfrak M=\mathcal B(\mathcal H),
\label{eq:main_typeI_factor}
\end{equation}
where states are represented by density matrices and relative entropy is
given by the usual trace formula. In local quantum field theory, by contrast,
the local algebras \(\mathcal A(\mathcal O)\) are generically type~III
factors. They do not possess an intrinsic trace, do not admit a canonical
density-matrix description of local state restrictions, and do not come with
a canonical Hilbert-space tensor factorization into a region and its
complement. This is the structural reason why the finite-dimensional
construction must be reformulated in modular terms.

The relevant modular input begins with a standard pair
\((\mathfrak M,\Omega)\), where \(\Omega\in\mathcal H\) is cyclic and
separating for \(\mathfrak M\). Cyclicity means that \(\mathfrak M\Omega\)
is dense in \(\mathcal H\), while separating means that
\begin{equation}
A\Omega=0,\quad A\in\mathfrak M
\qquad\Longrightarrow\qquad
A=0.
\label{eq:main_separating}
\end{equation}
For such a pair, one defines the Tomita operator on the dense domain
\(\mathfrak M\Omega\) by
\begin{equation}
S_0(A\Omega)=A^*\Omega,
\qquad
A\in\mathfrak M.
\label{eq:main_Tomita_operator}
\end{equation}
Its closure \(S\) has polar decomposition
\begin{equation}
S=J\Delta^{1/2},
\label{eq:main_Tomita_polar}
\end{equation}
where \(J\) is the modular conjugation and \(\Delta\) is the modular
operator. Tomita--Takesaki theory gives
\begin{equation}
J\mathfrak M J=\mathfrak M',
\label{eq:main_J_commutant}
\end{equation}
and the modular automorphism group
\begin{equation}
\sigma_s^\Omega(A)
=
\Delta^{is}A\Delta^{-is},
\qquad
A\in\mathfrak M,
\qquad
s\in\mathbb R,
\label{eq:main_modular_automorphism}
\end{equation}
satisfies \(\sigma_s^\Omega(\mathfrak M)=\mathfrak M\). Thus the modular
operator generates intrinsic dynamics of the algebra, while the
modular conjugation exchanges the algebra with its commutant.

In the local quantum-field-theoretic setting, the Reeh--Schlieder property
implies that the vacuum vector \(\Omega\) is cyclic and separating for the
local algebras associated with the regions considered here. Hence each pair
\begin{equation}
(\mathcal A(\mathcal O),\Omega)
\label{eq:main_local_standard_pair}
\end{equation}
is standard, and the modular data are available intrinsically.

For the local comparison problem below, it is useful to write the relevant
local algebra in the uniform form
\begin{equation}
M_t
:=
\mathcal A(O_t),
\qquad
J_t:=J_{M_t,\Omega},
\qquad
\Delta_t:=\Delta_{M_t,\Omega}.
\label{eq:main_local_modular_data}
\end{equation}
Here \(O_t\) denotes the localization region under consideration. In the
scalar-field example of Sec.~\ref{sec:wedge_coherent_states},
\(O_t=W_t\) is a wedge. In the chiral-current example of
Sec.~\ref{sec:chiral_u1}, \(O_t=(-\infty,t)\) is a half-line. For every such
standard pair, Tomita--Takesaki theory gives
\begin{equation}
J_tM_tJ_t=M_t'.
\label{eq:main_local_commutant}
\end{equation}
When \(O_t=W_t\) is a wedge and one identifies \(M_t'\) with the algebra of
the opposite wedge, one invokes wedge duality. When the modular objects are
given a geometric interpretation in the scalar wedge model, one additionally
invokes the Bisognano--Wichmann theorem.

A state on a von Neumann algebra \(\mathfrak M\) is a positive normalized
linear functional
\begin{equation}
\omega:\mathfrak M\to\mathbb C,
\qquad
\omega(A^*A)\ge0,
\qquad
\omega(\mathbf 1)=1.
\label{eq:main_state_functional}
\end{equation}
It is faithful if \(\omega(A^*A)=0\) implies \(A=0\), and normal if it is
ultraweakly continuous. In finite-dimensional type~I settings, normal states
are precisely those represented by density matrices. In type~III local
algebras, faithful normal states are still well defined, but there is no
intrinsic local density matrix with respect to a canonical trace.

To compare two faithful normal states \(\omega\) and \(\phi\) on
\(\mathfrak M\), one uses relative modular theory. In standard form, each
faithful normal state is represented by a vector \(\xi_\omega\) in the
natural cone. With the convention fixed in
Appendix~\ref{sec:RelativeModularAraki}, the relative modular operator
\(\Delta_{\phi\mid\omega}\) determines the Araki relative entropy
\cite{Araki1975,Araki1977}
\begin{equation}
S_{\mathfrak M}(\omega\Vert\phi)
:=
-\langle \xi_\omega,\log\Delta_{\phi\mid\omega}\,\xi_\omega\rangle .
\label{eq:main_Araki_entropy}
\end{equation}
This is the intrinsic notion of relative entropy appropriate to local
operator algebras. In the type~I case, if
\begin{equation}
\omega(A)=\operatorname{Tr}(\rho A),
\qquad
\phi(A)=\operatorname{Tr}(\sigma A),
\label{eq:main_typeI_states_density_matrices}
\end{equation}
with faithful density matrices \(\rho\) and \(\sigma\), then
\begin{equation}
S_{\mathfrak M}(\omega\Vert\phi)
=
\operatorname{Tr}\rho(\log\rho-\log\sigma),
\label{eq:main_Umegaki_reduction}
\end{equation}
so Araki relative entropy reduces to Umegaki relative entropy
\cite{Umegaki1962}.

The transition from the type~I construction to the local type~III setting can
now be stated conceptually. In finite dimensions, the reflected partner of a
state is another density matrix obtained by applying the chosen antiunitary
within the same matrix algebra. For a local type~III algebra, the vacuum
modular conjugation instead carries the algebra to its commutant. Thus the
reflected comparison cannot be formulated as an internal density-matrix
operation on the local algebra. One must restrict the ambient state to the
commutant and then pull that restriction back to the original local algebra by
the modular anti-isomorphism associated with the vacuum standard pair. This
construction is formulated in the next section for the local algebras relevant
to both examples: wedge algebras in the scalar-field model and half-line
algebras in the chiral-current model.

\section{Modular Self-Duality and Symmetrized Relative Entropy in Local Algebras}
\label{sec:reflection_paired_families}

Having fixed the operator-algebraic conventions, we now formulate the local
type~III analogue of the reflected construction of Sec.~\ref{sec:FiniteTypeI}.
The central replacement is that the finite-dimensional reflected density
matrix \(J\rho(g)J\) is replaced by a modularly pulled-back state on the same
local algebra. The comparison functional is then the symmetrized Araki
relative entropy, and its quadratic coefficient at a self-dual point is
governed by the local Bogoliubov--Kubo--Mori geometry. Table~\ref{tab:typeI_typeIII_correspondence} summarizes the correspondence
between the finite-dimensional reflected-state construction and the local
type~III modular-pullback construction, including the different
normalizations of the associated BKM susceptibility coefficients.

\begin{table*}[t]
\centering
\small
\renewcommand{\arraystretch}{1.35}
\begin{tabular}{p{0.18\textwidth}p{0.31\textwidth}p{0.41\textwidth}}
\hline\hline
Concept
&
Type~I / finite-dimensional system
&
Intrinsic type~III / local operator-algebraic extension
\\
\hline

State family
&
Density-matrix family
\(\rho(g)\)
&
\(
\begin{array}{l}
\displaystyle
M_t:=\mathcal A(O_t),\qquad (M_t,\Omega)\ \text{standard},
\\[1ex]
\displaystyle
\omega_{g,t}:=\varpi_g\!\restriction_{M_t}.
\end{array}
\)
\\[4mm]

Partner construction
&
Reflected state
\(
\rho_J(g):=J\rho(g)J
\)
&
\(
\begin{array}{l}
\displaystyle
\widetilde{\omega}_{g,t}
:=
\bigl(\varpi_g\!\restriction_{M_t'}\bigr)\circ j_t,
\\[1ex]
\displaystyle
j_t:M_t\to M_t',
\qquad
j_t(A):=J_tA^*J_t .
\end{array}
\)
\\[6mm]

Self-dual locus
&
\(\rho(g_\star)=\rho_J(g_\star)\)
&
\(
\omega_{0,t}
=
\widetilde{\omega}_{0,t}
=:
\omega_t
\)
\\[3mm]

Parameters
&
Deformation parameter \(g\)
&
\(
\begin{array}{l}
\text{Ambient deformation parameter } g,
\\
\text{localization parameter } t.
\end{array}
\)
\\[4mm]

Localization
&
No local-algebra parameter
&
\(
\begin{array}{l}
O_t=W_t
\quad\text{for wedge algebras},
\\
O_t=(-\infty,t)
\quad\text{for chiral half-lines}.
\end{array}
\)
\\[4mm]

Comparison functional
&
\(
\begin{array}{l}
\displaystyle
S_J(g)
:=
S(\rho(g)\Vert\rho_J(g))
\\[1ex]
\displaystyle\hspace{1.6cm}
+
S(\rho_J(g)\Vert\rho(g)).
\end{array}
\)
&
\(
\begin{array}{l}
\displaystyle
S(g,t)
:=
S_{M_t}(\omega_{g,t}\Vert\widetilde{\omega}_{g,t})
\\[1ex]
\displaystyle\hspace{1.15cm}
+
S_{M_t}(\widetilde{\omega}_{g,t}\Vert\omega_{g,t}).
\end{array}
\)
\\[7mm]

Vanishing linear term
&
\(
\begin{array}{l}
S_J(g_\star)=0,\qquad S_J(g)\ge 0,
\\
\partial_gS_J(g)|_{g=g_\star}=0
\quad\text{if differentiable}.
\end{array}
\)
&
\(
\begin{array}{l}
S(0,t)=0,\qquad S(g,t)\ge 0,
\\
\partial_gS(g,t)|_{g=0}=0
\quad\text{if differentiable}.
\end{array}
\)
\\[6mm]

Second-order coefficient
&
\(
\begin{array}{l}
\displaystyle
I_J(g_\star)
:=
\left.
\partial_g^2S_J(g)
\right|_{g=g_\star},
\\
\displaystyle
S_J(g)
=
\dfrac12 I_J(g_\star)(g-g_\star)^2
\\+
o\!\left((g-g_\star)^2\right).
\end{array}
\)
&
\(
\begin{array}{l}
\displaystyle
I_A(t)
:=
\left.
\partial_g^2S(g,t)
\right|_{g=0},
\\[1ex]
\displaystyle
S(g,t)
=
\dfrac12 I_A(t)g^2+o(g^2).
\end{array}
\)
\\[9mm]

Underlying BKM geometry
&
\(
\gamma_{\rho_\star}^{\mathrm{BKM}}(X,Y),
\qquad
\rho_\star:=\rho(g_\star)
\)
&
\(
\gamma_{\omega_t}^{\mathrm{BKM}}(u,v),
\qquad
\omega_t:=\omega_{0,t}
=
\widetilde{\omega}_{0,t}
\)
\\[4mm]

Family-selected tangent
&
\(
\dot\rho_\star
:=
\left.
\partial_g\rho(g)
\right|_{g=g_\star}
\)
&
\(
\begin{array}{l}
\displaystyle
u_t
:=
\left.
\partial_g\omega_{g,t}
\right|_{g=0}
-
\left.
\partial_g\widetilde{\omega}_{g,t}
\right|_{g=0},
\\[1ex]
\displaystyle
u_t\in (M_t)_{*,0}^{\mathrm{sa}} .
\end{array}
\)
\\[7mm]

Family-selected sensitivity
&
\(
\begin{array}{l}
\displaystyle
F_{\mathrm{BKM}}(g_\star)
:=
\gamma_{\rho_\star}^{\mathrm{BKM}}
(\dot\rho_\star,\dot\rho_\star),
\\[1ex]
\displaystyle
I_J(g_\star)=8\,F_{\mathrm{BKM}}(g_\star).
\end{array}
\)
&
\(
\begin{array}{l}
\displaystyle
F_{\mathrm{BKM}}^{\mathcal A}(t)
:=
\gamma_{\omega_t}^{\mathrm{BKM}}(u_t,u_t),
\\[1ex]
\displaystyle
I_A(t)=2\,F_{\mathrm{BKM}}^{\mathcal A}(t).
\end{array}
\)
\\

\hline\hline
\end{tabular}
\caption{
Structural correspondence between the finite-dimensional type~I prototype
and its intrinsic local type~III extension.  In the type~I theory,
modular self-duality compares a state with its reflected partner through
the symmetrized Umegaki relative entropy.  In the local type~III theory,
the comparison is between the local state \(\omega_{g,t}\) on
\(M_t=\mathcal A(O_t)\) and the modular pullback
\(\widetilde{\omega}_{g,t}\) of the commutant restriction.  The same
fixed-point structure holds in both settings, but the type~III partner
requires the modular anti-isomorphism \(j_t(A)=J_tA^*J_t\).  The table
also distinguishes the full BKM bilinear form from the Fisher-information
scalar obtained by evaluating that form on the tangent selected by the
one-parameter family.  The type~III BKM identification is understood under
the diagonal second-order regularity assumption stated in the text.
}
\label{tab:typeI_typeIII_correspondence}
\end{table*}

For the remainder of this section and in the explicit models below, we work
in the vacuum representation. Let \(O_t\) denote the localization region under
consideration and set
\begin{equation}
M_t
:=
\mathcal A(O_t)
\subset
\mathcal B(\mathcal H).
\label{eq:typeIII_Mt_def}
\end{equation}
In the scalar-field example of Sec.~\ref{sec:wedge_coherent_states},
\(O_t=W_t\) is a wedge. In the chiral-current example of
Sec.~\ref{sec:chiral_u1}, \(O_t=(-\infty,t)\) is a half-line. We assume
that \((M_t,\Omega)\) is a standard pair, so that the modular objects
\(J_t\) and \(\Delta_t\) are well defined. Thus
\begin{equation}
J_tM_tJ_t=M_t' .
\label{eq:typeIII_J_commutant}
\end{equation}
The induced local states considered below are assumed to be faithful and
normal on \(M_t\). When \(O_t=W_t\) is a wedge and we identify
\(M_t'\) with the algebra of the opposite wedge, we invoke wedge duality.
When the modular objects are given a geometric interpretation in the wedge
model, we additionally invoke the Bisognano--Wichmann theorem. At the
abstract level, the definition of the self-dual locus and of the
symmetrized comparison functional requires only the standard-pair
structure. The identification of the quadratic coefficient with a BKM
form requires the second-order differentiability assumption stated below.

In finite-dimensional type~I systems, the antiunitary \(J\) acts within a
single algebra and directly produces the reflected density matrix
\(J\rho(g)J\). In the local type~III setting, by contrast, there is no
intrinsic density matrix on \(M_t\), and the vacuum modular conjugation maps
\(M_t\) to its commutant. The local analogue therefore starts from one ambient state family and
uses \(J_t\) to pull the commutant restriction back to the original
local algebra \(M_t\).

Let \(\{\varpi_g\}_{g\in I}\) be a \(C^2\) family of normal states on
\(\mathcal B(\mathcal H)\), where \(g\) is the state-deformation parameter.
For each localization region \(O_t\), define the local restriction
\begin{equation}
\omega_{g,t}
:=
\varpi_g\!\restriction_{M_t}.
\label{eq:typeIII_local_restriction}
\end{equation}
Next introduce the canonical linear unital \(^{*}\)-anti-isomorphism
\begin{equation}
j_t:M_t\to M_t',
\qquad
j_t(A):=J_tA^*J_t.
\label{eq:typeIII_jt}
\end{equation}
The modularly pulled-back partner of \(\omega_{g,t}\) is then
\begin{equation}
\widetilde{\omega}_{g,t}
:=
\bigl(\varpi_g\!\restriction_{M_t'}\bigr)\circ j_t,
\label{eq:typeIII_partner_state}
\end{equation}
that is,
\begin{equation}
\widetilde{\omega}_{g,t}(A)
=
\varpi_g(J_tA^*J_t),
\qquad
A\in M_t.
\label{eq:typeIII_partner_state_explicit}
\end{equation}

Thus both \(\omega_{g,t}\) and \(\widetilde{\omega}_{g,t}\) are states on
the same local algebra \(M_t\). This is the local type~III replacement for
comparing two density matrices on a single type~I algebra. In the wedge
case \(O_t=W_t\), wedge duality identifies
\begin{equation}
M_t'
=
\mathcal A(W_t)'
=
\mathcal A(W_t'),
\label{eq:typeIII_wedge_duality_specialization}
\end{equation}
where \(W_t'\) denotes the causal complement, equivalently the opposite
wedge. In that case, \(\widetilde{\omega}_{g,t}\) may be viewed as the
restriction of the same ambient state to the opposite-wedge algebra,
pulled back to \(\mathcal A(W_t)\) by the modular anti-isomorphism
\(j_t(A)=J_tA^*J_t\). For other localization regions, such as the chiral
half-lines used below, the definition remains
\(\widetilde{\omega}_{g,t}:=(\varpi_g\!\restriction_{M_t'})\circ j_t\),
without requiring a separate wedge-duality interpretation.

We use the Hilbert-space convention that
\(\langle\cdot,\cdot\rangle\) is anti-linear in the first argument
and linear in the second. Thus, for an antiunitary involution \(J\),
\begin{equation}
\langle J\xi,J\eta\rangle=\langle \eta,\xi\rangle .
\label{eq:antiunitary-inner-product-convention}
\end{equation}
Consequently, for any bounded operator \(A\) and any vector \(\xi\),
\begin{equation}
\langle \xi,JA^*J\xi\rangle
=
\langle A^*J\xi,J\xi\rangle
=
\langle J\xi,A J\xi\rangle .
\label{eq:antiunitary-pullback-vector-state-identity}
\end{equation}
In particular, if the ambient state is represented by a vector
\(\xi_g\), so that
\begin{equation}
\varpi_g(B)=\langle \xi_g,B\xi_g\rangle ,
\qquad
B\in\mathcal B(\mathcal H),
\label{eq:ambient-vector-state-convention}
\end{equation}
then the modularly pulled-back state is the vector state on \(M_t\)
implemented by \(J_t\xi_g\):
\begin{equation}
\widetilde{\omega}_{g,t}(A)
=
\langle \xi_g,J_tA^*J_t\xi_g\rangle
=
\langle J_t\xi_g,A J_t\xi_g\rangle ,
\qquad
A\in M_t .
\label{eq:modular-pullback-vector-state-convention}
\end{equation}

The parameter \(g\) deforms the ambient state family, while the geometric
parameter \(t\) deforms the localization region and hence the algebra \(M_t\)
together with its modular conjugation \(J_t\). The local modular self-duality
condition is the fixed-point relation
\begin{equation}
\omega_{0,t}
=
\widetilde{\omega}_{0,t}.
\label{eq:typeIII_selfdual_condition}
\end{equation}
Equivalently,
\begin{equation}
\varpi_0(A)
=
\varpi_0(J_tA^*J_t),
\qquad
A\in M_t.
\label{eq:typeIII_selfdual_condition_explicit}
\end{equation}
At such a point, the local restriction of the ambient state coincides with
the modular pullback of its commutant restriction. This is the local type~III
analogue of the finite-dimensional fixed-point condition
\(\rho=J\rho J\). The self-dual locus is indexed by the geometric parameter
\(t\).

In standard form,
each faithful normal state has a unique natural-cone representative, and the
natural cone is fixed pointwise by the standard-form modular conjugation; see
Eq.~\eqref{eq:A-natural-cone-J-fixed}. The comparison used here
keeps the vacuum standard pair \((M_t,\Omega)\), its modular conjugation
\(J_t\), and the actual ambient preparation \(\varpi_g\) fixed. Self-duality
is therefore a relational condition on the region and commutant data of that
ambient state. The nontrivial response is the reflected-difference tangent
obtained from this fixed pairing, rather than the availability of a canonical
representative for the local restriction alone.

To compare the two local states at fixed \(t\), we use the Araki relative
entropy on the local algebra \(M_t\), with the convention fixed in
Appendix~\ref{sec:RelativeModularAraki}. Thus
\begin{equation}
S_{M_t}(\omega_{g,t}\Vert\widetilde{\omega}_{g,t})
=
-\bigl\langle
\xi_{\omega_{g,t}},
\log\Delta_{\widetilde{\omega}_{g,t}\mid\omega_{g,t}}
\,\xi_{\omega_{g,t}}
\bigr\rangle,
\label{eq:typeIII_Araki_entropy}
\end{equation}
where \(\xi_{\omega_{g,t}}\) is the natural-cone representative of
\(\omega_{g,t}\). This is the intrinsic local substitute for the
finite-dimensional Umegaki expression.

We define the symmetrized local comparison functional by
\begin{equation}
S(g,t)
:=
S_{M_t}(\omega_{g,t}\Vert\widetilde{\omega}_{g,t})
+
S_{M_t}(\widetilde{\omega}_{g,t}\Vert\omega_{g,t}).
\label{eq:typeIII_symmetrized_Araki}
\end{equation}
By construction, \(S(g,t)\geq 0\). At a modular self-dual point we write
\begin{equation}
\omega_{0,t}=\widetilde{\omega}_{0,t}=:\omega_t .
\end{equation}
Therefore
\begin{equation}
S(0,t)=0 .
\end{equation}
Thus \(g=0\) is a local minimum of the symmetrized Araki comparison functional. Whenever \(g\mapsto S(g,t)\) is differentiable at \(g=0\), one has
\begin{equation}
\left.\partial_g S(g,t)\right|_{g=0}=0 .
\end{equation}

We now make explicit the regularity assumption needed to identify the
quadratic coefficient with the local BKM form. Fix \(t\). Let
\((M_t)_*\) denote the predual of \(M_t\), and define
\begin{equation}
(M_t)_{*,0}^{\rm sa}
:=
\left\{\nu\in (M_t)_*^{\rm sa}:\nu(\mathbf 1)=0\right\}.
\label{eq:typeIII_tangent_space}
\end{equation}
Every \(C^1\) curve of normal states through \(\omega_t\) has tangent in this real Banach space. More explicitly, if \(\eta_g\) is such a curve with \(\eta_0=\omega_t\), then
\begin{equation}
\left.\partial_g\eta_g\right|_{g=0}\in (M_t)_{*,0}^{\rm sa}.
\end{equation}
We do not assume, however, that every element of \((M_t)_{*,0}^{\rm sa}\) is an entropy-admissible tangent direction. Instead, we restrict attention to directions realized by curves for which Araki relative entropy has a second-order expansion at the diagonal.

\medskip

We assume that the paired local curves
\begin{equation}
g\longmapsto \omega_{g,t},
\qquad
g\longmapsto \widetilde{\omega}_{g,t}
\end{equation}
are \(C^2\) as maps into \((M_t)_*\), take values in faithful normal states for \(|g|\) sufficiently small, and satisfy
\begin{equation}
\omega_{0,t}=\widetilde{\omega}_{0,t}=\omega_t .
\end{equation}
We further assume that, on the entropy-admissible tangent directions generated by the curves considered here, Araki relative entropy is twice differentiable at the diagonal and defines a positive symmetric bilinear form
\begin{equation}
\gamma_{\omega_t}^{\rm BKM}(\,\cdot\,,\,\cdot\,).
\end{equation}
Equivalently, for any admissible \(C^2\) curves
\begin{equation}
a\longmapsto \varphi_a,
\qquad
b\longmapsto \psi_b
\end{equation}
of faithful normal states on \(M_t\), with
\begin{align}
\varphi_0&=\omega_t, \\
\psi_0&=\omega_t, \\
\left.\partial_a\varphi_a\right|_{a=0}&=u, \\
\left.\partial_b\psi_b\right|_{b=0}&=v,
\end{align}
one has the diagonal expansion
\begin{equation}
S_{M_t}(\varphi_a\Vert\psi_b)
=
\frac{1}{2}\,
\gamma_{\omega_t}^{\rm BKM}(a u-b v,a u-b v)
+
o(a^2+b^2).
\end{equation}
Equivalently, the mixed Hessian is
\begin{equation}
\gamma_{\omega_t}^{\rm BKM}(u,v)
:=
-
\left.
\frac{\partial^2}{\partial a\,\partial b}
S_{M_t}(\varphi_a\Vert\psi_b)
\right|_{a=b=0}.
\end{equation}
The bilinear form is assumed to be independent of the admissible curves
used to realize the tangents. This is the local type III
Bogoliubov--Kubo--Mori bilinear form at \(\omega_t\), understood in the
sense of operator-algebraic information geometry and quasi-entropy
theory~\cite{Jencova2005,Petz1985Quasi, Hasegawa1993}.

Under this hypothesis, define the two tangents
\begin{align}
u_t^{(1)}
&:=
\left.\partial_g\omega_{g,t}\right|_{g=0}, \\
u_t^{(2)}
&:=
\left.\partial_g\widetilde{\omega}_{g,t}\right|_{g=0}.
\end{align}
The reflected-difference tangent is
\begin{equation}
u_t:=u_t^{(1)}-u_t^{(2)}
\in (M_t)_{*,0}^{\rm sa}.
\end{equation}
Applying the second-order expansion to
\begin{align}
\varphi_g&=\omega_{g,t}, \\
\psi_g&=\widetilde{\omega}_{g,t},
\end{align}
gives
\begin{equation}
S_{M_t}(\omega_{g,t}\Vert\widetilde{\omega}_{g,t})
=
\frac{1}{2}\,
\gamma_{\omega_t}^{\rm BKM}(u_t,u_t)\,g^2
+
o(g^2).
\end{equation}
Exchanging the two arguments gives
\begin{equation}
S_{M_t}(\widetilde{\omega}_{g,t}\Vert\omega_{g,t})
=
\frac{1}{2}\,
\gamma_{\omega_t}^{\rm BKM}(u_t,u_t)\,g^2
+
o(g^2).
\end{equation}
Therefore the symmetrized local comparison functional satisfies
\begin{equation}
S(g,t)
=
\gamma_{\omega_t}^{\rm BKM}(u_t,u_t)\,g^2
+
o(g^2).
\end{equation}
Consequently, the local modular Hessian
\begin{equation}
I_A(t):=
\left.\partial_g^2 S(g,t)\right|_{g=0}
\end{equation}
is
\begin{equation}
I_A(t)
=
2\,\gamma_{\omega_t}^{\rm BKM}(u_t,u_t).
\label{eq:typeIII-difference-tangent-hessian}
\end{equation}

It is useful to define the local BKM Fisher-information scalar of the
modularly paired family by
\begin{equation}
F_{\mathrm{BKM}}^{\mathcal A}(t)
:=
\gamma_{\omega_t}^{\mathrm{BKM}}(u_t,u_t).
\label{eq:typeIII_BKM_scalar}
\end{equation}
With this convention,
\begin{equation}
I_A(t)
=
2F_{\mathrm{BKM}}^{\mathcal A}(t).
\label{eq:typeIII_IA_FBKM_relation}
\end{equation}

Here \(\gamma_{\omega_t}^{\mathrm{BKM}}(u,v)\) is the local BKM bilinear
form, whereas \(F_{\mathrm{BKM}}^{\mathcal A}(t)\) is the scalar obtained
by evaluating that bilinear form on the particular tangent selected by the
modularly paired one-parameter family.
This convention differs from the finite-dimensional formula
\(I_J=8F_{\rm BKM}\) only because the finite-dimensional \(F_{\rm BKM}\)
there is defined using the single tangent \(\dot\rho_\star\), while the
symmetrized reflected comparison uses the tangent difference
\(2\dot\rho_\star\).

Operationally, \(I_A(t)\) is the local second-order distinguishability
susceptibility of the ambient deformation \(g\mapsto\varpi_g\) against
its modularly paired deformation at localization \(O_t\). It is not a
state function of \(\varpi_g\) alone: it depends on the chosen reference
standard pair \((M_t,\Omega)\), hence on the modular conjugation associated
with the region. This dependence is part of the construction. The quantity
\(I_A(t)\) measures the BKM norm of the reflected-difference tangent
\begin{equation}
u_t
=
\left.\partial_g\omega_{g,t}\right|_{g=0}
-
\left.\partial_g\widetilde{\omega}_{g,t}\right|_{g=0}.
\end{equation}
Only in the special case where the partner tangent equals the negative of
the original tangent does this become literally the odd component of the
state deformation. In the coherent-state examples below, the tangent
difference is represented by the reflected difference profile
\(\delta_t h=h-\vartheta_t h\).

This is the direct type~III counterpart of the type~I statement that the
Hessian of the symmetrized relative entropy is governed by the BKM metric;
the numerical factor depends on whether the Fisher information is defined
using a single tangent or the reflected tangent difference. For the
operator-algebraic information geometry of faithful normal states on a von
Neumann algebra, including the noncommutative divergence structure whose
second-order term yields the Kubo--Mori or Bogoliubov--Kubo--Mori geometry,
see~\cite{Jencova2005}. For the quasi-entropy framework on von Neumann
algebras and its relation to Araki relative entropy, see~\cite{Petz1985Quasi}.
In the present setting, the bilinear form
\(\gamma_{\omega_t}^{\mathrm{BKM}}(u,v)\) is the local type~III BKM metric at
the self-dual state \(\omega_t\), while \(F_{\mathrm{BKM}}^{\mathcal A}(t)\) is the
scalar obtained by evaluating that metric on the tangent difference \(u_t\)
selected by the modularly paired family. Thus \(F_{\mathrm{BKM}}^{\mathcal A}(t)\) is the appropriate
family-dependent BKM susceptibility associated with the modularly paired
local deformation, and \(I_A(t)=2F_{\mathrm{BKM}}^{\mathcal A}(t)\) is the
corresponding Hessian relation generated by the symmetrized Araki relative
entropy.

A useful special case occurs when the modularly paired family is locally odd
at the fixed point, in the sense that
\begin{equation}
\left.\partial_g\widetilde{\omega}_{g,t}\right|_{g=0}
=
-
\left.\partial_g\omega_{g,t}\right|_{g=0}.
\label{eq:typeIII_local_odd_pairing}
\end{equation}
Then
\begin{equation}
u_t
=
2\,\left.\partial_g\omega_{g,t}\right|_{g=0},
\label{eq:typeIII_local_odd_difference}
\end{equation}
and Eq.~\eqref{eq:typeIII-difference-tangent-hessian} reduces to
\begin{equation}
I_A(t)
=
8\,
\gamma_{\omega_t}^{\mathrm{BKM}}
\!\left(
\left.\partial_g\omega_{g,t}\right|_{g=0},
\left.\partial_g\omega_{g,t}\right|_{g=0}
\right).
\label{eq:typeIII_hessian_BKM_odd}
\end{equation}
In this odd-pairing case, the local type~III formula is exactly parallel to
the type~I reflected formula \(I_J=8F_{\mathrm{BKM}}\). In the explicit coherent-state models below, the more general
difference-tangent formula~\eqref{eq:typeIII-difference-tangent-hessian}
is the relevant one, with the reflected difference profile playing the role
of \(u_t\).

The local susceptibility \(I_A(t)\) depends both on the ambient state
family and on the localization geometry through the local algebra \(M_t\)
and its modular conjugation \(J_t\). The comparison problem is therefore naturally
organized as a two-parameter surface
\begin{equation}
(g,t)
\longmapsto
S(g,t),
\label{eq:typeIII_two_parameter_surface}
\end{equation}
where \(g\) measures departure from modular self-duality in state space and
\(t\) labels the localization geometry. Equivalently, \(t\) indexes a family
of local faithful state manifolds, self-dual base points \(\omega_t\), and
BKM bilinear forms \(\gamma_{\omega_t}^{\mathrm{BKM}}\). In the coherent-state
models studied below, \(S(g,t)\) is exactly quadratic in \(g\), so the
coefficients computed there are exact evaluations of the local BKM Fisher
information \(F_{\mathrm{BKM}}^{\mathcal A}(t)\). 

\section{Wedge Coherent States as an Exact Type III Model}
\label{sec:wedge_coherent_states}

We now present a concrete free-field realization of the local
type~III BKM--Araki fixed-point formalism introduced in
Sec.~\ref{sec:reflection_paired_families}. The conventions are those
fixed in the Appendices. The underlying coherent-state relative-entropy
formulas are known from
\cite{CasiniGrilloPontello2019,BostelmannCadamuroDelVecchio2022}. For the coherent family \(\omega_{gh}\), the modular pullback maps
the profile \(h\) to \(\vartheta_t h\).  The comparison therefore
depends only on the reflected difference profile
\(\delta_t h=h-\vartheta_t h\), and the known wedge entropy formula
evaluates the corresponding local BKM coefficient exactly. In this sense, the wedge
model furnishes an exact local-algebraic realization of the type~III
modular self-duality formalism.

We begin with the standard right wedge
\begin{equation}
W:=\{x\in\mathbb R^{1,d-1}:x^1>|x^0|\},
\end{equation}
and denote by
\begin{equation}
\mathcal A(W)\subset\mathcal B(\mathcal H)
\end{equation}
the corresponding represented local von Neumann algebra in the vacuum
representation. More generally, we write \(\mathcal A(W_t)\) for the
represented algebra of a wedge \(W_t\) in the family under
consideration. For the class of wedges used here, the vacuum vector
\(\Omega\) is cyclic and separating, so each pair
\begin{equation}
(\mathcal A(W_t),\Omega)
\end{equation}
is standard.

In the undeformed case \(W=W_0\), let \(U\) denote the vacuum
unitary representation of the proper orthochronous Poincar\'{e} group, and
let \(\Lambda_W(\chi)\) denote the one-parameter Lorentz boost
preserving the right wedge \(W\). We use the rapidity convention
\begin{equation}
\begin{array}{c}
\displaystyle
(\Lambda_W(\chi)x)^0
=
x^0\cosh\chi+x^1\sinh\chi,
\\[1ex]
\displaystyle
(\Lambda_W(\chi)x)^1
=
x^1\cosh\chi+x^0\sinh\chi,
\\[1ex]
\displaystyle
(\Lambda_W(\chi)x)^\perp
=
x^\perp .
\end{array}
\label{eq:wedge_boost_convention}
\end{equation}
With this convention, the Bisognano--Wichmann theorem identifies the
modular group of the vacuum standard pair
\((\mathcal A(W),\Omega)\) with the wedge-preserving boosts
\cite{BisognanoWichmann1975,BisognanoWichmann1976}:
\begin{equation}
\Delta_W^{is}
=
U(\Lambda_W(-2\pi s)),
\qquad
s\in\mathbb R .
\label{eq:BW_modular_group_wedge}
\end{equation}
Equivalently, the modular automorphism group acts geometrically on
wedge-localized observables \(A\in\mathcal A(W)\) by
\begin{equation}
\sigma_s^\Omega(A)
=
\Delta_W^{is}A\Delta_W^{-is}
=
U(\Lambda_W(-2\pi s))A
U(\Lambda_W(-2\pi s))^{-1}.
\label{eq:BW_modular_automorphism_wedge}
\end{equation}
The same theorem identifies the modular conjugation \(J_W\) with the
antiunitary operator implementing the corresponding wedge reflection.

More generally, for each wedge \(W_t\) in the translated family, we
write \(J_t\) for the vacuum modular conjugation of the standard pair
\((\mathcal A(W_t),\Omega)\). On Weyl operators, \(J_t\) induces a real
\textit{Weyl-label involution \(\vartheta_t\)}, characterized by
\begin{equation}
J_tW(f)J_t
=
W(\vartheta_t f),
\qquad
\vartheta_t^2=\mathbf 1 .
\label{eq:wedge_weyl_label_involution}
\end{equation}
The explicit form of \(\vartheta_t\) depends on the Weyl-label
convention and will be fixed below. These modular-theoretic ingredients
underlie the coherent-state formulas used in this section.

For the free scalar field, Sec.~\ref{smearedfields}  fixes the Weyl operators in terms of the smeared field \(\Phi(f)\). For
\(f\in C_0^\infty(\mathcal M,\mathbb R)\), we write
\begin{equation}
W(f):=e^{i\Phi(f)},
\qquad
\Omega_f:=W(f)\Omega,
\end{equation}
and denote by
\begin{equation}
\omega_f(B):=\langle \Omega_f,B\Omega_f\rangle,
\qquad
B\in\mathcal B(\mathcal H),
\end{equation}
the corresponding coherent state. Its restriction to a local algebra \(\mathcal A(W_t)\) is again denoted
by \(\omega_f\).  Since conjugation by \(W(f)\) acts on local Weyl
operators by a phase and hence preserves each local von Neumann algebra,
\(\omega_f|_{\mathcal A(W_t)}\) is obtained from the vacuum state by a
local automorphism.  Because the vacuum state is faithful on
\(\mathcal A(W_t)\), the coherent restriction is also faithful.  It is
normal in the vacuum representation, so the relative modular objects of
Appendix~\ref{sec:RelativeModularAraki} are well defined.

For coherent states, the Araki relative entropy satisfies the
difference identity \cite{BostelmannCadamuroDelVecchio2022}
\begin{equation}
S_{\mathcal A(W_t)}(\omega_f\Vert\omega_g)
=
S_{\mathcal A(W_t)}(\omega_{f-g}\Vert\omega_0)
\label{eq:wedge_coherent_difference}
\end{equation}
and, in particular,
\begin{equation}
S_{\mathcal A(W_t)}(\omega_f\Vert\omega_0)
=
S_{\mathcal A(W_t)}(\omega_0\Vert\omega_f).
\label{eq:wedge_coherent_vac_symmetry}
\end{equation}
Thus the comparison of two coherent states reduces to the comparison of
a single coherent excitation with the vacuum.

For the undeformed wedge \(W\), the relative entropy admits an exact
boost-energy formula. Let
\begin{equation}
P:=\Box+m^2,
\qquad
E:=E_{\mathrm{ret}}-E_{\mathrm{adv}},
\end{equation}
and let
\begin{equation}
\Psi_f:=Ef
\end{equation}
be the classical Klein--Gordon solution generated by \(f\), so that
\begin{equation}
(\Box+m^2)\Psi_f=0.
\end{equation}
For the wedge entropy formula, one introduces the left- and
right-wedge components of the excitation determined by \(f\). We write
this schematically as
\begin{equation}
f=f_L+f_R,
\qquad
\operatorname{supp}f_L\subset W',
\qquad
\operatorname{supp}f_R\subset W,
\label{eq:wedge_split_new}
\end{equation}
where \(W'\) is the causal complement of \(W\).  \(f_R\) denotes the right-wedge component of the associated
one-particle vector, equivalently of the Cauchy data of the classical
solution \(Ef\).  Also, this notation is
not meant as a literal decomposition of an arbitrary test function
\(f\in C_0^\infty(\mathcal M)\) inside \(C_0^\infty(\mathcal M)\).
Rather, it is the standard coherent-state wedge decomposition at the
level of the associated one-particle vector, equivalently of the
Cauchy data of the classical solution \(\Psi_f\) on the surface
\(x^0=0\). In this sense, the wedge algebra is sensitive only to the
right-wedge component, and the relative entropy depends only on
\(f_R\).

Let \(\mathcal H_1\) denote the one-particle Hilbert space of the
free scalar field, and let \(U_1\) be the corresponding one-particle
representation of the Poincar\'{e} group. The boost subgroup preserving
the right wedge acts on \(\mathcal H_1\) by a strongly continuous
one-parameter unitary group. With the boost convention fixed above, we
define its self-adjoint generator \(k_1\) by
\begin{equation}
U_1(\Lambda_W(\chi))
=
\exp(i\chi k_1).
\label{eq:one_particle_boost_generator_def}
\end{equation}
The induced Fock-space representation is the second quantization of the
one-particle representation,
\begin{equation}
U(\Lambda_W(\chi))
=
\Gamma\!\left(U_1(\Lambda_W(\chi))\right),
\label{eq:fock_boost_representation}
\end{equation}
and therefore
\begin{equation}
U(\Lambda_W(\chi))
=
\exp(i\chi K_1),
\qquad
K_1=d\Gamma(k_1).
\label{eq:fock_boost_generator_from_unitaries}
\end{equation}
Here \(d\Gamma(k_1)\) denotes the usual second quantization of the
one-particle generator \cite{DerezinskiGerard2013}. Thus \(k_1\) is the
one-particle boost generator, whereas \(K_1\) is the corresponding
boost generator on the symmetric Fock space.
With ~\eqref{eq:BW_modular_group_wedge} and
~\eqref{eq:fock_boost_generator_from_unitaries}, one has
\begin{equation}
\Delta_W^{is}
=
\exp(-2\pi i s K_1)
\label{eq:wedge_modular_group_fock_generator}
\end{equation}
or equivalently,
\begin{equation}
-\log\Delta_W
=
2\pi K_1.
\label{eq:wedge_modular_hamiltonian_positive}
\end{equation}

With this convention, the wedge relative entropy of a coherent
excitation is
\cite{CasiniGrilloPontello2019,BostelmannCadamuroDelVecchio2022}
\begin{equation}
S_{\mathcal A(W)}(\omega_f\Vert\omega_0)
=
2\pi\langle f_R,k_1 f_R\rangle .
\label{eq:wedge_boost_formula}
\end{equation}
This identity is understood as a quadratic-form identity on the form
domain of the one-particle boost generator. Equivalently, since
\(-\log\Delta_W=2\pi K_1\), it is the coherent-state expectation of the
wedge modular Hamiltonian, with the vacuum contribution subtracted. The
quantity \(\langle f_R,k_1 f_R\rangle\) is the corresponding
one-particle boost-energy quadratic form.

The relation between Eq.~\eqref{eq:wedge_boost_formula} and the
stress-tensor expression is the standard identification of the
one-particle boost quadratic form with the classical Noether charge for
the Lorentz boost preserving the right wedge. Let
\begin{equation}
\xi_W
:=
x^1\partial_0+x^0\partial_1
\label{eq:wedge_boost_killing_vector}
\end{equation}
be the boost Killing vector field associated with \(W\). For a classical
Klein--Gordon solution \(\Psi\), define the corresponding boost charge
on the wedge Cauchy surface by
\begin{equation}
Q_{\xi_W}[\Psi]
:=
\int_{x^0=0,\;x^1>0}
d\Sigma_\mu\,
T^\mu{}_\nu[\Psi]\xi_W^\nu .
\label{eq:wedge_classical_boost_charge}
\end{equation}
For the classical solution \(\Psi_{f_R}\) associated with the
right-wedge one-particle vector \(f_R\), the normalization used in
Eq.~\eqref{eq:wedge_boost_formula} is
\begin{equation}
\langle f_R,k_1f_R\rangle
=
Q_{\xi_W}[\Psi_{f_R}] .
\label{eq:wedge_one_particle_noether}
\end{equation}
Thus the word ``boost energy'' refers to the classical Noether charge
whose one-particle quantization gives the quadratic form of \(k_1\).

On the hypersurface \(x^0=0\), the future-directed surface element is
\begin{equation}
d\Sigma_\mu
=
(1,0,\ldots,0)\,d^{d-1}x,
\label{eq:wedge_surface_element}
\end{equation}
and the boost Killing vector restricts to
\begin{equation}
\xi_W\big|_{x^0=0}
=
x^1\partial_0 .
\label{eq:wedge_boost_vector_restriction}
\end{equation}
Therefore
\begin{equation}
T^\mu{}_\nu[\Psi]\xi_W^\nu d\Sigma_\mu
=
T^0{}_0[\Psi]\,x^1\,d^{d-1}x
=
x^1T_{00}[\Psi]\,d^{d-1}x ,
\label{eq:wedge_boost_charge_density}
\end{equation}
where in the last equality we used the mostly-minus metric convention.
Consequently,
\begin{equation}
\langle f_R,k_1f_R\rangle
=
\int_{x^1>0}
d^{d-1}x\,
x^1\,T_{00}[\Psi_{f_R}](0,\mathbf x).
\label{eq:wedge_boost_charge_slice}
\end{equation}
Since the wedge relative entropy depends only on the right-wedge component of the
excitation, one may write the resulting classical solution simply as \(\Psi_f\),
with the understanding that only its right-wedge component contributes. Substituting
\eqref{eq:wedge_boost_charge_slice} into \eqref{eq:wedge_boost_formula} gives
\begin{equation}
S_{\mathcal A(W)}(\omega_f\Vert\omega_0)
=
2\pi\int_{x^1>0}d^{d-1}x\,x^1\,T_{00}[\Psi_f](0,\mathbf x).
\label{eq:wedge_entropy_stresstensor}
\end{equation}

Here
\begin{equation}
T_{\mu\nu}[\Psi_f]
=
\partial_\mu\Psi_f\,\partial_\nu\Psi_f
-
\frac12\eta_{\mu\nu}
\bigl(
\partial^\alpha\Psi_f\,\partial_\alpha\Psi_f
-
m^2\Psi_f^{\,2}
\bigr)
\end{equation}
is the classical Klein--Gordon stress tensor. In particular, for the mostly-minus
metric convention,
\begin{equation}
T_{00}[\Psi_f]
=
\frac12
\Bigl(
(\partial_0\Psi_f)^2+|\nabla\Psi_f|^2+m^2\Psi_f^{\,2}
\Bigr),
\label{eq:wedge_T00_explicit_bridge}
\end{equation}
which yields the expanded form used below.
In particular,
\begin{equation}
\begin{array}{c}
S_{\mathcal A(W)}(\omega_f\Vert\omega_0)
=
2\pi\displaystyle\int_{x^1>0}d^{d-1}x\,x^1\\\\ \times
\dfrac12
\Bigl(
(\partial_0\Psi_f)^2+|\nabla\Psi_f|^2+m^2\Psi_f^{\,2}
\Bigr)\Big|_{x^0=0}.
\label{eq:wedge_entropy_stresstensor_expanded}
\end{array}
\end{equation}
By covariance, the analogous formula holds for any wedge \(W_t\) in the
family, with the corresponding modular boost generator and
wedge-adapted Cauchy slice.

We now specialize the modular self-duality construction of
Sec.~\ref{sec:reflection_paired_families} to coherent states. Fix
\(h\in C_0^\infty(\mathcal M,\mathbb R)\) and define the ambient
coherent-state family on \(\mathcal B(\mathcal H)\) by
\begin{equation}
\varpi_g:=\omega_{gh}.
\label{eq:wedge_varpi}
\end{equation}
For each wedge \(W_t\), its restriction to the local algebra is
\begin{equation}
\omega_{g,t}:=\varpi_g\!\restriction_{\mathcal A(W_t)}
=
\omega_{gh}\!\restriction_{\mathcal A(W_t)}.
\label{eq:wedge_local_family}
\end{equation}
In accordance with Sec.~\ref{sec:reflection_paired_families}, the local
partner is defined by modular pullback of the commutant restriction,
\begin{equation}
\widetilde\omega_{g,t}
:=
\bigl(\varpi_g\!\restriction_{\mathcal A(W_t)'}\bigr)\circ j_t,
\qquad
j_t(A):=J_tA^*J_t.
\label{eq:wedge_partner_def}
\end{equation}
Thus, for \(A\in\mathcal A(W_t)\),
\begin{equation}
\widetilde{\omega}_{g,t}(A)
=
\varpi_g(J_tA^*J_t)
=
\langle \Omega_{gh},J_tA^*J_t\Omega_{gh}\rangle .
\label{eq:scalar-pullback-before-antiunitary-identity}
\end{equation}
Using the antiunitary identity
\eqref{eq:antiunitary-pullback-vector-state-identity}, this becomes
\begin{equation}
\widetilde{\omega}_{g,t}(A)
=
\langle J_t\Omega_{gh},A J_t\Omega_{gh}\rangle .
\label{eq:scalar-pullback-vector-state}
\end{equation}
Using \(J_t\Omega=\Omega\), \(J_tW(f)J_t=W(\vartheta_t f)\), and
\(g\in\mathbb R\), one obtains
\begin{equation}
\begin{array}{c}
J_t\Omega_{gh}
=
J_tW(gh)\Omega
=
(J_tW(gh)J_t)J_t\Omega
\\\\=
W(g\vartheta_t h)\Omega
=
\Omega_{g\vartheta_t h}.
\label{eq:scalar-reflected-coherent-vector}
\end{array}
\end{equation}
Therefore
\begin{equation}
\widetilde{\omega}_{g,t}(A)
=
\langle \Omega_{g\vartheta_t h},A\Omega_{g\vartheta_t h}\rangle
=
\omega_{g\vartheta_t h}(A),
\;
A\in\mathcal A(W_t).
\label{eq:scalar-pulled-back-partner-coherent}
\end{equation}
So the modularly pulled-back partner is again a coherent state on the
same local algebra, now with profile \(\vartheta_t h\).
At \(g=0\), the two local states coincide,
\begin{equation}
\omega_{0,t}
=
\widetilde\omega_{0,t}
=
\omega_0\!\restriction_{\mathcal A(W_t)}.
\label{eq:wedge_selfdual}
\end{equation}
Thus \(g=0\) is the modular self-dual locus. 

Introduce the
\(t\)-dependent reflected difference profile
\begin{equation}
\delta_t h:=h-\vartheta_t h.
\label{eq:wedge_delta_profile}
\end{equation}
To connect with the general type~III BKM formalism, define for each
\(f\in C_0^\infty(\mathcal M,\mathbb R)\) the tangent functional
\begin{equation}
\nu_{f,t}(A)
:=
\left.\partial_g\,\omega_{gf}(A)\right|_{g=0},
\qquad
A\in\mathcal A(W_t).
\label{eq:wedge_tangent_functional_def}
\end{equation}
Equivalently,
\begin{equation}
\omega_{gf}(A)
=
\langle W(gf)\Omega,A W(gf)\Omega\rangle .
\label{eq:wedge_coherent_state_curve}
\end{equation}
Since \(\Omega\) lies in the domain of the smeared field
\(\Phi(f)\), the vector curve \(g\mapsto W(gf)\Omega\) is
norm differentiable at \(g=0\), with
\begin{equation}
\left.
\frac{d}{dg}W(gf)\Omega
\right|_{g=0}
=
i\Phi(f)\Omega .
\label{eq:wedge_coherent_vector_derivative}
\end{equation}
Therefore the derivative of the vector state is
\begin{equation}
\begin{array}{c}
\displaystyle
\nu_{f,t}(A)
=
\left.
\frac{d}{dg}
\langle W(gf)\Omega,A W(gf)\Omega\rangle
\right|_{g=0}
\\[2ex]
\displaystyle
=
\langle i\Phi(f)\Omega,A\Omega\rangle
+
\langle \Omega,A\,i\Phi(f)\Omega\rangle .
\end{array}
\label{eq:wedge_tangent_vector_state_derivative}
\end{equation}
Using the convention that the Hilbert-space inner product is antilinear
in the first argument and using the symmetry of \(\Phi(f)\), this becomes
\begin{equation}
\nu_{f,t}(A)
=
i\,\langle\Omega,[A,\Phi(f)]\Omega\rangle .
\label{eq:wedge_tangent_functional_comm}
\end{equation}
The last expression is understood as the normal functional obtained from
the differentiable coherent-state curve. Equivalently, the commutator
formula may first be read on the usual finite-particle core, or on
bounded Weyl-polynomial observables, and then extended by normality.

In particular, the tangent difference selected by the modularly paired
family is
\begin{align}
u_t(A)
&:=
\left.\partial_g\,\omega_{g,t}(A)\right|_{g=0}
-
\left.\partial_g\,\widetilde\omega_{g,t}(A)\right|_{g=0}
\nonumber\\[1ex]
&=
\nu_{h,t}(A)-\nu_{\vartheta_t h,t}(A)
=
\nu_{\delta_t h,t}(A).
\label{eq:wedge_difference_tangent}
\end{align}
Thus the local type~III BKM tangent is generated exactly by the
reflected difference profile \(\delta_t h\). In general,
\(\vartheta_t h\neq -h\), so the relevant formula is the general
difference-tangent relation of Sec.~\ref{sec:reflection_paired_families}
rather than the locally odd special case.

The coherent-state identities above now give
\begin{equation}
S_{\mathcal A(W_t)}(\omega_{g,t}\Vert\widetilde\omega_{g,t})
=
S_{\mathcal A(W_t)}(\omega_{g\delta_t h}\Vert\omega_0).
\label{eq:wedge_onesided_pair_reduction}
\end{equation}
By homogeneity of the coherent-state entropy,
\begin{equation}
S_{\mathcal A(W_t)}(\omega_{g\delta_t h}\Vert\omega_0)
=
g^2\,S_{\mathcal A(W_t)}(\omega_{\delta_t h}\Vert\omega_0),
\label{eq:wedge_homogeneity}
\end{equation}
and the same expression is obtained after exchanging the two arguments.
Therefore the symmetrized local comparison functional becomes
\begin{equation}
S(g,t)
=
2g^2\,S_{\mathcal A(W_t)}(\omega_{\delta_t h}\Vert\omega_0).
\label{eq:wedge_symmetrized_surface}
\end{equation}

Comparing \eqref{eq:wedge_symmetrized_surface} with the general
fixed-point expansion
\begin{equation}
S(g,t)=F_{\mathrm{BKM}}^{\mathcal A}(t)\,g^2+o(g^2),
\;
F_{\mathrm{BKM}}^{\mathcal A}(t)
:=
\gamma_{\omega_t}^{\mathrm{BKM}}(u_t,u_t),
\label{eq:wedge_general_BKM_expansion}
\end{equation}
where \(\omega_t=\omega_{0,t}\), we obtain the exact identification
\begin{equation}
F_{\mathrm{BKM}}^{\mathcal A}(t)
=
2\,S_{\mathcal A(W_t)}(\omega_{\delta_t h}\Vert\omega_0).
\label{eq:wedge_BKM_exact}
\end{equation}
Accordingly, the local Araki Hessian is
\begin{equation}
I_A(t)
:=
\partial_g^2 S(g,t)\big|_{g=0}
=
2\,F_{\mathrm{BKM}}^{\mathcal A}(t)
=
4\,S_{\mathcal A(W_t)}(\omega_{\delta_t h}\Vert\omega_0).
\label{eq:wedge_IA_exact}
\end{equation}
Thus, in the modularly defined coherent-state model, the coefficient
computed from the symmetrized Araki entropy is exactly the local BKM
Fisher information of the reflected-difference tangent, and the Hessian
is twice that Fisher information.

For the undeformed wedge \(W=W_0\), substitution of the boost-energy
formula gives
\begin{equation}
F_{\mathrm{BKM}}^{\mathcal A}(0)
=
4\pi\,\langle(\delta_0 h)_R,k_1(\delta_0 h)_R\rangle,
\label{eq:wedge_FBKM0_boost}
\end{equation}
where \((\delta_0 h)_R\) denotes the right-wedge component of
\(\delta_0 h=h-\vartheta_0 h\). Equivalently,
\begin{equation}
F_{\mathrm{BKM}}^{\mathcal A}(0)
=
4\pi\int_{x^1>0}d^{d-1}x\,x^1\,T_{00}[\Psi_{\delta_0 h}](0,\mathbf x),
\label{eq:wedge_FBKM0_stress}
\end{equation}
or, after expanding \(T_{00}\),
\begin{equation}
\begin{array}{c}
\; F_{\mathrm{BKM}}^{\mathcal A}(0)
=
2\pi\displaystyle\int_{x^1>0}d^{d-1}x\;x^1
\\\\ \times \Bigl(
(\partial_0\Psi_{\delta_0 h})^2
+
|\nabla\Psi_{\delta_0 h}|^2
+
m^2\Psi_{\delta_0 h}^{\,2}
\Bigr)\Big|_{x^0=0},
\end{array}
\label{eq:wedge_FBKM0_expanded}
\end{equation}
with
\begin{equation}
\Psi_{\delta_0 h}:=E(\delta_0 h).
\end{equation}
Hence
\begin{equation}
I_A(0)
=
2\,F_{\mathrm{BKM}}^{\mathcal A}(0)
=
8\pi\,\langle(\delta_0 h)_R,k_1(\delta_0 h)_R\rangle
\label{eq:wedge_IA0_boost}
\end{equation}
and
\begin{equation}
I_A(0)
=
8\pi\int_{x^1>0}d^{d-1}x\,x^1\,T_{00}[\Psi_{\delta_0 h}](0,\mathbf x).
\label{eq:wedge_IA0_stress}
\end{equation}
Thus both the local BKM Fisher information and the local Hessian are
governed by positive boost-energy forms of the reflected difference
profile.

To analyze the geometric dependence of these quantities, we now
restrict the localization parameter \(t\) to the one-parameter family of
wedges obtained by null translations of the right wedge,
\begin{equation}
W_t:=W+t\,v,
\qquad
v:=(1,1,0,\ldots,0).
\label{eq:wedge_null_family}
\end{equation}
Thus the edge of the wedge is translated along the null direction
\(v\). Let
\begin{equation}
\Sigma_t:=W_t\cap\{x^0=t\}
\end{equation}
denote the spatial section of the translated wedge on the Cauchy
surface \(x^0=t\). In coordinates,
\begin{equation}
\Sigma_t=\{(t,\mathbf x):x^1>t\}.
\end{equation}
For this family,
\begin{equation}
J_t=U(tv)\,J_W\,U(-tv),
\end{equation}
and the corresponding involution on Weyl labels is
\begin{equation}
\vartheta_t=\tau_{tv}\circ\vartheta_0\circ\tau_{-tv},
\end{equation}
where \(\tau_{tv}\) denotes spacetime translation by \(tv\). Here \(\tau_a\) acts on test functions by
\begin{equation}
(\tau_a f)(x):=f(x-a).
\label{eq:test_function_translation_convention}
\end{equation}

Equivalently, if
\begin{equation}
r_0(x^0,x^1,\mathbf x_\perp)
=
(-x^0,-x^1,\mathbf x_\perp)
\end{equation}
denotes the geometric wedge reflection associated with \(W\), then the
translated geometric reflection is
\begin{equation}
r_t(x)
=
tv+r_0(x-tv)
\end{equation}
or
\begin{equation}
r_t(x^0,x^1,\mathbf x_\perp)
=
(2t-x^0,\,2t-x^1,\,\mathbf x_\perp).
\label{eq:translated_wedge_reflection_coordinates}
\end{equation}
It satisfies \(r_t^2=\mathrm{id}\). 

It is important to distinguish the geometric action on the scalar field
from the induced action on Weyl labels. For the real scalar field there
is no additional internal sign in the geometric field transformation:
\begin{equation}
J_t\Phi(f)J_t
=
\Phi(f\circ r_t^{-1}).
\end{equation}
However, \(J_t\) is antiunitary, while the Weyl operators are defined by
\begin{equation}
W(f)=e^{i\Phi(f)}.
\end{equation}
Therefore
\begin{equation}
\begin{array}{c}
J_tW(f)J_t
=
J_t e^{i\Phi(f)}J_t
=
e^{-iJ_t\Phi(f)J_t}
\\\\=
e^{-i\Phi(f\circ r_t^{-1})}
=
W(-f\circ r_t^{-1}).
\end{array}
\end{equation}
Thus the involution \(\vartheta_t\) appearing in
\(J_tW(f)J_t=W(\vartheta_t f)\) is the Weyl-label
involution
\begin{equation}
(\vartheta_t h)(x)
:=
-h(r_t^{-1}x)
=
-h(r_t x),
\label{eq:scalar-translated-reflection-pullback}
\end{equation}
where the last equality uses \(r_t^2=\mathrm{id}\). The minus sign is
not an internal sign of the real scalar field; it is the sign forced on
the Weyl label by the antiunitarity of \(J_t\). 
Hence the reflected
difference profile entering the local susceptibility is
\begin{equation}
\delta_t h(x)
:=
h(x)-(\vartheta_t h)(x)
=
h(x)+h(r_t x).
\label{eq:scalar-translated-reflected-difference}
\end{equation}

By covariance of the coherent-state entropy formula, one has for each
fixed \(t\),
\begin{equation}
S_{\mathcal A(W_t)}(\omega_f\Vert\omega_0)
=
2\pi\int_{\Sigma_t}d^{d-1}x\,(x^1-t)\,T_{00}[\Psi_f](t,\mathbf x).
\label{eq:wedge_covariant_entropy}
\end{equation}
Applying this with \(f=\delta_t h\) yields
\begin{equation}
F_{\mathrm{BKM}}^{\mathcal A}(t)
=
4\pi\int_{\Sigma_t}d^{d-1}x\,(x^1-t)\,
T_{00}[\Psi_{\delta_t h}](t,\mathbf x),
\label{eq:wedge_FBKMt_stress}
\end{equation}
or equivalently
\begin{equation}
\begin{array}{c}
F_{\mathrm{BKM}}^{\mathcal A}(t)
=
2\pi\displaystyle\int_{x^1>t}d^{d-1}x\,(x^1-t)
\\\\ \times \Bigl(
(\partial_0\Psi_{\delta_t h})^2
+
|\nabla\Psi_{\delta_t h}|^2
+
m^2\Psi_{\delta_t h}^{\,2}
\Bigr)\Big|_{x^0=t}.
\end{array}
\label{eq:wedge_FBKMt_expanded}
\end{equation}
Therefore
\begin{equation}
I_A(t)
=
2\,F_{\mathrm{BKM}}^{\mathcal A}(t)
=
8\pi\int_{\Sigma_t}d^{d-1}x\,(x^1-t)\,
T_{00}[\Psi_{\delta_t h}](t,\mathbf x),
\label{eq:wedge_IAt_stress}
\end{equation}
or equivalently
\begin{equation}
\begin{array}{c}
I_A(t)
=
4\pi\displaystyle\int_{x^1>t}d^{d-1}x\,(x^1-t)
\\\\ \times \Bigl(
(\partial_0\Psi_{\delta_t h})^2
+
|\nabla\Psi_{\delta_t h}|^2
+
m^2\Psi_{\delta_t h}^{\,2}
\Bigr)\Big|_{x^0=t}.
\end{array}
\label{eq:wedge_IAt_expanded}
\end{equation}

This is the precise local type~III BKM--Araki coefficient selected by
the modular construction. The dependence on \(t\) enters simultaneously
through the translated wedge algebra \(\mathcal A(W_t)\) and through the
reflected profile \(\vartheta_t h\) appearing in \(\delta_t h\). The
geometric dependence is therefore encoded in the boost-energy of the
\(t\)-dependent reflected difference profile. Taken together, these
formulas furnish the wedge realization of the local type~III fixed-point
formalism.

\section{The Chiral \(U(1)\) Current as an Exact Type~III Model}
\label{sec:chiral_u1}

We next consider the chiral \(U(1)\) current, a standard exactly
solvable chiral conformal quantum field theory on the light ray. It may
be viewed as the left- or right-moving current sector of the
two-dimensional massless free boson. In light-cone coordinates one
retains a single null coordinate, denoted here by \(x\), and works with
a Hermitian current field \(j(x)\). Formally, this current is the
derivative of a chiral scalar field,
\begin{equation}
j(x)=\partial_x\phi(x).
\end{equation}
Thus only one chiral sector is retained; there are no independent left-
and right-moving degrees of freedom within the reduced model.

The local observables are generated by smeared currents
\begin{equation}
J(f):=\int_{\mathbb R} j(x)f(x)\,dx,
\qquad
f\in C^\infty_c(\mathbb R,\mathbb R).
\end{equation}
We use the symplectic convention
\begin{equation}
[J(f),J(g)]
=
2i\,\sigma(f,g)\mathbf 1,
\qquad
\sigma(f,g)
=
\int_{\mathbb R} f(x)g'(x)\,dx .
\label{eq:chiral_current_commutator}
\end{equation}
With the Weyl operators
\begin{equation}
\begin{array}{c}
W(f):=e^{iJ(f)},
\\\\
W(f)W(g)=e^{-i\sigma(f,g)}W(f+g),
\\\\
W(f)^*=W(-f),
\end{array}
\end{equation}
this is the CCR normalization used in the standard-subspace
relative-entropy formulas below.  For an interval or half-line
\(I\subset\mathbb R\), the local algebra \(\mathcal A(I)\) is generated
by those \(W(f)\) with \(\operatorname{supp}f\subset I\).

The coherent states used below are obtained by Weyl displacement of
the vacuum.  For \(h\in C^\infty_c(\mathbb R,\mathbb R)\), define
\begin{equation}
\omega_h(B)
:=
\langle \Omega,W(h)^*BW(h)\Omega\rangle,
\qquad
B\in\mathcal B(\mathcal H).
\end{equation}
Formally, this displacement shifts the smeared current by
\begin{equation}
\begin{array}{c}
W(h)^*J(f)W(h)
=
J(f)+2\sigma(h,f)\mathbf 1
\\\\=
J(f)-2\int_{\mathbb R}h'(x)f(x)\,dx\,\mathbf 1,
\end{array}
\end{equation}
where the last equality follows by integration by parts.  Thus, up to
the normalization fixed by the Weyl convention above, the Weyl label
\(h\) determines a classical current profile proportional to \(h'(x)\).
This is why the half-line relative-entropy formula below is naturally
expressed as a quadratic functional of \(h'\).

The restriction of \(\omega_h\) to each local algebra is faithful and
normal.  Indeed, by the Weyl relations,
\begin{equation}
W(h)^*W(f)W(h)
=
e^{2i\sigma(h,f)}W(f),
\qquad
\operatorname{supp}f\subset I,
\end{equation}
so conjugation by \(W(h)\) preserves the local algebra
\(\mathcal A(I)\).  Hence \(\omega_h\vert_{\mathcal A(I)}\) is obtained
from the faithful vacuum state by an automorphism of the local algebra.  Since it is a
vector state in the vacuum representation, it is normal.

The vacuum representation of the chiral current is a positive-energy
representation of the translation-dilation symmetry on the line. For
half-lines, the vacuum modular group acts geometrically, just as the
Bisognano--Wichmann modular group acts geometrically on wedge algebras
in the scalar-field example of Sec.~\ref{sec:wedge_coherent_states}.
This makes the chiral current a useful complementary exact model: the
same local type~III BKM--Araki fixed-point construction appears, but the
final sensitivity coefficient is a closed one-dimensional half-line
functional rather than a spacetime boost-energy integral.

We now recall the operator-algebraic realization of this model, following
the standard-subspace formulation of
~\cite{BostelmannCadamuroDelVecchio2022,GarbarzPalau2023}. In the
vacuum representation, the one-particle standard subspaces associated
with the half-lines \((-\infty,t)\) are
\begin{equation}
L_t^{U(1)}
:=
\overline{C_c^\infty((-\infty,t),\mathbb R)},
\end{equation}
where the closure is taken in the one-particle Hilbert space. Their
second quantization yields the represented local von Neumann algebras
\begin{equation}
\mathcal A((-\infty,t))
\subset
\mathcal B(\mathcal H),
\end{equation}
acting on the vacuum Hilbert space \(\mathcal H\) with vacuum vector
\(\Omega\). The modular flow acts geometrically on the standard
subspaces \(L_t^{U(1)}\), and hence on the half-line algebras. We write
\(J_t\) for the modular conjugation of the standard pair
\begin{equation}
(\mathcal A((-\infty,t)),\Omega).
\end{equation}

For the vacuum half-line
\begin{equation}
I_t:=(-\infty,t),
\label{eq:chiral-vacuum-half-line}
\end{equation}
the modular conjugation is the geometric orientation reversal of the light
ray about the endpoint \(t\). We write this reflection as
\begin{equation}
r_t(x):=2t-x,
\qquad
r_t(I_t)=(t,\infty).
\label{eq:chiral-vacuum-light-ray-reflection}
\end{equation}
The chiral current is treated as a one-form under this orientation reversal.

In the following formulas \(J_t\) denotes the modular conjugation,
whereas \(J(f)\) denotes the smeared current.
With the symplectic convention fixed above, this means that the current
distribution transforms as
\begin{equation}
J_t j(x)J_t=-j(r_t x).
\label{eq:chiral-vacuum-current-reflection}
\end{equation}
Consequently, for smeared currents,
\begin{equation}
J_t J(f)J_t=-J(f\circ r_t).
\label{eq:chiral-vacuum-smeared-current-reflection}
\end{equation}
Since \(J_t\) is antiunitary, \(J_t i J_t=-i\). Therefore the induced action
on Weyl operators is
\begin{equation}
J_t W(f)J_t=W(\vartheta_t f),
\;
(\vartheta_t f)(x):=(f\circ r_t)(x)=f(2t-x).
\label{eq:chiral-vacuum-theta-explicit}
\end{equation}
This convention is orientation-sensitive: the map \(\vartheta_t\) is
anti-symplectic,
\begin{equation}
\sigma(\vartheta_t f,\vartheta_t g)=-\sigma(f,g),
\label{eq:chiral-vacuum-theta-antisymplectic}
\end{equation}
which is precisely compensated by the antiunitarity of \(J_t\) in the Weyl
relations. Thus the vacuum reflected difference profile is
\begin{equation}
\delta_t h(x):=h(x)-(\vartheta_t h)(x)=h(x)-h(2t-x),
\label{eq:chiral-vacuum-delta-explicit}
\end{equation}
and hence
\begin{equation}
(\delta_t h)'(x)=h'(x)+h'(2t-x).
\label{eq:chiral-vacuum-delta-derivative-explicit}
\end{equation}

In the standard-subspace literature, the quantity often denoted
\(S_{L_t^{U(1)}}(h,h)\) is, in the present notation, the Araki relative
entropy
\begin{equation}
S_{\mathcal A((-\infty,t))}(\omega_h\Vert\omega_0),
\end{equation}
where \(\omega_0\) is the vacuum state restricted to the half-line
algebra. The key simplification in this model is that this quantity is explicitly
computable \cite{BostelmannCadamuroDelVecchio2022,GarbarzPalau2023},
\begin{equation}
S_{\mathcal A((-\infty,t))}(\omega_h\Vert\omega_0)
=
2\pi\int_{-\infty}^{t}(t-x)\,[h'(x)]^2\,dx.
\label{eq:chiral_vacuum_relent}
\end{equation}
Equivalently, this formula may be viewed as the coherent-state increase
of the vacuum modular Hamiltonian for the half-line. In traditional CFT
language, the vacuum modular Hamiltonian for \((-\infty,t)\) is the
geometric generator
\begin{equation}
K_t
=
2\pi\int_{-\infty}^{t}(t-x)\,T(x)\,dx,
\end{equation}
where \(T(x)\) is the chiral stress tensor. For the \(U(1)\) current, \(T(x)\) is the Sugawara energy density,
quadratic in the current, with the overall normalization chosen
consistently with the symplectic convention in
Eq.~\eqref{eq:chiral_current_commutator} and the relative-entropy
formula~\eqref{eq:chiral_vacuum_relent}. The
coherent displacement shifts the current by the classical profile
proportional to \(h'(x)\), and hence the relative entropy reduces to the
positive boundary-weighted quadratic form in
\eqref{eq:chiral_vacuum_relent}.

We now implement the local type~III construction of
Sec.~\ref{sec:reflection_paired_families}. Fix
\(h\in C_c^\infty(\mathbb R,\mathbb R)\) and define the ambient
coherent-state family on \(\mathcal B(\mathcal H)\) by
\begin{equation}
\varpi_g
:=
\omega_{gh}.
\label{eq:chiral_varpi}
\end{equation}
For each half-line \((-\infty,t)\), its restriction to the local
algebra is
\begin{equation}
\omega_{g,t}
:=
\varpi_g\!\restriction_{\mathcal A((-\infty,t))}
=
\omega_{gh}\!\restriction_{\mathcal A((-\infty,t))}.
\label{eq:chiral_local_family}
\end{equation}
As in Sec.~\ref{sec:reflection_paired_families}, the local partner is
defined by modular pullback of the commutant restriction,
\begin{equation}
\widetilde\omega_{g,t}
:=
\bigl(
\varpi_g\!\restriction_{\mathcal A((-\infty,t))'}
\bigr)\circ j_t,
\qquad
j_t(A):=J_tA^*J_t.
\label{eq:chiral_partner_def}
\end{equation}
Writing
\begin{equation}
\Omega_{gh}:=W(gh)\Omega ,
\label{eq:chiral-vacuum-coherent-vector}
\end{equation}
and using \eqref{eq:antiunitary-pullback-vector-state-identity}, we obtain
\begin{equation}
\widetilde{\omega}_{g,t}(A)
=
\langle J_t\Omega_{gh},A J_t\Omega_{gh}\rangle ,
\qquad
A\in\mathcal A((-\infty,t)).
\label{eq:chiral-vacuum-pullback-vector-state}
\end{equation}
Since \(J_t\Omega=\Omega\) and \(J_tW(f)J_t=W(\vartheta_t f)\),
\begin{equation}
J_t\Omega_{gh}
=
J_tW(gh)\Omega
=
W(g\vartheta_t h)\Omega
=
\Omega_{g\vartheta_t h}.
\label{eq:chiral-vacuum-reflected-coherent-vector}
\end{equation}
Hence, one has
\begin{equation}
\widetilde{\omega}_{g,t}
=
\omega_{g\vartheta_t h}
\!\restriction_{\mathcal A((-\infty,t))}.
\label{eq:chiral-vacuum-pulled-back-partner-coherent}
\end{equation}
Thus the modularly pulled-back partner is again a coherent state on
the same half-line algebra, now with reflected profile
\(\vartheta_t h\).

At \(g=0\), the two local states coincide:
\begin{equation}
\omega_{0,t}
=
\widetilde\omega_{0,t}
=
\omega_0\!\restriction_{\mathcal A((-\infty,t))}.
\label{eq:chiral_selfdual}
\end{equation}
Thus \(g=0\) is the modular self-dual locus. Introduce the reflected
difference profile
\begin{equation}
\delta_t h
:=
h-\vartheta_t h.
\label{eq:chiral_delta}
\end{equation}
To connect with the general type~III BKM formalism, define for each
\(f\in C_c^\infty(\mathbb R,\mathbb R)\) the tangent functional
\begin{equation}
\nu_{f,t}(A)
:=
\left.\partial_g\,\omega_{gf}(A)\right|_{g=0},
\qquad
A\in\mathcal A((-\infty,t)).
\label{eq:chiral_tangent_def}
\end{equation}
Then the tangent difference selected by the modularly paired family is
\begin{align}
u_t(A)
&:=
\left.\partial_g\,\omega_{g,t}(A)\right|_{g=0}
-
\left.\partial_g\,\widetilde\omega_{g,t}(A)\right|_{g=0}
\nonumber\\[1ex]
&=
\nu_{h,t}(A)-\nu_{\vartheta_t h,t}(A)
=
\nu_{\delta_t h,t}(A).
\label{eq:chiral_difference_tangent}
\end{align}
Thus the local type~III BKM tangent is generated exactly by the
reflected difference profile \(\delta_t h\).

The coherent-state identities now reduce the local comparison problem
to a one-sided vacuum entropy. For coherent states of the CCR net,
\begin{align}
S_{\mathcal A((-\infty,t))}(\omega_f\Vert\omega_g)
&=
S_{\mathcal A((-\infty,t))}
(\omega_{f-g}\Vert\omega_0),
\nonumber\\
S_{\mathcal A((-\infty,t))}(\omega_f\Vert\omega_0)
&=
S_{\mathcal A((-\infty,t))}
(\omega_0\Vert\omega_f).
\label{eq:chiral_coherent_difference_identity}
\end{align}
Therefore
\begin{equation}
S_{\mathcal A((-\infty,t))}(\omega_{g,t}\Vert\widetilde\omega_{g,t})
=
S_{\mathcal A((-\infty,t))}(\omega_{g\delta_t h}\Vert\omega_0).
\label{eq:chiral_one_sided_reduction}
\end{equation}
By homogeneity of the coherent-state relative entropy,
\begin{equation}
S_{\mathcal A((-\infty,t))}(\omega_{g\delta_t h}\Vert\omega_0)
=
g^2\,
S_{\mathcal A((-\infty,t))}(\omega_{\delta_t h}\Vert\omega_0).
\label{eq:chiral_homogeneity}
\end{equation}
The same expression is obtained after exchanging the two arguments.
Consequently the symmetrized local comparison functional becomes
\begin{equation}
S(g,t)
=
2g^2\,
S_{\mathcal A((-\infty,t))}(\omega_{\delta_t h}\Vert\omega_0).
\label{eq:chiral_symmetrized_surface}
\end{equation}

Comparing \eqref{eq:chiral_symmetrized_surface} with the general
fixed-point expansion
\begin{equation}
S(g,t)
=
F_{\mathrm{BKM}}^{\mathcal A}(t)\,g^2
+
o(g^2),
\;
F_{\mathrm{BKM}}^{\mathcal A}(t)
:=
\gamma_{\omega_t}^{\mathrm{BKM}}(u_t,u_t),
\label{eq:chiral_general_BKM}
\end{equation}
where \(\omega_t=\omega_{0,t}\), yields the exact identification
\begin{equation}
F_{\mathrm{BKM}}^{\mathcal A}(t)
=
2\,
S_{\mathcal A((-\infty,t))}(\omega_{\delta_t h}\Vert\omega_0).
\label{eq:chiral_BKM_exact}
\end{equation}
Accordingly,
\begin{equation}
I_A(t)
:=
\partial_g^2 S(g,t)\big|_{g=0}
=
2\,F_{\mathrm{BKM}}^{\mathcal A}(t)
.
\label{eq:chiral_IA_exact}
\end{equation}
Thus, in the modularly defined chiral model, the coefficient extracted
from the symmetrized Araki entropy is exactly the local BKM Fisher
information of the reflected-difference tangent, and the Hessian is
twice that Fisher information.

Substituting the explicit reflection convention
\eqref{eq:chiral-vacuum-delta-derivative-explicit} into the half-line
entropy formula gives
\begin{equation}
F_{\mathrm{BKM}}^{\mathcal A}(t)
=
4\pi
\int_{-\infty}^{t}
(t-x)\,
\bigl[h'(x)+h'(2t-x)\bigr]^2\,dx .
\label{eq:chiral-vacuum-F-explicit-reflection}
\end{equation}
Accordingly,
\begin{equation}
I_A(t)
=
8\pi
\int_{-\infty}^{t}
(t-x)\,
\bigl[h'(x)+h'(2t-x)\bigr]^2\,dx .
\label{eq:chiral-vacuum-I-explicit-reflection}
\end{equation}
The vacuum chiral \(U(1)\) current therefore provides an exact
one-dimensional realization of the local BKM--Araki fixed-point
principle. The modular susceptibility is the positive half-line quadratic
form in Eq.~\eqref{eq:chiral-vacuum-I-explicit-reflection}, obtained from
the reflected difference profile
\(\delta_t h(x)=h(x)-h(2t-x)\) and its derivative
\((\delta_t h)'(x)=h'(x)+h'(2t-x)\). 

The thermal half-line
extension is described in Appendix~\ref{app:thermal-chiral-u1}, where the
reflected profile is defined by the Borchers--Yngvason \cite{BorchersYngvason1999} standard-subspace
modular conjugation rather than by a pointwise light-ray reflection.

\section{MODULAR REFLECTION AND CANONICAL ENERGY IN AdS/CFT}
\label{sec:holography}
The anti-de Sitter/conformal field theory correspondence,
abbreviated as
AdS/CFT, relates a conformal field theory in \(d\) spacetime
dimensions to a gravitational theory in an asymptotically
\({\rm AdS}_{d+1}\) spacetime
\cite{Maldacena1998,GubserKlebanovPolyakov1998,Witten1998}.
The conformal field theory is defined on the conformal boundary
of the higher-dimensional spacetime.  A boundary state and a
boundary subregion determine a corresponding bulk region known
as the entanglement wedge. For a ball-shaped boundary region in the conformal field theory
vacuum, the modular susceptibility is twice the canonical energy of
the reflected-difference bulk perturbation in the associated
AdS--Rindler wedge. For such a region, conformal symmetry makes the modular flow of the
vacuum state geometric.  A conformal
transformation maps the causal development of the ball to a
hyperbolic cylinder, where the restricted vacuum is a thermal
state.  The corresponding bulk region is an AdS--Rindler wedge,
and the modular flow is generated by an AdS--Rindler Killing
field that vanishes on the associated extremal surface
\cite{CasiniHuertaMyers}.  The canonical energy is the quadratic
energy of a linearized bulk perturbation associated with this
Killing flow.  The Jafferis--Lewkowycz--Maldacena--Suh (JLMS) relation and the holographic
quantum Fisher information construction identify the second
variation of boundary relative entropy with this bulk canonical
energy
\cite{JLMS,FaulknerEtAl,LashkariVanRaamsdonk}.

The canonical energy does not by itself determine which bulk
perturbation represents the modularly paired comparison studied here.  The boundary modular conjugation does not act
internally on the ball algebra.  It maps the ball algebra to its
commutant, after which the commutant restriction is pulled back
to the original algebra.  When bulk and boundary modular
conjugations intertwine, the corresponding bulk operation is the
AdS--Rindler modular reflection.  The tangent selected by the
boundary comparison is therefore represented by the difference
between a bulk perturbation and its AdS--Rindler reflection.
Equivalently, the susceptibility depends only on the modular-odd
part of the bulk perturbation, which fixes the response direction
on which the canonical energy is evaluated.

In an exact complementary-recovery model for a linear scalar field
on a fixed AdS background, relative entropy is preserved and the
bulk and boundary modular conjugations intertwine
\cite{FaulknerConditional2020}.  The modular partner is then
exactly the reflected bulk coherent state, and the
symmetrized relative entropy is exactly quadratic in the
deformation parameter.  For coupled matter and metric
perturbations, the corresponding second-order semiclassical
quadratic form is the Hollands--Wald canonical energy of the
reflected-difference solution.

\subsection{Holographic background for the ball construction}
\label{subsec:holographic-background}

Anti-de Sitter space in \(d+1\) dimensions, denoted by
\({\rm AdS}_{d+1}\), is the maximally symmetric Lorentzian spacetime
with constant negative curvature.  In Poincar\'e coordinates
\((z,x^\mu)\), where \(z>0\) and \(\mu=0,\ldots,d-1\), its metric is
\begin{equation}
\begin{aligned}
ds^2
&=
\frac{L_{\rm AdS}^2}{z^2}
\left(
\eta_{\mu\nu}dx^\mu dx^\nu-dz^2
\right),
\\
\eta_{\mu\nu}
&=
{\rm diag}(1,-1,\ldots,-1).
\end{aligned}
\label{eq:ads-poincare-metric}
\end{equation}
Here \(L_{\rm AdS}\) is the AdS curvature radius and \(z\) is a
spatial coordinate.  The limit \(z\rightarrow0\) defines the
conformal boundary.  Although the metric diverges there, the rescaled
metric \(z^2ds^2/L_{\rm AdS}^2\) has the finite boundary
representative \(\eta_{\mu\nu}dx^\mu dx^\nu\).  Only its conformal
class is intrinsic.  The isometry group of \({\rm AdS}_{d+1}\) is
locally \(SO(2,d)\), which is also the global conformal group of
\(d\)-dimensional Minkowski spacetime.

For a scalar bulk field \(\phi\), the near-boundary expansion in
standard quantization has the form
\begin{equation}
\phi(z,x)
=
z^{d-\Delta}\phi_{(0)}(x)
+
z^\Delta\phi_{(\Delta)}(x)
+
\cdots,
\qquad
z\rightarrow0,
\label{eq:bulk-scalar-boundary-behavior}
\end{equation}
with the usual logarithmic modifications at exceptional values of
\(\Delta\).  The coefficient \(\phi_{(0)}\) is the boundary source
for a scalar CFT operator \(\mathcal O\) of scaling dimension
\(\Delta\), while \(\phi_{(\Delta)}\) contains the normalizable
response.  The bulk mass and boundary scaling dimension obey
\begin{equation}
m^2L_{\rm AdS}^2
=
\Delta(\Delta-d).
\label{eq:ads-mass-dimension-relation}
\end{equation}
The generating-functional relation is schematically
\begin{equation}
Z_{\rm CFT}[\phi_{(0)}]
=
Z_{\rm bulk}
\left[
\lim_{z\rightarrow0}
z^{\Delta-d}\phi(z,x)
=
\phi_{(0)}
\right].
\label{eq:ads-cft-generating-functional}
\end{equation}
The semiclassical regime used here evaluates the bulk functional
perturbatively around a classical AdS solution.  A Euclidean
boundary source may be present on the contour that prepares the
state.  After that source is switched off, the Lorentzian
perturbations obey the normalizable reflecting
boundary condition \(\phi_{(0)}=0\).  The preparation source and the
subsequent Lorentzian boundary condition therefore play distinct
roles.

Let \(B\) be a ball of radius \(R_B\), centered at
\(\mathbf{x}_B\) on the boundary time slice \(x^0=0\), and let
\(D(B)\) denote its causal development.  The correspondence also
associates bulk regions with boundary subregions.  The bulk region
associated with \(B\) is its entanglement wedge.  In the vacuum, a
constant-time bulk hypersurface is bounded by \(B\) and by the
Ryu--Takayanagi extremal surface \(\chi_B\), which is anchored on
\(\partial B\)
\cite{RyuTakayanagi2006}.  The Lorentzian entanglement wedge is the
domain of dependence of this hypersurface.  Within the semiclassical
reconstruction sector, bulk observables in the wedge can be
represented by boundary observables localized in \(B\)
\cite{DongHarlowWall}.

For the ball \(B\) in the CFT vacuum, the entanglement wedge is an
AdS--Rindler wedge.  It is a coordinate patch of pure AdS adapted to
the modular flow of the ball.  In suitable coordinates,
\begin{equation}
\begin{aligned}
ds^2
&=
\left(
\frac{\rho^2}{L_{\rm AdS}^2}-1
\right)d\tau^2
-
\frac{d\rho^2}{
\displaystyle
\frac{\rho^2}{L_{\rm AdS}^2}-1
}
-
\rho^2dH_{d-1}^2,
\end{aligned}
\label{eq:ads-rindler-metric}
\end{equation}
where \(dH_{d-1}^2\) is the metric of unit
\((d-1)\)-dimensional hyperbolic space.  The surface
\(\rho=L_{\rm AdS}\) is a Killing horizon whose bifurcation surface
is \(\chi_B\).  The coordinate \(\tau\) is the bulk counterpart of
the modular-flow parameter of the boundary ball.

The Casini--Huerta--Myers conformal transformation maps the causal
development of the ball to a hyperbolic cylinder
\cite{CasiniHuertaMyers}.  Under this map, the vacuum restricted to
the ball becomes a thermal state, and the AdS--Rindler wedge becomes
the exterior of the massless hyperbolic black hole described by
Eq.~\eqref{eq:ads-rindler-metric}.

Write \(x=(x^0,\mathbf{x})\) for the boundary coordinates.  Repeated
spatial indices \(i=1,\ldots,d-1\) are summed,
\(\partial_\mu:=\partial/\partial x^\mu\), and \(|\cdot|\) denotes
the Euclidean norm on a constant-time slice.  The modular flow of the
vacuum state restricted to \(B\) is generated by the conformal
Killing field
\begin{equation}
\begin{array}{rcl}
\zeta_B
&=&
\displaystyle
\frac{\pi}{R_B}
\left[
\left(
R_B^2-(x^0)^2-|\mathbf{x}-\mathbf{x}_B|^2
\right)\partial_0
\right.
\\[1mm]
&&
\displaystyle
\left.
\hspace{2.0cm}
-2x^0(x^i-x_B^i)\partial_i
\right].
\end{array}
\label{eq:ball-conformal-killing-field}
\end{equation}
On the slice \(x^0=0\), the nontrivial part of the modular
Hamiltonian of the vacuum state restricted to \(B\) is
\begin{equation}
K_B^{\rm CFT}
=
2\pi
\int_B
d^{d-1}x\,
\frac{
R_B^2-|\mathbf{x}-\mathbf{x}_B|^2
}{
2R_B
}
T_{00}(0,\mathbf{x})
\label{eq:ball-vacuum-modular-hamiltonian}
\end{equation}
up to an additive scalar \cite{CasiniHuertaMyers}.  Here \(T_{00}\)
is the boundary CFT energy density.  In the hyperbolic-cylinder
frame, the vacuum state restricted to \(B\) is a
Kubo--Martin--Schwinger (KMS) state with
\begin{equation}
\beta_B
=
2\pi R_B,
\qquad
T_B
=
\frac{1}{2\pi R_B}.
\label{eq:ball-hyperbolic-temperature}
\end{equation}
The AdS--Rindler Killing field \(\xi_B\) extends \(\zeta_B\) into
the bulk and vanishes on the bifurcation surface \(\chi_B\)
\cite{CasiniHuertaMyers,FaulknerEtAl}.

The bulk and boundary relative entropies are connected by the
Jafferis--Lewkowycz--Maldacena--Suh (JLMS) relation
\cite{JLMS}.  At the semiclassical order under consideration, it
identifies the relative entropy of two boundary states restricted to
\(B\) with the relative entropy of their bulk states restricted to
the entanglement wedge.  The gravitational area contribution appears
separately in the modular Hamiltonian and in the holographic entropy
but cancels from their relative-entropy difference.  Consequently,
the second variation of boundary relative entropy can be calculated
from the corresponding bulk perturbations.  For perturbations about
the vacuum, this bulk quadratic form is the canonical energy
associated with the AdS--Rindler Killing flow
\cite{FaulknerEtAl,LashkariVanRaamsdonk}.

Consider a neutral linear scalar field on the fixed AdS background.
It is described by the same
Weyl-algebra construction used in Sec.~\ref{sec:vN_TT}.  Let
\begin{equation}
P
:=
\Box_g+m^2
\label{eq:bulk-klein-gordon-operator}
\end{equation}
be its Klein--Gordon operator.  Since the conformal boundary of AdS
is timelike, a boundary condition must be specified in addition to
the field equation.  A normalizable reflecting boundary condition
compatible with the chosen AdS vacuum is imposed, with a well-posed
initial-boundary-value problem.

The real symplectic space underlying the bulk Weyl algebra is
\begin{equation}
\mathcal K_{\mathbb R}^{\rm bulk}
:=
C_0^\infty({\rm AdS},\mathbb R)
\big/
P C_0^\infty({\rm AdS},\mathbb R),
\label{eq:bulk-real-symplectic-space}
\end{equation}
with the chosen AdS boundary condition understood. As in
Sec.~\ref{smearedfields}, \(f\) denotes a real test function and
\([f]\) its class in \(\mathcal K_{\mathbb R}^{\rm bulk}\). Since
\(\Phi(Pk)=0\) for every
\(k\in C_0^\infty({\rm AdS},\mathbb R)\), the test functions \(f\)
and \(f+Pk\) define the same smeared quantum field and the same Weyl
operator,
\begin{equation}
W(f)
:=
e^{i\Phi(f)}.
\label{eq:bulk-weyl-operator}
\end{equation}
The brackets on \([f]\) will be suppressed when no ambiguity can
arise.  The notation also includes the admissible
vacuum-representation completion of
\(\mathcal K_{\mathbb R}^{\rm bulk}\).  Throughout, labels in this
completion are restricted to the finite domain of the AdS--Rindler
coherent-state relative-entropy form, equivalently, after the
identification below, of the canonical-energy form.

Let
\begin{equation}
E
:=
E_{\rm ret}-E_{\rm adv}
\label{eq:bulk-causal-propagator}
\end{equation}
be the causal propagator determined by the same boundary condition.
The test function \(f\) determines the real classical solution
\begin{equation}
\Psi_f
:=
Ef.
\label{eq:bulk-classical-solution-from-weyl-label}
\end{equation}
Because \(EPk=0\), the solution depends only on \([f]\).  The
test-function class \([f]\) specifies the Weyl displacement, while
\(\Psi_f\) is the corresponding classical bulk perturbation.

The Klein--Gordon symplectic current and symplectic form are
\begin{equation}
\omega_{\rm KG}^{A}
\left(
\Psi_1,\Psi_2
\right)
:=
\Psi_1\nabla^A\Psi_2
-
\Psi_2\nabla^A\Psi_1
\label{eq:bulk-klein-gordon-symplectic-current}
\end{equation}
and
\begin{equation}
\sigma_{\rm KG}
\left(
\Psi_1,\Psi_2
\right)
:=
\int_{\Sigma}
d\Sigma_A\,
\omega_{\rm KG}^{A}
\left(
\Psi_1,\Psi_2
\right).
\label{eq:bulk-klein-gordon-symplectic-form}
\end{equation}
The field equation and the reflecting boundary condition make the
integral independent of \(\Sigma\).

For \(g\in\mathbb R\), the vector
\begin{equation}
\Omega_{g,f}^{\rm bulk}
:=
W(gf)\Omega_{\rm bulk}
\label{eq:bulk-coherent-vector}
\end{equation}
is obtained by applying the Weyl operator to the bulk vacuum and
represents the field-theoretic counterpart of a harmonic-oscillator
coherent state.  In the oscillator construction
\(\lvert\alpha\rangle=D(\alpha)\lvert0\rangle\), the displacement
operator \(D(\alpha)\) translates the oscillator phase-space
variables.  Here \(W(gf)\) plays the corresponding role for the bulk
field.  The Weyl label \(gf\) corresponds to the classical solution
\(\Psi_{gf}=g\Psi_f\), while its connected correlation functions of
order two and higher agree with those of the vacuum.  The
test-function class \([f]\) need not belong entirely to the real
standard subspace of the AdS--Rindler wedge or to its symplectic
complement.

\subsection{The ball algebra and its exact modular partner}
\label{subsec:boundary-ball-reflection}

Let the boundary CFT be represented on its vacuum Hilbert space
\(\mathcal H_{\rm CFT}\), with vacuum vector \(\Omega\).  The
represented algebra associated with the causal development \(D(B)\)
is
\begin{equation}
M_B
:=
\mathcal A_{\rm CFT}(D(B)).
\label{eq:boundary-ball-algebra}
\end{equation}
Under the standard assumptions of local quantum field theory, the
vacuum is cyclic and separating for \(M_B\)
\cite{Haag1,Guido2011}.  Hence \((M_B,\Omega)\) is standard, and its
modular conjugation \(J_B\) satisfies
\begin{equation}
J_BM_BJ_B
=
M_B',
\qquad
J_B\Omega
=
\Omega.
\label{eq:ball-standard-pair}
\end{equation}
The associated modular anti-isomorphism is
\begin{equation}
j_B(A)
:=
J_BA^*J_B,
\qquad
j_B
:
M_B
\longrightarrow
M_B'.
\label{eq:ball-modular-anti-isomorphism}
\end{equation}
When Haag duality holds, \(M_B'\) is the algebra of the causal
complement of \(D(B)\), understood on the conformal compactification.

Let \(\psi\in\mathcal H_{\rm CFT}\) be a unit vector, and let
\(\varpi_\psi\) denote the corresponding normal vector state on
\(\mathcal B(\mathcal H_{\rm CFT})\), defined by
\[
\varpi_\psi(X)
:=
\langle\psi,X\psi\rangle,
\qquad
X\in\mathcal B(\mathcal H_{\rm CFT}).
\]
Its restriction to the ball algebra is
\begin{equation}
\omega_{\psi,B}(A)
:=
\varpi_\psi(A)
=
\langle\psi,A\psi\rangle,
\qquad
A\in M_B.
\label{eq:ambient-ball-state}
\end{equation}
Its modular partner is the pullback of the commutant restriction,
\begin{equation}
\widetilde{\omega}_{\psi,B}
:=
\left(
\varpi_\psi|_{M_B'}
\right)
\circ j_B.
\label{eq:exact-ball-partner-definition}
\end{equation}
Antiunitarity of \(J_B\) gives
\begin{equation}
\begin{aligned}
\widetilde{\omega}_{\psi,B}(A)
&=
\langle\psi,J_BA^*J_B\psi\rangle
\\
&=
\langle J_B\psi,A J_B\psi\rangle.
\end{aligned}
\label{eq:exact-ball-partner-vector}
\end{equation}
Thus the modular partner is the restriction to \(M_B\) of the vector
state implemented by \(J_B\psi\). For the vacuum,
\begin{equation}
\omega_{\Omega,B}
=
\widetilde{\omega}_{\Omega,B}
=:
\omega_B.
\label{eq:ball-vacuum-fixed-point}
\end{equation}

The geometric meaning of \(J_B\) follows from conformal modular
covariance.  Let \(C_B\) map a Rindler wedge \(W\) to \(D(B)\), and
let \(r_W\) be the Bisognano--Wichmann wedge reflection.  Up to the
theory-dependent charge conjugation, parity, and time reversal (CPT)
action on spin and internal indices,
\begin{equation}
J_B
=
U(C_B)J_WU(C_B)^*,
\qquad
r_B
:=
C_B\circ r_W\circ C_B^{-1}.
\label{eq:ball-geometric-modular-reflection}
\end{equation}
The map \(r_B\) sends \(D(B)\) to its causal complement.  This
double-cone modular action was established explicitly for the free
massless scalar field in \cite{HislopLongo1982} and
operator-algebraically for conformal nets in
\cite{BrunettiGuidoLongo1993}.

Equation~\eqref{eq:exact-ball-partner-vector} also fixes the
reflected source without a separate pointwise prescription. Let
\(\eta\) be a real Lorentzian boundary source, and let
\(X_\eta=X_\eta^*\) denote the corresponding smeared boundary
operator. For \(g\in\mathbb R\), the associated unitary preparation
of the boundary CFT vacuum \(\Omega\) is
\begin{equation}
\Omega_{g,\eta}
=
e^{igX_\eta}\Omega.
\label{eq:lorentzian-source-preparation}
\end{equation}
The reflected source operator is defined by
\begin{equation}
X_{\eta^{J_B}}
:=
-J_BX_\eta J_B.
\label{eq:lorentzian-reflected-source-operator}
\end{equation}
The minus sign follows from \(J_BiJ_B=-i\), and
\begin{equation}
J_Be^{igX_\eta}\Omega
=
e^{igX_{\eta^{J_B}}}\Omega.
\label{eq:lorentzian-source-partner}
\end{equation}
For a Euclidean preparation, the reflected source is obtained by the
corresponding contour reflection, complex conjugation, and CPT action
on the inserted operator.  Conformal weights, Jacobians, and spin
matrices are included by defining the reflected smeared operator
through Eq.~\eqref{eq:lorentzian-reflected-source-operator}.

If \(U_g\in M_B\) is unitary, then
\(J_BU_gJ_B\in M_B'\) and commutes with every \(A\in M_B\).
Consequently,
\begin{equation}
\widetilde{\omega}_{U_g\Omega,B}(A)
=
\langle\Omega,A\Omega\rangle,
\qquad
A\in M_B.
\label{eq:ball-local-unitary-partner}
\end{equation}
A unitary excitation localized entirely in the ball therefore has
the vacuum as its modular partner on \(M_B\). A nonvacuum modular
partner requires a preparation whose restriction to \(M_B'\) differs
from that of the vacuum.

\subsection{AdS--Rindler representation of the reflected-difference tangent}
\label{subsec:grav-content}

Let
\begin{equation}
\mathcal N_B
:=
\mathcal A_{\rm bulk}(\mathcal W_B)
\label{eq:bulk-wedge-algebra}
\end{equation}
be the represented von Neumann algebra generated by the linear bulk
field in the AdS--Rindler wedge \(\mathcal W_B\), acting on
\(\mathcal H_{\rm bulk}\). The bulk vacuum
\(\Omega_{\rm bulk}\) is cyclic and separating for
\(\mathcal N_B\), so
\((\mathcal N_B,\Omega_{\rm bulk})\) is standard. We denote its
modular conjugation by \(J_B^{\rm bulk}\).

For the AdS--Rindler vacuum, the bulk modular conjugation implements
a geometric wedge reflection. Let
\(\mathcal R_B^{\rm bulk}\) denote the AdS isometry induced by the
boundary reflection \(r_B\). It reverses the two normal directions
to the bifurcation surface and exchanges the wedge with its causal
complement,
\begin{equation}
\mathcal R_B^{\rm bulk}(\mathcal W_B)
=
\mathcal W_B',
\qquad
\left(
\mathcal R_B^{\rm bulk}
\right)^2
=
1.
\label{eq:geometric-bulk-modular-reflection}
\end{equation}
For a neutral scalar, the corresponding real-linear action on
classical solutions is
\begin{equation}
\left(
\mathcal J_B^{\rm cl}\Psi
\right)(X)
:=
\Psi
\left(
\left(
\mathcal R_B^{\rm bulk}
\right)^{-1}X
\right),
\qquad
\left(
\mathcal J_B^{\rm cl}
\right)^2
=
1.
\label{eq:classical-bulk-modular-reflection}
\end{equation}
For fields with spin or internal quantum numbers, the appropriate
tensor, spin, and CPT actions are included in
\(\mathcal J_B^{\rm cl}\).

Let
\begin{equation}
V
:
\mathcal H_{\rm bulk}
\longrightarrow
\mathcal H_{\rm CFT},
\qquad
V\Omega_{\rm bulk}
=
\Omega,
\label{eq:bulk-boundary-isometry}
\end{equation}
be an isometry from the bulk reconstruction sector into the boundary
CFT Hilbert space. Assume that \(\mathcal N_B\) is standardly
\(c\)-reconstructable from \(M_B\) in the sense of
Ref.~\cite{FaulknerConditional2020}.  Theorem~2 of that reference
then implies preservation of relative entropy for normal states in
the reconstruction sector and the bulk--boundary
modular-intertwining relation
\begin{equation}
VJ_B^{\rm bulk}
=
J_BV.
\label{eq:bulk-boundary-modular-intertwining}
\end{equation}
These conclusions are exact in the complementary-recovery model.
Their extension to coupled matter and metric perturbations,
discussed in Subsec.~\ref{subsec:semiclassical-gravity}, requires
additional assumptions at the retained semiclassical order.

The action of \(J_B^{\rm bulk}\) on the Weyl algebra induces an
involution \(\vartheta_B\) on
\(\mathcal K_{\mathbb R}^{\rm bulk}\),
\begin{equation}
J_B^{\rm bulk}W(f)J_B^{\rm bulk}
=
W(\vartheta_Bf),
\qquad
\vartheta_B^2
=
1.
\label{eq:bulk-weyl-label-reflection}
\end{equation}
As in the Minkowski-wedge construction of
Sec.~\ref{sec:wedge_coherent_states}, \(\vartheta_B\) includes the
sign required by
\[
J_B^{\rm bulk}iJ_B^{\rm bulk}=-i.
\]
Let \(\mathcal R_{B*}^{\rm bulk}\) denote the pullback induced by
\(\mathcal R_B^{\rm bulk}\) on scalar functions.  We use the same
symbol for its action on test functions and on classical solutions.
For the Weyl convention \(W(f)=e^{i\Phi(f)}\), antiunitarity gives
\begin{equation}
\vartheta_Bf
=
-\mathcal R_{B*}^{\rm bulk}f
\end{equation}
for a CPT-even neutral scalar.  Since
\(\mathcal R_B^{\rm bulk}\) reverses time orientation, it exchanges
the retarded and advanced propagators.  Consequently,
\begin{equation}
E\mathcal R_{B*}^{\rm bulk}
=
-\mathcal R_{B*}^{\rm bulk}E.
\end{equation}
The two signs cancel when the reflected Weyl label is mapped to its
classical solution:
\begin{equation}
\begin{aligned}
\Psi_{\vartheta_Bf}
&=
E\vartheta_Bf
\\
&=
\mathcal R_{B*}^{\rm bulk}Ef
\\
&=
\mathcal J_B^{\rm cl}\Psi_f.
\end{aligned}
\label{eq:weyl-label-to-classical-reflection}
\end{equation}
For fields with nontrivial CPT parity, spin, or internal quantum
numbers, the corresponding actions are included in both
\(\vartheta_B\) and \(\mathcal J_B^{\rm cl}\).

The difference between a Weyl label and its modular reflection is
\begin{equation}
\delta_Bf
:=
f-\vartheta_Bf.
\label{eq:bulk-reflected-label}
\end{equation}
It vanishes for a modular-even label and is twice the
\(\vartheta_B\)-odd part of \(f\).

If the one-particle vector determined by \([f]\) lies in the real
standard subspace associated with \(\mathcal W_B\), then
\[
W(f)\in\mathcal N_B,
\qquad
W(\vartheta_Bf)\in\mathcal N_B'.
\]
Consequently, for every \(A\in\mathcal N_B\),
\begin{equation}
\begin{aligned}
&
\left\langle
W(g\vartheta_Bf)\Omega_{\rm bulk},
A W(g\vartheta_Bf)\Omega_{\rm bulk}
\right\rangle
\\
&\hspace{2.0cm}
=
\langle\Omega_{\rm bulk},A\Omega_{\rm bulk}\rangle.
\end{aligned}
\label{eq:bulk-localized-partner-vacuum}
\end{equation}
The reflected coherent state therefore restricts to the vacuum on
\(\mathcal N_B\). In this localized case, the modular comparison
reduces to the usual comparison between an excitation in the ball
and the vacuum. A nonvacuum modular partner requires a Weyl label
that is not confined to the wedge standard subspace.

For \(f\in\mathcal K_{\mathbb R}^{\rm bulk}\) in the reconstruction
sector, define the corresponding boundary vector by
\begin{equation}
\Omega_{g,f}^{\rm CFT}
:=
V\Omega_{g,f}^{\rm bulk}
=
VW(gf)\Omega_{\rm bulk}
\label{eq:encoded-bulk-coherent-vector}
\end{equation}
and its restriction to the boundary ball by
\begin{equation}
\omega_{g,f;B}(A)
:=
\left\langle
\Omega_{g,f}^{\rm CFT},
A\Omega_{g,f}^{\rm CFT}
\right\rangle,
\qquad
A\in M_B.
\label{eq:encoded-ball-coherent-state}
\end{equation}
Since \(J_B^{\rm bulk}\Omega_{\rm bulk}=\Omega_{\rm bulk}\), the
modular-intertwining relation gives
\begin{equation}
\begin{aligned}
J_B\Omega_{g,f}^{\rm CFT}
&=
J_BVW(gf)\Omega_{\rm bulk}
\\
&=
VJ_B^{\rm bulk}W(gf)\Omega_{\rm bulk}
\\
&=
VW(g\vartheta_Bf)\Omega_{\rm bulk}.
\end{aligned}
\label{eq:encoded-coherent-partner-vector}
\end{equation}
Together with Eq.~\eqref{eq:exact-ball-partner-vector}, this yields
the exact boundary modular partner
\begin{equation}
\widetilde{\omega}_{g,f;B}
=
\omega_{g,\vartheta_Bf;B}.
\label{eq:exact-encoded-ball-partner}
\end{equation}
Thus the two boundary families entering the modular comparison are
the images under \(V\) of the bulk coherent families with labels
\(f\) and \(\vartheta_Bf\). At \(g=0\), both families reduce to the
restricted boundary vacuum,
\begin{equation}
\omega_{0,f;B}
=
\widetilde{\omega}_{0,f;B}
=
\omega_B.
\label{eq:encoded-ball-self-dual-point}
\end{equation}

Let
\begin{equation}
v_{B,f}
:=
\left.
\partial_g\omega_{g,f;B}
\right|_{g=0}
\label{eq:single-family-ball-tangent}
\end{equation}
denote the tangent generated by the coherent family with label \(f\).
The tangent selected by the modularly paired comparison is
\begin{equation}
u_{B,f}(A)
:=
\left.
\frac{d}{dg}
\left[
\omega_{g,f;B}(A)
-
\widetilde{\omega}_{g,f;B}(A)
\right]
\right|_{g=0}.
\label{eq:encoded-ball-difference-tangent}
\end{equation}
The exact partner relation
\(\widetilde{\omega}_{g,f;B}
=\omega_{g,\vartheta_Bf;B}\), together with the real linearity of
the first variation in the Weyl label, gives
\begin{equation}
\begin{aligned}
u_{B,f}
&=
v_{B,f}
-
v_{B,\vartheta_Bf}
\\
&=
v_{B,\delta_Bf}.
\end{aligned}
\label{eq:boundary-tangent-difference-relation}
\end{equation}

Under the real-linear correspondence
\(f\mapsto\Psi_f=Ef\), the difference label \(\delta_Bf\) corresponds
to the reflected-difference solution
\begin{equation}
\begin{aligned}
\delta_J\Psi_f
&:=
\Psi_f
-
\mathcal J_B^{\rm cl}\Psi_f
\\
&=
\Psi_f
-
\Psi_{\vartheta_Bf}
\\
&=
\Psi_{\delta_Bf}.
\end{aligned}
\label{eq:bulk-represented-tangent}
\end{equation}
Since \(\vartheta_B^2=1\),
\begin{equation}
\vartheta_B\delta_Bf
=
-\delta_Bf,
\qquad
\mathcal J_B^{\rm cl}\Psi_{\delta_Bf}
=
-\Psi_{\delta_Bf}.
\label{eq:reflected-difference-oddness}
\end{equation}
Thus the reflected-difference perturbation is modular odd.

Once the ambient coherent family is fixed, the label \(f\) is fixed.
The modular pullback fixes its partner label \(\vartheta_Bf\), and
the difference of the two families fixes the tangent
\(\delta_Bf=f-\vartheta_Bf\). The response direction is therefore a
consequence of the modular pairing rather than an independently
chosen bulk perturbation.

\subsection{Exact relative entropy and the canonical energy of the
reflected-difference tangent}
\label{subsec:exact-holographic-hessian}

Let \(\omega_{g,f}^{\rm bulk}\) denote the restriction to
\(\mathcal N_B\) of the coherent vector state represented by
\(\Omega_{g,f}^{\rm bulk}\), and let
\(\omega_0^{\rm bulk}\) denote the restricted bulk vacuum. The
diagonal normalization of the scalar canonical-energy form is fixed
by
\begin{equation}
E_{\rm can}^{\mathcal W_B}
\left(
\Psi_f,\Psi_f
\right)
:=
\left.
\frac{d^2}{dg^2}
S_{\mathcal N_B}
\left(
\omega_{g,f}^{\rm bulk}
\middle\|
\omega_0^{\rm bulk}
\right)
\right|_{g=0}.
\label{eq:canonical-energy-hessian-normalization}
\end{equation}
Its symmetric bilinear extension is obtained by polarization. In
terms of the Klein--Gordon symplectic current defined in
Eq.~\eqref{eq:bulk-klein-gordon-symplectic-current}, it is
\begin{equation}
\begin{aligned}
E_{\rm can}^{\mathcal W_B}
\left(
\Psi_{f_1},\Psi_{f_2}
\right)
&=
-
\int_{\Sigma_B}
d\Sigma_A\,
\omega_{\rm KG}^{A}
\left(
\Psi_{f_1},
\mathcal L_{\xi_B}\Psi_{f_2}
\right),
\end{aligned}
\label{eq:scalar-canonical-energy-form}
\end{equation}
where \(\Sigma_B\) is a Cauchy surface of \(\mathcal W_B\) and
\(d\Sigma_A\) is future-directed. Conservation of the symplectic
current, the Killing property of \(\xi_B\), and the reflecting
boundary condition make this form symmetric on the stated domain.
The minus sign follows from the metric and symplectic-current
conventions and makes the diagonal form positive. With \(\xi_B\)
normalized as the bulk extension of the boundary modular-flow
generator \(\zeta_B\), Eq.~\eqref{eq:scalar-canonical-energy-form}
agrees with the relative-entropy Hessian in
Eq.~\eqref{eq:canonical-energy-hessian-normalization}. The
identification of the ball-region quantum Fisher form with
AdS--Rindler canonical energy was established in
\cite{FaulknerEtAl,LashkariVanRaamsdonk}.

We now apply relative-entropy preservation to the modularly paired
boundary states. Equation~\eqref{eq:exact-encoded-ball-partner}
identifies them with the boundary images of the bulk coherent states
labelled by \(f\) and \(\vartheta_Bf\). Consequently,
\begin{equation}
\begin{aligned}
&S_{M_B}
\left(
\omega_{g,f;B}
\middle\|
\widetilde{\omega}_{g,f;B}
\right)
\\
&\qquad
=
S_{\mathcal N_B}
\left(
\omega_{g,f}^{\rm bulk}
\middle\|
\omega_{g,\vartheta_Bf}^{\rm bulk}
\right),
\\[1mm]
&S_{M_B}
\left(
\widetilde{\omega}_{g,f;B}
\middle\|
\omega_{g,f;B}
\right)
\\
&\qquad
=
S_{\mathcal N_B}
\left(
\omega_{g,\vartheta_Bf}^{\rm bulk}
\middle\|
\omega_{g,f}^{\rm bulk}
\right).
\end{aligned}
\label{eq:boundary-to-bulk-paired-entropies}
\end{equation}

Relative entropy between equal-covariance coherent states of a
canonical commutation relation (CCR) algebra depends only on the
difference of their test-function classes
\cite{BostelmannCadamuroDelVecchio2022}. The two bulk relative
entropies in
Eq.~\eqref{eq:boundary-to-bulk-paired-entropies} therefore reduce to
\begin{equation}
\begin{aligned}
S_{\mathcal N_B}
\left(
\omega_{g,f}^{\rm bulk}
\middle\|
\omega_{g,\vartheta_Bf}^{\rm bulk}
\right)
&=
S_{\mathcal N_B}
\left(
\omega_{g,\delta_Bf}^{\rm bulk}
\middle\|
\omega_0^{\rm bulk}
\right),
\\
S_{\mathcal N_B}
\left(
\omega_{g,\vartheta_Bf}^{\rm bulk}
\middle\|
\omega_{g,f}^{\rm bulk}
\right)
&=
S_{\mathcal N_B}
\left(
\omega_{g,-\delta_Bf}^{\rm bulk}
\middle\|
\omega_0^{\rm bulk}
\right).
\end{aligned}
\label{eq:bulk-paired-entropy-difference-labels}
\end{equation}
The coherent-state relative entropy is exactly quadratic in its
difference label. It is therefore unchanged when
\(\delta_Bf\) is replaced by \(-\delta_Bf\). Equations
\eqref{eq:canonical-energy-hessian-normalization} and
\eqref{eq:bulk-paired-entropy-difference-labels} give
\begin{equation}
\begin{aligned}
S_{M_B}
\left(
\omega_{g,f;B}
\middle\|
\widetilde{\omega}_{g,f;B}
\right)
&=
S_{M_B}
\left(
\widetilde{\omega}_{g,f;B}
\middle\|
\omega_{g,f;B}
\right)
\\
&=
\frac{g^2}{2}
E_{\rm can}^{\mathcal W_B}
\left(
\Psi_{\delta_Bf},
\Psi_{\delta_Bf}
\right).
\end{aligned}
\label{eq:both-ball-relative-entropies}
\end{equation}

The symmetrized comparison is consequently
\begin{equation}
\begin{aligned}
\mathcal S_J(g;B,f)
&:=
S_{M_B}
\left(
\omega_{g,f;B}
\middle\|
\widetilde{\omega}_{g,f;B}
\right)
\\
&\quad+
S_{M_B}
\left(
\widetilde{\omega}_{g,f;B}
\middle\|
\omega_{g,f;B}
\right)
\\
&=
g^2
E_{\rm can}^{\mathcal W_B}
\left(
\Psi_{\delta_Bf},
\Psi_{\delta_Bf}
\right).
\end{aligned}
\label{eq:exact-encoded-symmetrized-comparison}
\end{equation}
The corresponding bilinear BKM form on coherent-state tangents
follows by allowing the two coherent states to vary independently.
For \(f_1,f_2\in\mathcal K_{\mathbb R}^{\rm bulk}\) in the
reconstruction sector and \(a,b\in\mathbb R\), define
\begin{equation}
f_{a,b}
:=
af_1-bf_2.
\label{eq:two-parameter-difference-label}
\end{equation}
Linearity of the causal propagator gives
\begin{equation}
\Psi_{f_{a,b}}
=
a\Psi_{f_1}
-
b\Psi_{f_2}.
\label{eq:two-parameter-difference-solution}
\end{equation}
Relative-entropy preservation and the coherent-state difference
formula then give
\begin{equation}
\begin{aligned}
S_{M_B}
\left(
\omega_{a,f_1;B}
\middle\|
\omega_{b,f_2;B}
\right)
&=
S_{\mathcal N_B}
\left(
\omega_{1,f_{a,b}}^{\rm bulk}
\middle\|
\omega_0^{\rm bulk}
\right)
\\
&=
\frac{1}{2}
E_{\rm can}^{\mathcal W_B}
\left(
\Psi_{f_{a,b}},
\Psi_{f_{a,b}}
\right).
\end{aligned}
\label{eq:exact-two-parameter-ball-entropy}
\end{equation}

For the tangents defined in
Eq.~\eqref{eq:single-family-ball-tangent}, the mixed-derivative
characterization of the BKM form gives
\begin{equation}
\begin{aligned}
&
\gamma_{\omega_B}^{\rm BKM}
\left(
v_{B,f_1},
v_{B,f_2}
\right)
\\
&\quad=
-
\left.
\partial_a\partial_b
S_{M_B}
\left(
\omega_{a,f_1;B}
\middle\|
\omega_{b,f_2;B}
\right)
\right|_{a=b=0}
\\
&\quad=
E_{\rm can}^{\mathcal W_B}
\left(
\Psi_{f_1},
\Psi_{f_2}
\right).
\end{aligned}
\label{eq:ball-bkm-equals-canonical-energy}
\end{equation}
Using Eq.~\eqref{eq:boundary-tangent-difference-relation}, the
susceptibility evaluates the BKM form, and hence the canonical
energy, on the reflected-difference tangent,
\begin{equation}
\begin{aligned}
\chi_J(B;f)
&:=
\left.
\frac{d^2}{dg^2}
\mathcal S_J(g;B,f)
\right|_{g=0}
\\
&=
2
\gamma_{\omega_B}^{\rm BKM}
\left(
u_{B,f},
u_{B,f}
\right)
\\
&=
2
E_{\rm can}^{\mathcal W_B}
\left(
\Psi_{\delta_Bf},
\Psi_{\delta_Bf}
\right).
\end{aligned}
\label{eq:bulk-reflected-difference-energy}
\end{equation}
Equations~\eqref{eq:boundary-tangent-difference-relation},
\eqref{eq:reflected-difference-oddness}, and
\eqref{eq:bulk-reflected-difference-energy} show that the modular
pullback fixes a modular-odd reflected-difference tangent whose
susceptibility is exactly twice its AdS--Rindler canonical energy.
Because the coherent-state relative entropy is exactly quadratic,
Eq.~\eqref{eq:exact-encoded-symmetrized-comparison} gives the complete
dependence on \(g\), rather than only its Hessian at \(g=0\).

The general reflected-difference result reduces to the standard
ball-region formula for a wedge-localized label. Let
\(f_{\rm loc}\) satisfy
\[
W(f_{\rm loc})\in\mathcal N_B,
\qquad
W(\vartheta_Bf_{\rm loc})\in\mathcal N_B'.
\]
The commutant displacement in
\(W(g\delta_Bf_{\rm loc})\) does not change the state restricted to
\(\mathcal N_B\). Hence
\begin{equation}
\omega_{g,\delta_Bf_{\rm loc}}^{\rm bulk}
=
\omega_{g,f_{\rm loc}}^{\rm bulk}
\qquad
\text{on }\mathcal N_B.
\label{eq:localized-bulk-state-reduction}
\end{equation}
It follows that
\begin{equation}
\begin{aligned}
E_{\rm can}^{\mathcal W_B}
\left(
\Psi_{\delta_Bf_{\rm loc}},
\Psi_{\delta_Bf_{\rm loc}}
\right)
&=
E_{\rm can}^{\mathcal W_B}
\left(
\Psi_{f_{\rm loc}},
\Psi_{f_{\rm loc}}
\right),
\\
\chi_J(B;f_{\rm loc})
&=
2
E_{\rm can}^{\mathcal W_B}
\left(
\Psi_{f_{\rm loc}},
\Psi_{f_{\rm loc}}
\right).
\end{aligned}
\label{eq:localized-bulk-coherent-reduction}
\end{equation}
Thus the localized case recovers the canonical energy of the original
wedge perturbation, whereas an ambient perturbation is evaluated on
the reflected-difference solution
\(\Psi_f-\mathcal J_B^{\rm cl}\Psi_f\).

\subsection{Modular-odd decomposition and boundary sources}
\label{subsec:boundary-source-realization}

Decompose the bulk test-function class into its modular-even and
modular-odd parts,
\begin{equation}
f_\pm^{(B)}
:=
\frac{1}{2}
\left(
f\pm\vartheta_Bf
\right).
\label{eq:bulk-modular-even-odd-labels}
\end{equation}
These classes satisfy
\begin{equation}
\vartheta_Bf_\pm^{(B)}
=
\pm f_\pm^{(B)},
\qquad
\delta_Bf
=
2f_-^{(B)}.
\label{eq:modular-odd-bulk-source}
\end{equation}
Substitution into
Eq.~\eqref{eq:bulk-reflected-difference-energy} gives
\begin{equation}
\chi_J(B;f)
=
8
E_{\rm can}^{\mathcal W_B}
\left(
\Psi_{f_-^{(B)}},
\Psi_{f_-^{(B)}}
\right).
\label{eq:modular-odd-source-susceptibility}
\end{equation}
The modular-even component of \(f\) therefore makes no contribution
to the susceptibility.  In particular,
\(\chi_J(B;f)=0\) if \(\vartheta_Bf=f\).  For a modular-odd
test-function class \(f_{\rm odd}\),
\(\vartheta_Bf_{\rm odd}=-f_{\rm odd}\), one has
\begin{equation}
\chi_J(B;f_{\rm odd})
=
8
E_{\rm can}^{\mathcal W_B}
\left(
\Psi_{f_{\rm odd}},
\Psi_{f_{\rm odd}}
\right).
\label{eq:fully-modular-odd-susceptibility}
\end{equation}

The same decomposition can be expressed directly in terms of the
boundary preparation. Let
\(f_\eta\in\mathcal K_{\mathbb R}^{\rm bulk}\) denote the bulk
test-function class produced by a boundary source \(\eta\). Assume
that the preparation map
\begin{equation}
\eta
\longmapsto
f_\eta
\end{equation}
is real linear and equivariant under modular reflection,
\begin{equation}
f_{\eta^{J_B}}
=
\vartheta_Bf_\eta.
\label{eq:source-bulk-reflection-equivariance}
\end{equation}
Here \(\eta^{J_B}\) denotes the reflected boundary preparation. For
a Lorentzian unitary preparation, it is defined at the operator level
by Eq.~\eqref{eq:lorentzian-reflected-source-operator}. For a
Euclidean preparation, it is defined by the corresponding reflected
contour and operator insertions.

On the real vector space of source preparations, define
\begin{equation}
\eta_\pm^{(B)}
:=
\frac{1}{2}
\left(
\eta\pm\eta^{J_B}
\right).
\label{eq:boundary-source-even-odd-parts}
\end{equation}
Real linearity and reflection equivariance imply
\begin{equation}
\begin{aligned}
f_\eta
&=
f_{\eta_+^{(B)}}
+
f_{\eta_-^{(B)}},
\\
\vartheta_Bf_{\eta_\pm^{(B)}}
&=
\pm f_{\eta_\pm^{(B)}},
\\
\delta_Bf_\eta
&=
2f_{\eta_-^{(B)}}.
\end{aligned}
\label{eq:source-induced-bulk-decomposition}
\end{equation}
Writing
\(\chi_J(B;\eta):=\chi_J(B;f_\eta)\), the source-prepared
susceptibility is therefore
\begin{equation}
\chi_J(B;\eta)
=
8
E_{\rm can}^{\mathcal W_B}
\left(
\Psi_{f_{\eta_-^{(B)}}},
\Psi_{f_{\eta_-^{(B)}}}
\right).
\label{eq:reflection-paired-source-susceptibility}
\end{equation}
Thus only the modular-odd part of the boundary preparation generates
the reflected-difference bulk solution on which the canonical energy
is evaluated.

For a neutral Hermitian scalar primary prepared on a
reflection-compatible Euclidean contour \(\mathcal C\), let
\(\mathcal R_B^{\mathcal C}\) denote the contour involution induced
by the boundary reflection \(r_B\). The reflected source is defined
exactly by equality of the corresponding smeared operators. In
reflection-adapted coordinates, after the conformal weight and
Jacobian have been absorbed into the source profile, this relation
takes the form
\begin{equation}
\eta^{J_B}(x)
=
\epsilon_{\mathcal O}
\eta^*
\left(
\mathcal R_B^{\mathcal C}x
\right),
\qquad
x\in\mathcal C,
\label{eq:explicit-reflected-source}
\end{equation}
where \(\epsilon_{\mathcal O}=\pm1\) is the CPT or internal parity of
the scalar operator. For a real source coupled to a CPT-even scalar,
\begin{equation}
\eta^{J_B}(x)
=
\eta
\left(
\mathcal R_B^{\mathcal C}x
\right).
\end{equation}
For the Lorentzian unitary preparation, the sign required by
antiunitarity is instead included in
Eq.~\eqref{eq:lorentzian-reflected-source-operator}.

\subsection{Semiclassical extension to matter and gravity}
\label{subsec:semiclassical-gravity}

For the linear bulk field under exact complementary recovery, the
coherent-state identities in
Subsecs.~\ref{subsec:grav-content}
and~\ref{subsec:exact-holographic-hessian} are exact.  For coupled
matter and metric perturbations, we work to quadratic order in the
perturbation within the semiclassical approximation.  At this order,
we assume the reconstruction-sector bulk--boundary
modular-intertwining relation and reflection equivariance of the
linearized boundary-source-to-bulk-solution map.  The boundary
reflected-difference tangent is then represented by the corresponding
reflected-difference solution of the linearized bulk equations.  The
JLMS relation, the bulk constraints, and the gauge-invariant
Hollands--Wald prescription identify its BKM quadratic form with the
bulk canonical energy.

Let \(\delta\mathcal X\) denote a combined linearized matter and
metric perturbation satisfying the bulk constraints and the required
asymptotic and extremal-surface boundary conditions. The real-linear
modular reflection \(\mathcal J_B^{\rm cl}\) acts geometrically on
the matter and metric fields, including the tensor, spin, and CPT
actions appropriate to each field. Since the reflection preserves
the linearized equations and boundary conditions, it maps an
admissible perturbation to another admissible perturbation. The
reflected-difference solution is
\begin{equation}
\delta_J\mathcal X
:=
\delta\mathcal X
-
\mathcal J_B^{\rm cl}\delta\mathcal X.
\label{eq:combined-reflected-difference-solution}
\end{equation}

Let
\(E_{\rm can}^{\mathcal W_B}\) denote the Hollands--Wald
canonical-energy form on the constrained perturbation space. The
ball-region BKM form is represented by this canonical energy
\cite{FaulknerEtAl,LashkariVanRaamsdonk,HollandsWald2013}.
Consequently, at the retained semiclassical order, the
modular-reflection susceptibility is
\begin{equation}
\chi_J(B)
=
2
E_{\rm can}^{\mathcal W_B}
\left(
\delta_J\mathcal X,
\delta_J\mathcal X
\right).
\label{eq:semiclassical-reflected-canonical-energy}
\end{equation}
The factor of two is the same factor relating the symmetrized
relative-entropy Hessian to the BKM form in the exact linear-field
construction.

For a scalar perturbation on the fixed AdS background,
Eq.~\eqref{eq:semiclassical-reflected-canonical-energy} reduces to
the scalar canonical-energy expression in
Eq.~\eqref{eq:scalar-canonical-energy-form}. For a purely
gravitational perturbation, it becomes
\begin{equation}
\chi_J^{\rm grav}(B)
=
2
\int_{\Sigma_B}
\omega_{\rm grav}
\left(
\delta_Jh,
\mathcal L_{\xi_B}\delta_Jh
\right),
\label{eq:gravitational-canonical-energy}
\end{equation}
where \(\delta_Jh\) is the metric component of
\(\delta_J\mathcal X\), and \(\omega_{\rm grav}\) is the covariant
gravitational symplectic current in the Hollands--Wald convention.

The gravitational expression is evaluated in Hollands--Wald gauge,
or with the equivalent extremal-surface boundary-and-corner
completion that makes the subregion canonical energy gauge invariant
\cite{HollandsWald2013,IyerWald1994,DonnellyFreidel2016,
HarlowWu2020}. Proper gauge transformations preserving the
prescribed boundary conditions are quotiented out. Transformations
carrying nonzero boundary or edge charges are not null directions of
the completed symplectic form. The gravitational extension therefore
preserves the modular selection of the reflected-difference
perturbation while replacing the fixed-background scalar quadratic
form by the gauge-invariant canonical energy of the coupled
linearized solution.

\section{DISCUSSION AND OUTLOOK}

The central conclusion of this work is a fixed-point principle for
state comparison that extends from finite-dimensional type~I systems
to local type~III algebras. In the type~I setting, the coincidence
point is the equality of a density matrix with its modularly reflected
state. In the local quantum-field-theoretic setting, it is the equality
of a local state with the modular pullback of its commutant
restriction. In finite dimensions the antiunitary reflection acts
within a matrix algebra. In the local theory the vacuum modular
conjugation carries the algebra to its commutant. The pullback is
therefore the operator-algebraic operation that makes the comparison
intrinsic to the original local algebra.

The fixed representative alone is not the main result. For a vacuum
standard pair, vacuum coincidence follows from invariance of the
vacuum under modular conjugation. More generally, every normal
positive functional has a unique natural-cone representative fixed by
the standard-form \(J\). The present construction instead holds fixed
the reference standard pair and the ambient preparation. These data
determine the commutant restriction, its pullback, and the
reflected-difference tangent. Choosing a new canonical representative
for each local state would remove the ambient comparison measured by
the susceptibility.

At coincidence, the symmetrized relative entropy has no linear term
because the two compared states agree and relative entropy is
nonnegative. Its leading nontrivial coefficient is governed in finite
dimensions by BKM metric evaluated on the
reflected-difference tangent. For local type~III algebras, the
corresponding coefficient is the local BKM bilinear form obtained from
the second-order expansion of Araki relative entropy at the diagonal.
The resulting susceptibility is attached to the tangent selected by
the modular pairing rather than to an arbitrary curve of states. It
also depends on the localization region because changing the region
changes the local algebra, its modular conjugation, and the tangent
tested by the BKM form.

The coherent-state examples realize this structure exactly. For the
free scalar field on a wedge, the susceptibility is a positive
boost-energy quadratic form, equivalently a stress-tensor integral, of
the reflected-difference profile. Translations of the wedge change the
algebra and its modular reflection, so the localization parameter
enters the response intrinsically. For the chiral \(U(1)\) current, the
construction reduces to a half-line quadratic form. In the thermal
half-line theory, the vacuum geometric reflection is replaced by the
Borchers--Yngvason standard-subspace reflection. These examples show
that the fixed-point susceptibility is well defined in local type~III
quantum field theory and is not merely a finite-dimensional analogy.

The vacuum-ball construction has a related equilibrium
interpretation. The Casini--Huerta--Myers conformal transformation maps
the ball vacuum to a KMS state on the hyperbolic cylinder, while the
bulk geometry becomes the AdS--Rindler exterior
\cite{CasiniHuertaMyers}. For regular exponential source deformations
that admit a Kubo--Mori score representation, the BKM geometry has the
corresponding equilibrium-response form
\cite{PetzToth1993,Bratteli1997}. The ball vacuum is therefore a
geometrically selected equilibrium state, and the
reflected-difference tangent measures a departure from its modular
reflection symmetry.

For the vacuum ball and a chosen ambient perturbation, modular
conjugation determines its reflected comparison and hence the bulk
perturbation on which the canonical energy is evaluated. In the
encoded linear coherent family, this perturbation is the difference
between a bulk excitation and its AdS--Rindler modular reflection. The
symmetrized Araki relative entropy is exactly quadratic, and its
Hessian equals twice the canonical energy of this
reflected-difference perturbation. At the retained semiclassical
order, the same prescription extends to admissible matter and metric
perturbations. Unlike ordinary ball-region Fisher information, the
modular-reflection susceptibility uses \(B\) and \(J_B\) to determine
the perturbation tested by the quadratic form.

This construction also differs from holographic entanglement
asymmetry \cite{BeniniGodetSingh2025}. An internal-symmetry twirl and
the present modular pair can both produce a BKM coefficient represented
by a bulk quadratic form, but they select different tangent
directions. The present susceptibility isolates the component odd
under the AdS--Rindler modular/CPT reflection rather than the component
that breaks an internal symmetry. The coherent scalar construction in
Subsecs.~\ref{subsec:grav-content}--%
\ref{subsec:boundary-source-realization}
gives an explicit realization of this selection.
Subsection~\ref{subsec:semiclassical-gravity} extends the same modular
selection principle to admissible coupled matter and metric
perturbations at the retained second-order semiclassical level.

The main technical limitation is the assumed differentiability of
Araki relative entropy at the diagonal. The BKM interpretation applies
to deformations for which the required second-order expansion exists,
and the explicit quantum-field-theoretic calculations in this paper
use coherent-state deformations. An important extension is to
determine broader classes of faithful normal deformations with the
same local BKM Hessian. These may include perturbations not generated
by Weyl translations, perturbations produced by interacting dynamics,
and families whose reflected tangent has no elementary description.

Localization provides a second direction for further work. In the
examples considered here, it appears through boost weights, half-line
weights, thermal standard-subspace reflections, and AdS--Rindler
entanglement-wedge reflections. A more intrinsic treatment could
relate this dependence to modular inclusions, translated standard
pairs, and variations of local algebras under changes of region. Such
a formulation may clarify the relation between the present
susceptibility and other local response quantities, including modular
energy variations, relative-entropy inequalities, and holographic
canonical-energy constraints.

The vacuum-ball analysis relies on its geometric modular flow. For
more general regions or excited reference states, modular flow is
typically nonlocal, and no simple bulk Killing description analogous
to the AdS--Rindler flow is available. Extending the construction
would require control of the relevant standard pair, the corresponding
bulk--boundary modular data, and the domain of the reflected
perturbation.

It would also be useful to compare the intrinsic type~III coefficient
with regulated type~I approximations. Cutoff density matrices should
be treated as approximations to the Araki comparison rather than as
definitions of the local quantity. Higher-order terms in the
symmetrized Araki functional may define additional local response
coefficients. In holographic settings, they may also encode nonlinear
gravitational response or backreaction associated with the modularly
reflected bulk comparison.

\begin{acknowledgments}
The author thanks an anonymous referee for suggesting that the modular
construction be examined in a holographic setting.
\end{acknowledgments}
\section*{Data Availability}

No data were generated or analyzed in this work.

\appendix

\section{Relative Modular Operators and Araki Relative Entropy}
\label{sec:RelativeModularAraki}

This appendix fixes the relative modular-operator convention used in
the main text. Let \(\mathfrak M\) be a von Neumann algebra in standard
form on a Hilbert space \(\mathcal H\), with modular conjugation \(J\)
and natural cone \(\mathcal P^\natural\subset\mathcal H\). Standard-form
theory implies that each faithful normal state \(\omega\) on
\(\mathfrak M\) is represented by a unique vector
\(\xi_\omega\in\mathcal P^\natural\)
\cite{Bratteli1997,DerezinskiGerard2013}. Its defining relation is
\begin{equation}
\omega(A)
=
\langle \xi_\omega,A\xi_\omega\rangle,
\qquad
A\in\mathfrak M .
\label{eq:A-natural-cone-representative}
\end{equation}
One of the defining standard-form properties is that the natural cone
is fixed pointwise by \(J\):
\begin{equation}
J\xi=\xi,
\qquad
\xi\in\mathcal P^\natural.
\label{eq:A-natural-cone-J-fixed}
\end{equation}
In particular, \(J\xi_\omega=\xi_\omega\). The vector
\(\xi_\omega\) is therefore the canonical \(J\)-symmetric
standard-form representative of \(\omega\). This statement concerns
the canonical representative of a state on \(\mathfrak M\); it does
not identify an arbitrary ambient vector implementing the same local
state with that representative.
This representation is intrinsic to the standard form and does not
require a density matrix or a trace on \(\mathfrak M\).

Given two faithful normal states \(\omega\) and \(\phi\), with natural
cone representatives \(\xi_\omega\) and \(\xi_\phi\), the relative
Tomita operator is defined on the dense domain
\(\mathfrak M\xi_\phi\) by
\begin{equation}
S_{\omega\mid\phi}(A\xi_\phi)
=
A^*\xi_\omega,
\qquad
A\in\mathfrak M .
\label{eq:A-relative-Tomita}
\end{equation}
This antilinear operator is closable. Its closure, denoted again by
\(S_{\omega\mid\phi}\), determines the relative modular operator
\begin{equation}
\Delta_{\omega\mid\phi}
:=
S_{\omega\mid\phi}^*S_{\omega\mid\phi}.
\label{eq:A-relative-modular}
\end{equation}
The operator \(\Delta_{\omega\mid\phi}\) is positive and self-adjoint.
When \(\omega=\phi\), this construction reduces to the ordinary modular
operator of the state \(\omega\).

With these conventions, the Araki relative entropy is
\cite{Araki1975,Araki1977}
\begin{equation}
S_{\mathfrak M}(\omega\Vert\phi)
:=
-\langle
\xi_\omega,
\log\Delta_{\phi\mid\omega}\,
\xi_\omega
\rangle .
\label{eq:A-Araki-relative-entropy}
\end{equation}
The order of the relative modular operator in
Eq.~\eqref{eq:A-Araki-relative-entropy} is part of the convention:
the entropy \(S_{\mathfrak M}(\omega\Vert\phi)\) is expressed in terms
of \(\Delta_{\phi\mid\omega}\). The logarithm is understood through
the spectral calculus, with the usual extended-value convention. This
definition applies directly to type~III local algebras, where there is
no intrinsic trace and no canonical local density matrix.

In the finite-dimensional type~I case,
Eq.~\eqref{eq:A-Araki-relative-entropy} reduces to the usual Umegaki
relative entropy. Let \(\mathfrak M=\mathcal B(\mathcal K)\), and let
\begin{equation}
\omega(A)
=
\operatorname{Tr}(\rho A),
\qquad
\phi(A)
=
\operatorname{Tr}(\sigma A),
\qquad
A\in\mathcal B(\mathcal K),
\label{eq:A-type-I-states}
\end{equation}
where \(\rho\) and \(\sigma\) are faithful density matrices on
\(\mathcal K\). In the Hilbert--Schmidt standard form, the natural cone
representatives are \(\rho^{1/2}\) and \(\sigma^{1/2}\), and left and
right multiplication define
\begin{equation}
L_\sigma(X)=\sigma X,
\qquad
R_\rho(X)=X\rho .
\label{eq:A-left-right-multiplication}
\end{equation}
The relative modular operator appearing in
Eq.~\eqref{eq:A-Araki-relative-entropy} is then
\begin{equation}
\Delta_{\phi\mid\omega}
=
L_\sigma R_\rho^{-1}.
\label{eq:A-type-I-relative-modular}
\end{equation}
Since \(L_\sigma\) and \(R_\rho\) commute, one obtains
\begin{equation}
\log\Delta_{\phi\mid\omega}
=
L_{\log\sigma}-R_{\log\rho}.
\label{eq:A-log-relative-modular-type-I}
\end{equation}
Substitution into Eq.~\eqref{eq:A-Araki-relative-entropy} gives
\begin{equation}
\begin{array}{c}
\displaystyle
S_{\mathfrak M}(\omega\Vert\phi)
=
-\operatorname{Tr}
\left[
\rho^{1/2}
\left(
\log\sigma\,\rho^{1/2}
-
\rho^{1/2}\log\rho
\right)
\right]
\\[2ex]
\displaystyle
=
\operatorname{Tr}\rho(\log\rho-\log\sigma).
\end{array}
\label{eq:A-Umegaki-reduction}
\end{equation}
Thus the Araki expression used for local algebras is the
operator-algebraic extension of the finite-dimensional relative entropy
used in the type~I part of the paper \cite{Umegaki1962}.

The local construction uses the following pullback fact. Let
\((\mathfrak M,\Omega)\) be a standard pair, and let \(J\) be its
modular conjugation. Tomita--Takesaki theory gives
\begin{equation}
J\mathfrak M J
=
\mathfrak M' .
\label{eq:A-Tomita-commutant}
\end{equation}
Define
\begin{equation}
j(A)
:=
J A^* J,
\qquad
A\in\mathfrak M .
\label{eq:A-modular-pullback-map}
\end{equation}
Then \(j\) is a linear unital \(*\)-anti-isomorphism
\begin{equation}
j:\mathfrak M\longrightarrow\mathfrak M' .
\label{eq:A-anti-isomorphism}
\end{equation}
Indeed,
\begin{equation}
j(AB)
=
j(B)j(A),
\qquad
j(A^*)
=
j(A)^*,
\qquad
j(\mathbf 1)=\mathbf 1 .
\label{eq:A-anti-isomorphism-properties}
\end{equation}
Since \(j\) is a von Neumann algebra anti-isomorphism, it is ultraweakly
continuous. Consequently, if \(\psi\) is a normal state on
\(\mathfrak M'\), then \(\psi\circ j\) is a normal state on
\(\mathfrak M\). Positivity follows from
\begin{equation}
(\psi\circ j)(A^*A)
=
\psi\!\left(j(A^*A)\right)
=
\psi\!\left(j(A)j(A)^*\right)
\ge 0 ,
\label{eq:A-pullback-positivity}
\end{equation}
and normalization follows from
\begin{equation}
(\psi\circ j)(\mathbf 1)
=
\psi(\mathbf 1)
=
1 .
\label{eq:A-pullback-normalization}
\end{equation}
This is the abstract mechanism used in Sec.~\ref{sec:reflection_paired_families}: the commutant
restriction of an ambient state is pulled back to the original local
algebra by the modular anti-isomorphism induced by the vacuum modular
conjugation.

\section{Thermal chiral \(U(1)\) half-line}
\label{app:thermal-chiral-u1}

This appendix formulates the finite-temperature analogue of the chiral
half-line formulas of Sec.~\ref{sec:chiral_u1}.  The modular pullback
construction and the coherent-state reduction are the same as in
Secs.~\ref{sec:reflection_paired_families} and \ref{sec:chiral_u1};
only the thermal modular reflection and the thermal entropy kernel are
different.

Let \(\omega_\beta\) be the thermal equilibrium state at inverse
temperature \(\beta>0\), and let
\((\pi_\beta,\mathcal H_\beta,\Omega_\beta)\) be its GNS
representation.  For a real test function
\(f\in C_c^\infty(\mathbb R,\mathbb R)\), write
\begin{equation}
W_\beta(f):=\pi_\beta(W(f)).
\label{eq:B-thermal-Weyl}
\end{equation}
Here \(\mathfrak A((-\infty,t))\) denotes the abstract chiral Weyl
\(C^*\)-algebra of the half-line.  The represented thermal half-line
von Neumann algebra is
\begin{equation}
\mathcal A_\beta((-\infty,t))
:=
\pi_\beta\!\left(
\mathfrak A((-\infty,t))
\right)'' .
\label{eq:B-thermal-halfline-algebra}
\end{equation}
The pair
\((\mathcal A_\beta((-\infty,t)),\Omega_\beta)\) is standard.  Let
\(J_t^\beta\) denote its modular conjugation.  Then
\begin{equation}
J_t^\beta\,
\mathcal A_\beta((-\infty,t))\,
J_t^\beta
=
\mathcal A_\beta((-\infty,t))' .
\label{eq:B-thermal-commutant}
\end{equation}
The associated modular pullback map is
\begin{equation}
j_t^\beta(A)
:=
J_t^\beta A^*J_t^\beta,
\qquad
A\in\mathcal A_\beta((-\infty,t)).
\label{eq:B-thermal-j-map}
\end{equation}

In the vacuum half-line case, the modular conjugation is implemented by
the geometric reflection \(x\mapsto 2t-x\).  In the thermal
representation, this pointwise reflection is replaced by the
Borchers--Yngvason standard-subspace modular conjugation
\cite{BorchersYngvason1999}.  Its induced action on Weyl labels is
defined by
\begin{equation}
J_t^\beta W_\beta(f)J_t^\beta
=
W_\beta(\vartheta_t^\beta f).
\label{eq:B-thermal-reflection-weyl}
\end{equation}
The map \(\vartheta_t^\beta\) is an involutive anti-symplectic
reflection on the thermal one-particle space:
\begin{equation}
(\vartheta_t^\beta)^2=1,
\qquad
\sigma(\vartheta_t^\beta f,\vartheta_t^\beta g)
=
-\sigma(f,g).
\label{eq:B-thermal-antisymplectic}
\end{equation}
Unlike the vacuum reflection \(f(x)\mapsto f(2t-x)\), the thermal
reflection is not represented here by a pointwise formula on the light
ray.

For a real test function \(h\), define the thermal reflected difference
profile
\begin{equation}
\delta_t^\beta h
:=
h-\vartheta_t^\beta h .
\label{eq:B-thermal-delta}
\end{equation}
This is the finite-temperature analogue of the vacuum profile
\(h-h(2t-\cdot)\).  Although \(h\) is initially taken in
\(C_c^\infty(\mathbb R,\mathbb R)\), the vectors
\(\vartheta_t^\beta h\) and \(\delta_t^\beta h\) are understood in the
thermal one-particle completion.

Let \(\omega_h^\beta\) denote the coherent thermal state obtained from
\(\omega_\beta\) by Weyl displacement:
\begin{equation}
\omega_h^\beta(B)
:=
\langle
W_\beta(h)\Omega_\beta,
B\,W_\beta(h)\Omega_\beta
\rangle,
\qquad
B\in\mathcal B(\mathcal H_\beta).
\label{eq:B-thermal-coherent-state}
\end{equation}
With the normalization conventions of Sec.~\ref{sec:chiral_u1}, the
thermal half-line relative entropy is
\cite{BorchersYngvason1999,GarbarzPalau2023}
\begin{equation}
S_{\mathcal A_\beta((-\infty,t))}
(\omega_h^\beta\Vert\omega_\beta)
=
\int_{-\infty}^{t}
\bigl(h'(x)\bigr)^2\,
\beta
\left(
1-e^{2\pi(x-t)/\beta}
\right)
dx .
\label{eq:B-thermal-entropy}
\end{equation}
The zero-temperature limit agrees with the vacuum half-line result,
since
\begin{equation}
\beta
\left(
1-e^{2\pi(x-t)/\beta}
\right)
=
2\pi(t-x)+O(\beta^{-1}),
\qquad
\beta\to\infty .
\label{eq:B-vacuum-limit}
\end{equation}

Now take the coherent thermal family
\begin{equation}
\varpi_g^\beta:=\omega_{gh}^\beta .
\label{eq:B-thermal-family}
\end{equation}
Repeating the modular pullback construction of
Sec.~\ref{sec:reflection_paired_families}, with \(J_t\) replaced by
\(J_t^\beta\), gives
\begin{equation}
\begin{aligned}
\omega_{g,t}^\beta
&=
\omega_{gh}^\beta
\!\restriction_{\mathcal A_\beta((-\infty,t))},
\\
\widetilde\omega_{g,t}^\beta
&=
\omega_{g\vartheta_t^\beta h}^\beta
\!\restriction_{\mathcal A_\beta((-\infty,t))}.
\end{aligned}
\label{eq:B-thermal-local-pair}
\end{equation}
At \(g=0\), the two local states coincide:
\begin{equation}
\omega_{0,t}^\beta
=
\widetilde\omega_{0,t}^\beta
=
\omega_\beta
\!\restriction_{\mathcal A_\beta((-\infty,t))}.
\label{eq:B-thermal-selfdual}
\end{equation}
Thus \(g=0\) is the thermal self-dual point.

The coherent-state difference identity and homogeneity of the
relative entropy give
\cite{BostelmannCadamuroDelVecchio2022}
\begin{equation}
\begin{array}{c}
\displaystyle
S_\beta(g,t)
:=
S_{\mathcal A_\beta((-\infty,t))}
(\omega_{g,t}^\beta\Vert\widetilde\omega_{g,t}^\beta)
+
S_{\mathcal A_\beta((-\infty,t))}
(\widetilde\omega_{g,t}^\beta\Vert\omega_{g,t}^\beta)
\\[2ex]
\displaystyle
=
2g^2\,
S_{\mathcal A_\beta((-\infty,t))}
(\omega_{\delta_t^\beta h}^\beta\Vert\omega_\beta).
\end{array}
\label{eq:B-thermal-symmetrized}
\end{equation}
Comparing with the local BKM expansion gives
\begin{equation}
\begin{array}{c}
\displaystyle
F_{\mathrm{BKM},\beta}^{\mathcal A}(t)
=
2\,
S_{\mathcal A_\beta((-\infty,t))}
(\omega_{\delta_t^\beta h}^\beta\Vert\omega_\beta),
\\[2ex]
\displaystyle
I_{A,\beta}(t)
=
2F_{\mathrm{BKM},\beta}^{\mathcal A}(t).
\end{array}
\label{eq:B-thermal-F-I}
\end{equation}

Substituting Eq.~\eqref{eq:B-thermal-entropy} yields
\begin{equation}
F_{\mathrm{BKM},\beta}^{\mathcal A}(t)
=
2
\int_{-\infty}^{t}
\bigl((\delta_t^\beta h)'(x)\bigr)^2\,
\beta
\left(
1-e^{2\pi(x-t)/\beta}
\right)
dx ,
\label{eq:B-thermal-F-explicit}
\end{equation}
and therefore
\begin{equation}
I_{A,\beta}(t)
=
4
\int_{-\infty}^{t}
\bigl((\delta_t^\beta h)'(x)\bigr)^2\,
\beta
\left(
1-e^{2\pi(x-t)/\beta}
\right)
dx .
\label{eq:B-thermal-I-explicit}
\end{equation}

The derivative in Eqs.~\eqref{eq:B-thermal-F-explicit} and
\eqref{eq:B-thermal-I-explicit} is understood in the entropy form domain
of the thermal half-line.  Equivalently, one may view it as the image
under the quadratic-form map
\begin{equation}
\phi_{t,\beta}(f)
:=
f'\!\restriction_{(-\infty,t)}
\in
L^2\!\left(
(-\infty,t),
\beta
\left(
1-e^{2\pi(x-t)/\beta}
\right)
dx
\right).
\label{eq:B-thermal-form-domain}
\end{equation}
Thus, when \(f=\delta_t^\beta h\), the notation
\((\delta_t^\beta h)'(x)\) denotes
\begin{equation}
(\delta_t^\beta h)'(x)
:=
\phi_{t,\beta}(\delta_t^\beta h)(x).
\label{eq:B-thermal-derivative-convention}
\end{equation}
For ordinary compactly supported test functions this agrees with the
usual derivative.  For the reflected vector
\(\vartheta_t^\beta h\), it denotes the corresponding element of the
thermal entropy form domain.

\bibliographystyle{apsrev4-2}

\end{document}